\documentclass[iop,revtex4,lscape]{emulateapj-rtx4}
\usepackage{multirow}
\usepackage{amsmath}
\usepackage{lscape}

\makeatletter

\newenvironment{inlinefigure}{%
\def\@captype{figure}%
\noindent\begin{minipage}{0.999\linewidth}\begin{center}}
{\end{center}\end{minipage}\smallskip}
\makeatother

\begin{document}

\title{Is There a Maximum Star Formation Rate in High-Redshift Galaxies?\altaffilmark{1,2,3,4}}

\author{
A.~J.~Barger\altaffilmark{5,6,7}, 
L.~L.~Cowie\altaffilmark{7},
C.-C.~Chen\altaffilmark{7,8},
F.~N.~Owen\altaffilmark{9},
W.-H.~Wang\altaffilmark{10},
C.~M.~Casey\altaffilmark{7,11},
N.~Lee\altaffilmark{7},
D.~B.~Sanders\altaffilmark{7},
J.~P. Williams\altaffilmark{7}
}

\altaffiltext{1}{The James Clerk Maxwell Telescope is operated by the Joint 
Astronomy Centre on behalf of the Science and Technology 
Facilities Council of the United Kingdom, the National Research 
Council of Canada, and (until 31 March 2013) the Netherlands Organisation
for Scientific Research.}
\altaffiltext{2}{The National Radio Astronomy Observatory is a facility of 
the National Science Foundation operated under cooperative agreement by 
Associated Universities, Inc.}
\altaffiltext{3}{The Submillimeter Array is a joint project between the
Smithsonian Astrophysical Observatory and the Academia Sinica Institute
of Astronomy and Astrophysics and is funded by the Smithsonian Institution
and the Academia Sinica.}
\altaffiltext{4}{The W.~M.~Keck Observatory is operated as a scientific
partnership among the the California Institute of Technology, the University
of California, and NASA, and was made possible by the generous financial
support of the W.~M.~Keck Foundation.}
\altaffiltext{5}{Department of Astronomy, University of Wisconsin-Madison,
475 N. Charter Street, Madison, WI 53706, USA}
\altaffiltext{6}{Department of Physics and Astronomy, University of Hawaii,
2505 Correa Road, Honolulu, HI 96822, USA}
\altaffiltext{7}{Institute for Astronomy, University of Hawaii,
2680 Woodlawn Drive, Honolulu, HI 96822, USA}
\altaffiltext{8}{Institute for Computational Cosmology, Durham University, 
South Road, Durham DH1 3LE, UK}
\altaffiltext{9}{National Radio Astronomy Observatory, P.O. Box O, 
Socorro, NM 87801, USA}
\altaffiltext{10}{Academia Sinica Institute of Astronomy and Astrophysics, 
P.O. Box 23-141, Taipei 10617, Taiwan}
\altaffiltext{11}{Department of Physics and Astronomy, University of California
at Irvine, 2162 Frederick Reines Hall, Irvine, CA 92697, USA}

\slugcomment{Accepted by The Astrophysical Journal}


\begin{abstract}

We use the James Clerk Maxwell Telescope's SCUBA-2 camera 
to image a 400~arcmin$^2$ area surrounding the GOODS-N field.
The 850~$\mu$m rms noise ranges from a value of 0.49~mJy
in the central region to 3.5~mJy at the outside edge. 
From these data, we construct an 850~$\mu$m source
catalog to 2~mJy  containing 49 sources detected above the $4~\sigma$ level.
We use an ultradeep ($11.5~\mu$Jy at 5$\sigma$) 1.4~GHz image 
obtained with the Karl G. Jansky Very Large Array together with
observations made with the Submillimeter Array to identify
counterparts to the submillimeter galaxies. For most cases of 
multiple radio counterparts, we can identify the correct counterpart
from new and existing Submillimeter Array data.
We have spectroscopic redshifts for 62\% of the radio sources in the
$9'$ radius highest sensitivity region (556/894) and 67\% of the
radio sources in the GOODS-N region (367/543).
We supplement these with a modest number of additional 
photometric redshifts in the GOODS-N region (30).
We measure millimetric redshifts from the radio to submillimeter flux ratios
for the unidentified submillimeter sample,
assuming an Arp~220 spectral energy distribution.
We find a radio flux dependent $K-z$ relation for the radio sources,
which we use to estimate redshifts for the remaining radio sources.
We determine the star formation rates (SFRs) of the submillimeter sources based 
on their radio powers and their submillimeter and find that they agree well.
The radio data are deep enough to detect star-forming galaxies with SFRs
$>2000~M_\odot$~yr$^{-1}$ to $z\sim6$. We find galaxies with 
SFRs up to $\sim6,000~M_\odot$~yr$^{-1}$ over the redshift range
$z=1.5-6$, but we see evidence for a turn-down in the SFR distribution 
function above 2000~$M_\odot$~yr$^{-1}$.

\end{abstract}

\keywords{
cosmology: observations 
--- galaxies: starburst
--- galaxies: active
--- galaxies: evolution
--- galaxies: distances and redshifts 
--- galaxies: high-redshift
}


\section{Introduction}
\label{secintro}

Finding the galaxies that have the highest star formation rates (SFRs)
at high redshifts has been a difficult problem.
Such galaxies cannot be easily picked out in rest-frame
ultraviolet (UV) or optical samples due to their very 
large and highly variable extinctions 
(e.g., Bouwens et al.\ 2009; Reddy et al.\ 2012), 
and though they can be found in far-infrared (FIR) or submillimeter 
selected galaxy samples, the poor resolution of single-dish submillimeter
telescopes makes their interpretation
complex. In particular, recent follow-up surveys
with submillimeter interferometers have shown that at least
some of the brightest submillimeter galaxies (SMGs) are blends of
fainter sources with lower SFRs (e.g., Wang et al.\ 2011;
Smol\v{c}i\'{c} et al.\ 2012;
Barger et al.\ 2012; Hodge et al.\ 2013b). In fact, Karim et al.\ (2013)
suggested that almost all bright ($>9$~mJy) SMGs are blends
and that there is a natural upper limit of $\sim 1000~M_\sun$ yr$^{-1}$
on the SFRs of galaxies.

One thing all SMG studies agree on is that there is a large 
fraction of cosmic star formation hidden by dust 
(e.g., Barger et al.\ 2000, 2012; Lagache et al.\ 2005;
Chapman et al.\ 2005; Wang et al.\ 2006; 
Serjeant et al.\ 2008; Wardlow et al.\ 2011; Casey et al.\ 2013), most of
which is occurring in the most massively star-forming galaxies in the universe.
Thus, the construction of a complete picture
of galaxy evolution requires a full understanding of galaxies 
at both optical and FIR/submillimeter wavelengths. However, in order to 
develop this understanding of the dusty universe, we need large, 
uniformly selected samples with well determined star-forming 
properties.

In this paper, we work towards this goal using a combination 
of powerful new data on the heavily studied Great Observatories Origins Deep 
Survey-North (GOODS-N; Giavalisco et al.\ 2004)/{\em Chandra\/} Deep Field-North 
(CDF-N; Alexander et al.\ 2003) field.
We begin with a new, uniformly selected sample of 850~$\mu$m galaxies
observed with the SCUBA-2 camera (Holland et al.\ 2013)
on the 15~m James Clerk Maxwell Telescope (JCMT). 
Then, using a combination of extremely deep Karl G. Jansky Very Large 
Array (VLA) 1.4~GHz observations 
(F. Owen 2014, in preparation; hereafter, Owen14) and
high-resolution Submillimeter Array (SMA; Ho et al.\ 2004) 860~$\mu$m 
observations, we analyze the fraction of SMGs that are single sources 
and estimate the SFR distribution function for the most massively
star-forming galaxies in the universe.

The principal underpinning of this work is the 
new submillimeter imaging capability provided by SCUBA-2.
SCUBA-2 covers 16 times the area of SCUBA (Holland et al.\ 1999)
and has a mapping speed that is considerably faster than SCUBA,
which means that large samples of SMGs
can be obtained in only a few nights of observations. Previous
samples of SMGs in the GOODS-N/CDF-N field were based on mosaics
of SCUBA fields that only partially covered
the area and that had widely varying sensitivities. SCUBA-2 enables
the development of uniform, deep SMG samples over the entire field.

In order to construct the SFR distribution function,
we need to determine how many of the SCUBA-2 selected 
sources are multiples, where the observed flux is a blend of two or 
more individual galaxies. High spatial resolution
follow-up of bright SMGs at the same wavelength is now possible 
with the SMA or the Atacama Large Millimeter/submillimeter Array (ALMA), 
but the level of multiplicity is still somewhat controversial, 
particularly for the brightest SMGs (see Chen et al.\ 2013 for a full discussion).

As an illustration of the different results that have been obtained,
Karim et al.\ (2013) found that all of the brightest ($>12$~mJy) sources in their 
LABOCA Extended {\em Chandra\/} 
Deep Field-South Submillimeter Survey (LESS) 
that they successfully detected with their targeted ALMA observations
were composed of emission from multiple fainter SMGs, each with 
870~$\mu$m fluxes of $\lesssim9$~mJy.   
(Note that of the 88 ``best" ALMA maps used in their analysis, 19 contained no
$>3.5\sigma$ detections.)
They also did not find any ALMA sources with fluxes $>9$~mJy.
In contrast, Barger et al.\ (2012) confirmed with the SMA three single sources in the 
GOODS-N field with fluxes $>9$~mJy, two of which had fluxes $\gtrsim12$~mJy.
The differences may be partly explainable as a purely observational blending effect 
due to the different beam sizes of the single-dish submillimeter telescopes 
used to construct the SMG samples ($14''$ for SCUBA versus $19.2''$ for LABOCA). 
However, this emphasizes the importance
of determining the multiplicity level for the specific sample that is being used.

In this paper, we approach the multiplicity issue in two ways. 
Our first approach is the most direct:  submillimeter interferometric
imaging of the SCUBA-2 selected SMGs. 
Such follow-up observations can localize the submillimeter emission 
extremely accurately and allow the determination of whether the SMG is
a single source or a blend
(e.g., Iono et al.\ 2006; Wang et al.\ 2007, 2011; 
Younger et al.\ 2008a,b; Cowie et al.\ 2009; Hatsukade et al.\ 2010;
Knudsen et al.\ 2010; Chen et al.\ 2011; Barger et al.\ 2012; Karim et al.\ 2013;
Hodge et al.\ 2013b). We have used the SMA to measure the properties
of a very large fraction of the SMGs in the SCUBA-2 sample, including many of
the brightest ones.

Our second approach is to use 1.4~GHz observations to identify the
counterparts to the SMGs. Historically, this approach has been less than 
ideal, because it introduced a strong bias against high-redshift SMGs due to the 
positive $K$-correction of the radio synchrotron emission and the negative 
$K$-correction of the submillimeter thermal dust emission. 
However, with the upgraded VLA, 1.4~GHz images (Owen14) 
are now deep enough to find counterparts to nearly all of the SCUBA
(Barger et al.\ 2012) or SCUBA-2 (this paper) sources, removing the radio bias. 
Thus, for most of the SCUBA-2 galaxies without SMA data, we can
identify single radio sources as their counterparts.

It is also possible to identify
high SFR galaxies directly in the radio, though this is complicated 
by the fact that  many high radio power sources are AGNs rather than star formers.
Radio galaxies pick out high-mass galaxies, as can be seen from
the ``$K-z$ relation'', a well-known tight 
correlation between the $K$-band magnitudes of 
radio host galaxies and their redshifts that was discovered by Lilly \& Longair (1984) 
using the bright 3CRR survey (Laing et al.\ 1983) and
confirmed with lower power radio surveys 
(e.g., Eales et al.\ 1997 using the 6CE;
Jarvis et al.\ 2001 using the 6C*;
Lacy et al.\ 2000 using the 7C-III). 
At an order of magnitude even fainter than the above surveys,
De Breuck et al.\ (2002) found that their $S_{\rm 1.4~GHz}>10$~mJy 
sample of ultra-steep-spectrum radio sources
also followed the $K-z$ relation
and traced the bright $K$-band envelope of field galaxies out to 
$z\lesssim1$ before becoming $\gtrsim2$~mag brighter at higher redshifts.
This led them to conclude that the radio galaxies were pinpointing the most  
massive systems at all redshifts.

Clearly, the $K-z$ relation has important implications for the formation of 
massive galaxies. 
Willott et al.\ (2003; hereafter, Willott03) combined the 7C Redshift Survey
(7CRS) with the 3CRR, 6CE, and 6C* samples
and found that the shape of the $K-z$ relation
is closely approximated by the magnitude-redshift relation for an 
elliptical galaxy that formed at $z=10$. 
Rocca-Volmerange et al.\ (2004) modeled the $K-z$
relation using magnitudes computed for a fixed baryonic
mass galaxy with an elliptical (fast conversion of gas to
stars) star formation history starting at high redshift ($z=10$).
In their interpretation, the most powerful radio sources are
in galaxies with baryonic masses of $\sim10^{12}~M_\sun$,
while lower radio power sources are in galaxies with slightly smaller
baryonic masses. Rocca-Volmerange et al.'s  adopted elliptical star 
formation timescale for this model (1~Gyr) would imply SFRs
$\sim1000~M_\sun$~yr$^{-1}$ in all luminous radio galaxies 
at $z\gtrsim4$.

The primary science goals of the present paper are to estimate for the 
most massively star-forming galaxies in the universe
(1) the highest SFRs, (2) the distribution of SFRs, and (3) the contribution 
to the universal star formation history and how that contribution
compares to the contribution from extinction-corrected UV selected samples.
The structure of the paper is as follows. In Section~\ref{secdata}, we discuss 
the GOODS-N/CDF-N imaging data sets that we use, 
including SCUBA-2, radio, SMA, $K_s$, and X-ray.
We construct both a $>4\sigma$ SCUBA-2 850~$\mu$m 
catalog and an SMA 860~$\mu$m catalog
using all available data. We identify radio counterparts
to the SMGs and give their $K_s$ magnitudes, as well as their spectroscopic
and photometric redshifts, where available. We measure millimetric redshifts 
from the radio to submillimeter flux ratios assuming an Arp~220 spectral
energy distribution (SED).
In Section~\ref{seczdist}, we show the redshift distribution of the radio sample.
In Section~\ref{seckz}, we present the $K-z$ relation for our sample and
use it to estimate additional redshifts.
In Section~\ref{secradiopower}, we focus on the high radio power sources in the
sample and how the SMGs are drawn from this population.
We also describe our conversions of the 1.4~GHz powers and 
submillimeter fluxes into SFRs.
In Section~\ref{secsfh}, we use our SCUBA-2 sample to determine the
SFR distribution function at $z=1.5-6$ and to construct the star formation history,
which we compare with the history determined from extinction-corrected UV samples.
In Section~\ref{secdisc}, we discuss the implications of our results, and
in Section~\ref{secsum}, we summarize the paper.

We adopt the AB magnitude system for the optical and NIR
photometry, and we assume the Wilkinson Microwave
Anisotropy Probe cosmology of $H_0=70.5$~km~s$^{-1}$~Mpc$^{-1}$,
$\Omega_{\rm M}=0.27$, and $\Omega_\Lambda=0.73$ 
(Larson et al.\ 2011) throughout.

\section{Data}
\label{secdata}

\subsection{SCUBA-2 Imaging}
\label{secscuba2obs}

We obtained 25.4~hr of observations on the CDF-N with SCUBA-2 on the
JCMT during observing runs in
2012 and 2013. The data were obtained using a mixture of scanning
modes and under a variety of weather conditions.
Using the CV Daisy scanning mode
(detailed information about the SCUBA-2 scan patterns can be found in
Holland et al.\ 2013),
we obtained a 2.2~hr observation
in band 1 weather (225~GHz opacity $<0.05$) and a 16.5~hr observation
in band 2 weather (225~GHz opacity $\sim0.05-0.08$).
We also obtained a 6.7~hr observation in band 2 weather
using the pong-900 scanning mode. While SCUBA-2 observes
at both 450~$\mu$m and 850~$\mu$m simultaneously,  
there are too few sources directly detected at 450~$\mu$m in our data
to be interesting. Thus, we only use the 850~$\mu$m data in our subsequent 
analysis.

In terms of the two scanning modes,
CV Daisy is optimal for going deep on small areas
($<4'$ radius), while pong-900 provides a uniform rms over a large area ($<12'$ radius). 
In Figure~\ref{sma_detected}, we compare the CV Daisy field (dark green shading 
indicates the area with the highest sensitivity, and light green shading indicates the area 
where the rms noise is less than 4 times the central sensitivity) with the pong-900 field
(yellow shading). To illustrate the size of the SCUBA-2 images, the black rectangle shows
the GOODS-N {\em HST\/} ACS coverage.

\begin{inlinefigure}
\vskip 0.5cm
\centerline{\includegraphics[width=3.2in,angle=180]{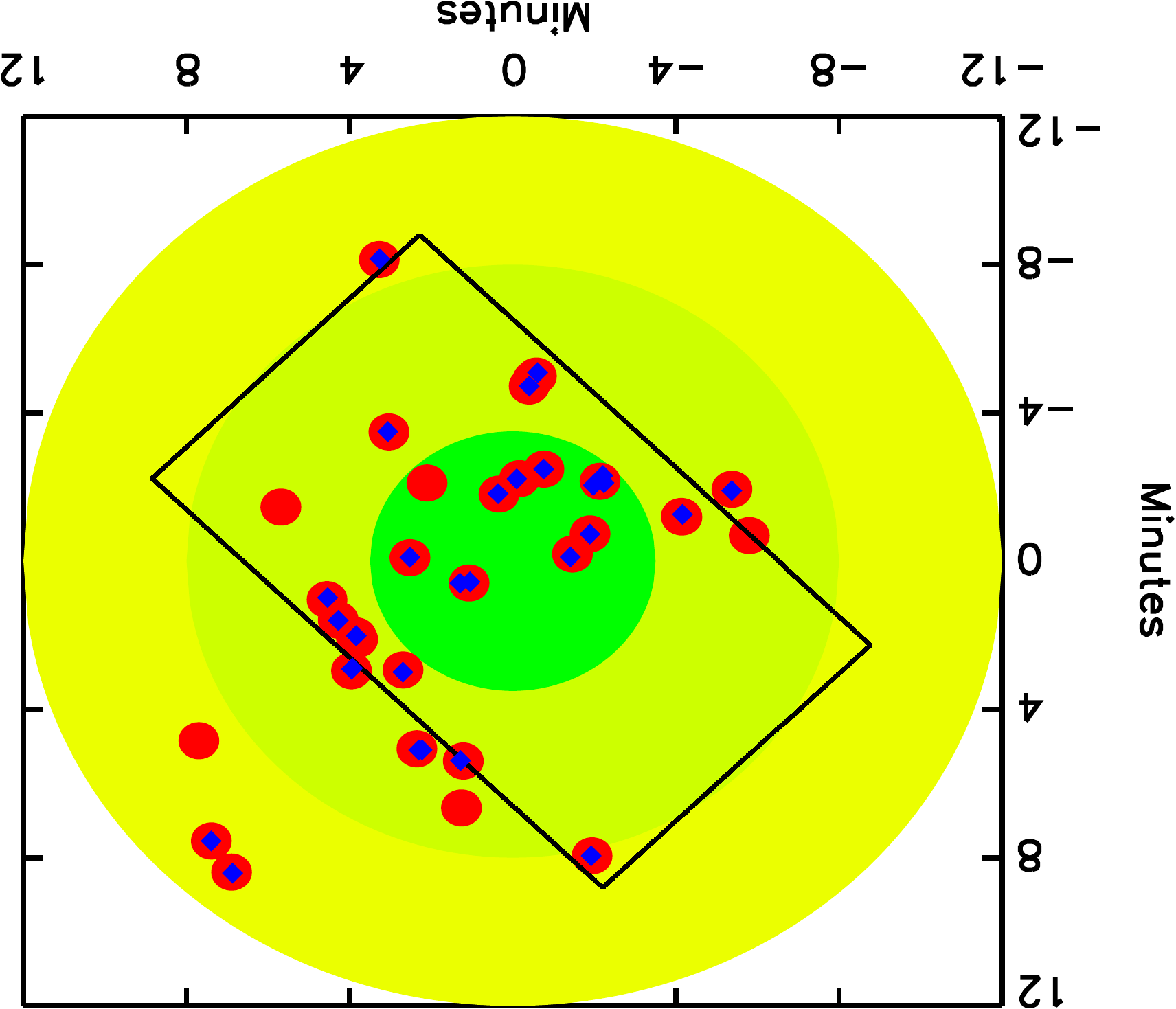}}
\caption{
GOODS-N/CDF-N
SCUBA-2 field obtained with the CV Daisy scanning mode (dark green --- area
with the highest sensitivity; light green --- area where the rms noise is $<4\times$
the central sensitivity) and the pong-900 scanning mode (yellow shading).
The black rectangle shows the GOODS-N {\em HST\/} ACS field.
The red circles ($24''$ radius to show the area where the rms noise is 
$<3\times$ the central sensitivity) denote the 28 SMA fields in the region
(see Section~\ref{secsma}); note that some of the SMA fields overlap.
The blue diamonds mark the $>4\sigma$ SMA detections in the SMA fields.
\label{sma_detected}
}
\end{inlinefigure}

We took darks and flat fields at
the beginning and end of each scan. We did Skydips at least twice per night in order
to calculate the opacity factors.
We reduced the data using the Dynamic Iterative 
Map-Maker (DIMM) in the SMURF 
package from the STARLINK software developed by the Joint Astronomy Centre.
We calibrated the fluxes using the standard Flux Conversion Factor for 850\,$\mu$m 
of 537\,Jy\,pW$^{-1}$. The relative calibration accuracy is expected to be
stable and good to 5\% at 850\,$\mu$m (Dempsey et al.\ 2013).
For more details on the GOODS-N/CDF-N SCUBA-2 data reduction and flux 
calibration, we refer the reader to Chen et al.\ (2013).

We combined the pong-900 map with only the portion of the CV Daisy map contained within
the dark and light green shaded regions of Figure~\ref{sma_detected} to avoid damaging the
sensitivity of the pong-900 map further out where the Daisy coverage is sparse.
Nearly all of the SMGs are expected to be compact relative to the
beam size of the JCMT at 850~$\mu$m.  In order to increase the 
detectability of these point sources, we applied a matched-filter
to our maps, which is a maximum likelihood estimator of the source strength 
(e.g., Serjeant et al.\ 2003a). The point spread function (PSF) for the matched-filter 
algorithm should ideally be a Gaussian normalized 
to a peak of unity with full-width half-maximum (FWHM) equal to the JCMT beam size
($14''$ at 850\,$\mu$m). However, the map produced from DIMM 
typically has low spatial frequency structures that
need to be subtracted off before source extraction can be done. 
Thus, before running the matched-filter, we convolved the map with a broad
Gaussian ($30''$ FWHM) normalized to a sum of unity, and we subtracted this 
convolved map from the original map. The resulting PSF is a Gaussian with a
broader Gaussian subtracted off, giving a Mexican hat-like wavelet. 
We show the area versus rms noise in the matched-filter 
850~$\mu$m image in Figure~\ref{area}.

\begin{inlinefigure}
\vskip 0.5cm
\centerline{\includegraphics[width=3.2in,angle=180]{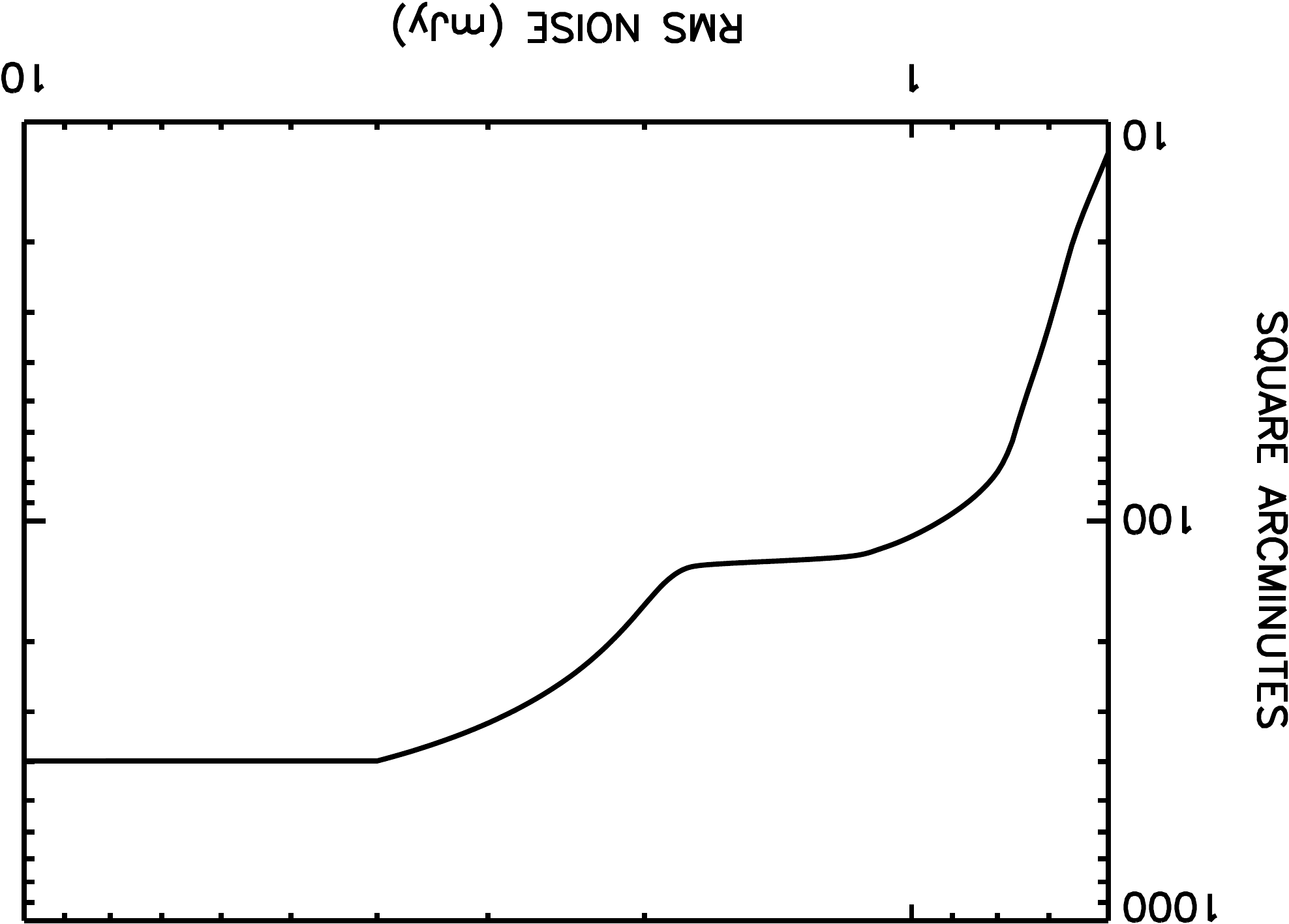}}
\caption{
Area vs. rms noise of the combined GOODS-N/CDF-N SCUBA-2 pong-900 
and CV Daisy 850~$\mu$m maps.
\label{area}
}
\end{inlinefigure}

We first extracted sources having a peak
S/N greater than 3.0. We did this by finding the maximum pixel in the matched-filter 
image. We then used the position and flux of the source at that peak
to subtract a PSF that we centered at that
position and scaled to match the flux. We iterated this process of identifying
and removing sources until there were no remaining peak S/N values greater than 3.0.
We treated the $>3\sigma$ peaks as the preliminary catalog. In forming the 
final catalog, we kept every $>4\sigma$ source in the preliminary catalog. 

In Figure~\ref{scuba2}, we mark on the SCUBA-2 map the 
850~$\mu$m $>4\sigma$ catalog sources (large circles). 
Hereafter, we refer to this as our SCUBA-2 sample.
In Table~1, we present the sample in tabular form. 
In Column~1, we give our CDFN name for 
the source. Where a GOODS~850 number from Wang et al.\ (2004)
or a GN number from Pope et al.\ (2005) exists, we give that name in parentheses.
In Columns~2 and 3, we give the J2000 right ascensions and declinations
from the SCUBA-2 data.
We have ordered the catalog by decreasing 850~$\mu$m flux, which we give in 
Column~4. In Columns 5 and 6, we list the 850~$\mu$m $1\sigma$ errors and signal-to-noise
ratios, respectively. For the sources detected at 860~$\mu$m with the SMA, in Column~7, 
we give the SMA fluxes with the $1\sigma$ errors in parentheses. 
In Column~8, we give the 1.4~GHz
fluxes with the $1\sigma$ errors in parentheses, and in Columns~9 and 10, 
we give the J2000 right ascensions and declinations from the radio data.
Where more than one radio source lies within the SCUBA-2 beam, and where
there are no SMA observations to determine the true radio counterpart, we list
both radio sources. We list the $K_s$ magnitudes in Column~11, and we give
the spectroscopic (including CO), photometric, and millimetric redshifts in 
Columns~$12-14$.

\begin{inlinefigure}
\centerline{\includegraphics[width=3.5in]{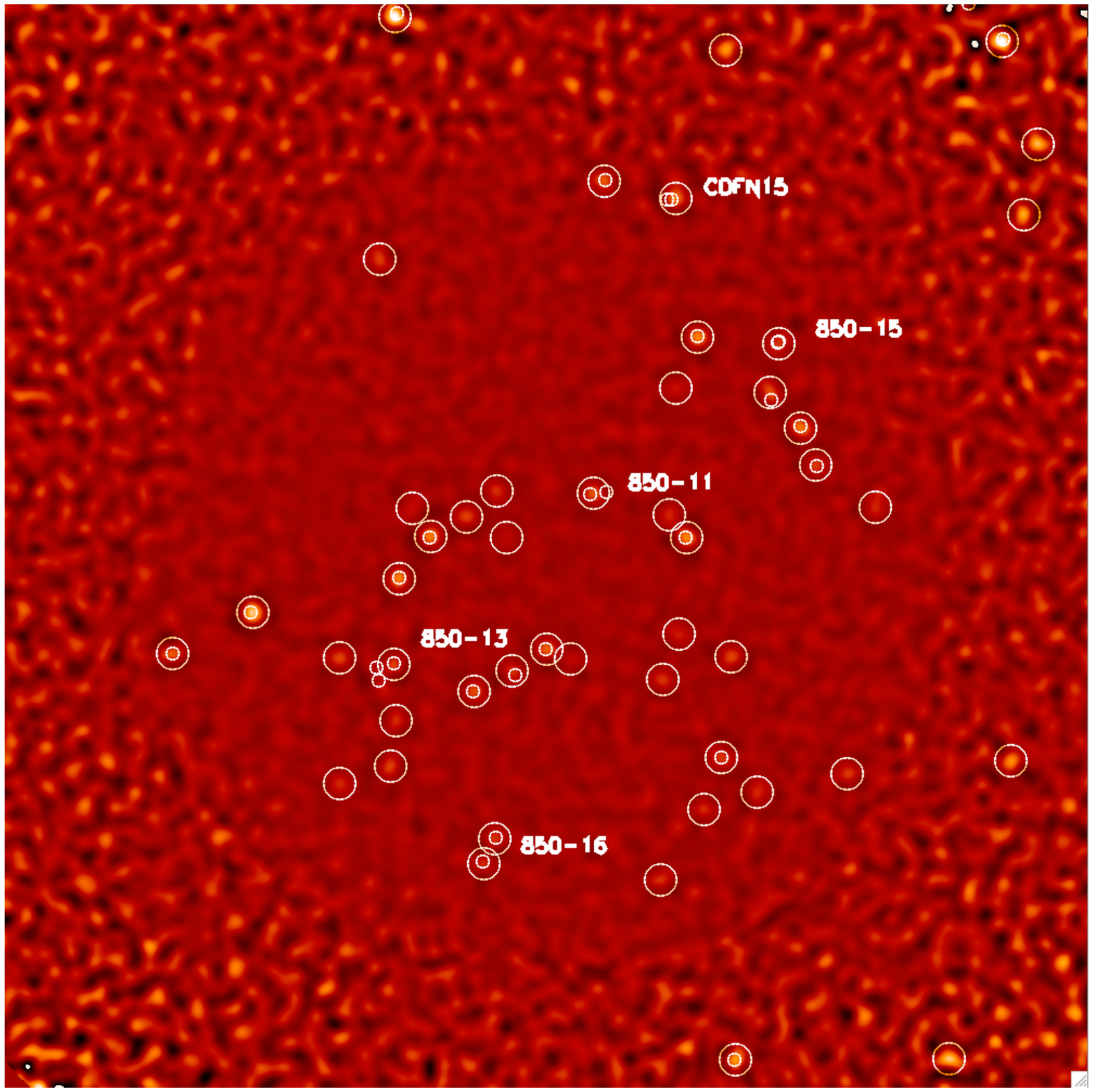}}
\vskip -1.35cm
\caption{
The 850~$\mu$m $>4\sigma$ catalog sources marked on the SCUBA-2 map 
(large circles). The portion of the map shown is $16.7'$ on a side.  
The SMA detections (see Section~\ref{secsma}) are shown with smaller circles.
Sources that were single detections in the SCUBA survey of 
Wang et al.\ (2004) but then were observed to
split into multiples in the SMA data are labeled with the Wang
et al.\ name. (Note that the GOODS~850-15 multiple is very close; see
Figure~14 in Barger et al.\ 2012.)
In some cases (GOODS~850-13 and GOODS~850-16),
the new SCUBA-2 data also separate the sources. (Note, however, that two
of the three SMA sources making up GOODS~850-13 lie below the SCUBA-2 
detection threshold.)
\label{scuba2}
}
\end{inlinefigure}

\subsection{Radio Imaging}
\label{secradioobs}

Owen14 constructed a catalog of 1.4~GHz sources detected in the VLA image of the
GOODS-N/CDF-N field. The image covers a $40'$ diameter region with an effective resolution of
$1\farcs8$. The absolute radio positions are known to $0\farcs1-0\farcs2$ rms.
The highest sensitivity region is about $9'$ in radius, producing a relatively uniform radio map
with an rms of 2.3~$\mu$Jy. We refer to this region as the full field in the rest of the
paper. There are 894 distinct radio sources in this region, excluding sources that appear
to be parts of other sources.

\subsection{SMA Imaging}
\label{secsma}

There are 28 fields within $\pm9\farcm5$ of the GOODS-N center that have been
observed with the SMA. Most of these observations were targeted on SCUBA
850~$\mu$m sources or, more recently, SCUBA-2 850~$\mu$m sources, primarily 
by our own group (24 of the 28 fields).
Twelve of these were presented in Wang et al.\ (2011)
and Barger et al.\ (2012), while Chen et al.\ (2013) used the full sample to analyze
the effects of multiplicity on the number counts.
We show all 28 fields (some of which are overlapped) in Figure~\ref{sma_detected} 
as the red circles. Taken together, they cover $\sim14$~arcmin$^2$.

There are 16 images not already analyzed in Wang et al.\ (2011) and
Barger et al.\ (2012). For one of these images (GN20), 
we simply adopted the SMA 890~$\mu$m
flux and error presented in Iono et al.\ (2006).
For the others, we calibrated and inspected the SMA 860~$\mu$m data using the 
IDL-based Caltech package MIR modified for the SMA, as in our previous work.

Considering only the regions in each image where the noise was less than four times the 
minimum noise, we searched all of the SMA images (except for the one containing GN20)
for sources detected above the $4\sigma$ threshold. Including GN20,
there are 29 such sources, which we mark in Figure~\ref{sma_detected} 
with blue diamonds. These sources were all the intended targets of the SMA observations.
Apart from the multiples, we found no serendipitous sources in the fields. 
Chen et al.\ (2013) compared the SMA and SCUBA-2 fluxes for the sources and found that most
agree statistically (see their Figure~10).

We compare the SMA (small circles)
and SCUBA-2 (large circles) detections in Figure~\ref{scuba2}.
All of the SCUBA-2 sources observed with the SMA
were detected.
We find that only three of the SCUBA-2 sources, CDFN16 (GOODS~850-11), CDFN37
(GOODS~850-15), and CDFN15,
have multiple counterparts in the SMA images. Two previous SCUBA
sources (GOODS~850-13 and GOODS~850-16) that were blends of SMA sources 
are separated in the SCUBA-2 data into individual sources. (Note, however, that two
of the three SMA sources making up GOODS~850-13 lie below the SCUBA-2 detection threshold.)

\begin{inlinefigure}
\centerline{\includegraphics[width=4.5in,angle=180]{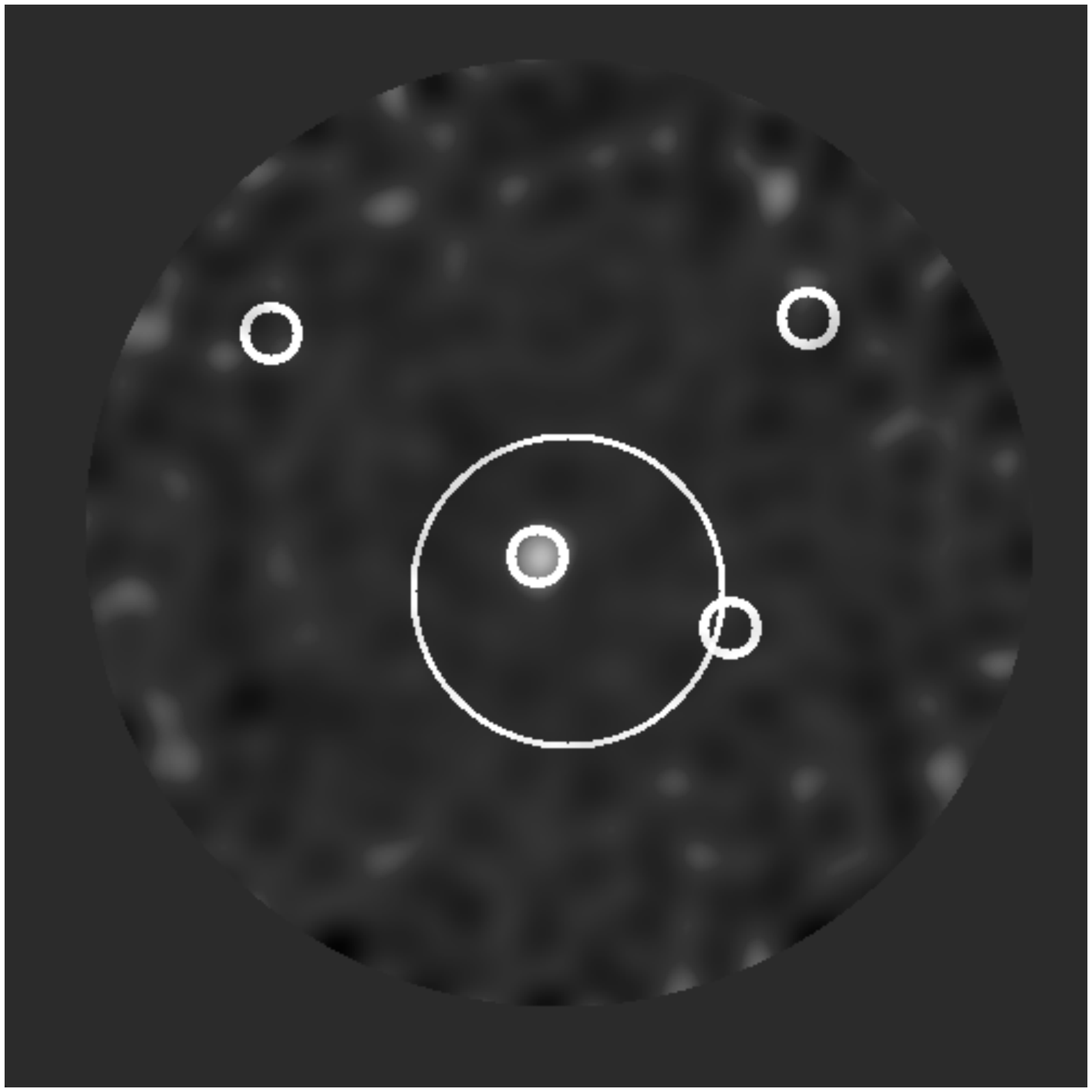}}
\vskip -0.5cm
\caption{SMA image of CDFN8 with a measured
860~$\mu$m flux of $11.5\pm0.7$~mJy from the SMA and an 850~$\mu$m
flux of $9.5\pm2.3$~mJy from SCUBA-2. 
The large circle shows the
SCUBA-2 beam centered at the SCUBA-2 position. The small circles
show the positions of 1.4~GHz sources in the field. The submillimeter
source is a single unresolved source at the position of a 1.4~GHz source
with a flux of $34.2\pm2.9$~$\mu$Jy. None of the other three radio
sources in the field have a significant submillimeter flux.
\label{sma_example}
}
\end{inlinefigure}

With our accurate SMA positions, we can unambiguously determine the radio 
counterparts to the SMGs.  Indeed, we find radio counterparts above the  
1.4~GHz threshold of $\sim 11.5~\mu$Jy ($5\sigma$) for all of the SMA sources,
except CDFN15a and CDFN15b.
We show one example in Figure~\ref{sma_example}. There are four radio 
sources in the SMA field (small circles), but only one of them is the clear
counterpart to the SMA source.  The other three have no significant submillimeter flux. 

We tested the astrometric accuracy of the SMA observations 
relative to the Owen14 1.4~GHz sample and found
that they are in perfect astrometric agreement.
The dispersion in the positional offsets is $0\farcs5$.

In Table~2, we summarize the properties of the 29 SMA sources.
In Column~1, we give the name from the literature or, for new sources,
the name from our SCUBA-2 catalog (Table~1).
In Columns~2 and 3, we give the J2000 right ascensions and declinations
for each source as measured from the SMA data.
In Column~4, we list the SMA $860~\mu$m fluxes and $1\sigma$ errors.
In Column~5, we give the references for the SMA data.
In Column~6, we list the  1.4~GHz  fluxes and $1\sigma$ errors from Owen14.
These are peak fluxes when the peak flux equals the extended
flux and extended fluxes otherwise.
In Column~7, we give the spectroscopic redshifts as found in the literature.
In Column~8, we give the references for those spectroscopic measurements.

\subsection{Near-Infrared Imaging}
\label{secnirobs}

Wang et al.\ (2010) constructed a $K_s$ catalog of the GOODS-N/CDF-N field, which they publicly
released along with the extremely deep Canada-France-Hawaii Telescope (CFHT) $K_s$
image from which the catalog was extracted.  In the GOODS-N region, the image has a 
$1\sigma$ depth of $0.12~\mu$Jy.
We measured $3''$ diameter aperture $K_s$ magnitudes corrected to total magnitudes
at the positions of the radio sources using the $K_s$ image. For sources
brighter than $K_s=19$, we used an isophotal magnitude computed using an
aperture corresponding to 1$\%$ of the central surface brightness
in the galaxy.

\subsection{X-ray Imaging}
\label{secxrobs}

Alexander et al.\ (2003) presented the 2~Ms X-ray image of the 
CDF-N, which they aligned with the Richards (2000) 
radio image. Near the aim point, the X-ray data reach limiting fluxes of
$f_{\rm 2-8~keV}\approx 1.4\times 10^{-16}$ and 
$f_{\rm 0.5-2~keV}\approx 1.5\times 10^{-17}$~erg~cm$^{-2}$~s$^{-1}$. 
We assume a conservative $L_X>10^{42}$~erg~s$^{-1}$ as the threshold for a source
to be classified as an X-ray active galactic nucleus (AGN) on energetic grounds 
(Zezas et al.\ 1998; Moran et al.\ 1999), 
and we assume $L_X>10^{44}$~erg~s$^{-1}$ as the threshold for a source to be
classified as an X-ray quasar.

\subsection{Spectroscopic Redshifts}
\label{secspecoobs}

Many redshifts have been obtained of galaxies in the GOODS-N/CDF-N field
using either the Low-Resolution Imaging Spectrograph (LRIS; Oke et al.\ 1995) 
on the Keck~I 10~m telescope
or the large-format DEep Imaging Multi-Object Spectrograph (DEIMOS; Faber et al.\ 2003) 
on the Keck II 10~m telescope. These include large magnitude-selected samples
(Cohen et al.\ 2000; Cowie et al.\ 2004b; Wirth et al.\ 2004; 
Barger et al.\ 2008; Cooper et al.\ 2011) or targeted
samples looking for interesting galaxy populations 
(Reddy et al.\ 2006; Chapman et al.\ 2003, 2004a, 2005; Swinbank et al.\ 2004;
Treu et al.\ 2005; Barger et al.\ 2002, 2003, 2005, 2007; Trouille et al.\ 2008).
There are also a small number of CO redshifts that have been measured for SMGs 
in the region (Daddi et al.\ 2009a,b; Bothwell et al.\ 2010; Walter et al.\ 2012).

We targeted new radio sources detected in the Owen14 catalog
during DEIMOS runs in 2012 and 2013. 
We used the 600~line~mm$^{-1}$ grating, giving a resolution of 3.5~\AA\ and a wavelength 
coverage of 5300~\AA. The spectra were centered at an average wavelength of 7200~\AA, 
although the exact wavelength range for each spectrum depends on the slit position. 
Each $\sim 1$~hr exposure was broken into three subsets, with the objects stepped along 
the slit by $1\farcs5$ in each direction. Unidentified sources were continuously re-observed,
giving maximum exposure times of up to 7~hr. We reduced the spectra in the same way 
as with previous LRIS spectra (Cowie et al.\ 1996). We only used spectra that could be  
identified confidently based on multiple emission and/or absorption lines. We identified
a number of spectra using the doublet structure of the [OII] 3727~\AA\ line, which is 
resolved in the spectra.

We also searched the infrared grism spectra obtained by P.I.~B.~Weiner 
({\em HST\/} Proposal ID \#11600)
using the G140 grism on the WFC3 camera on {\em HST}. We formed the
spectra of 5709 galaxies with F140W $< 24.5$ and identified
607, mostly using the [OIII]$\lambda$5007 doublet and
H$\beta$ or H$\alpha$. The galaxies primarily lie in the redshift
interval $z=0.8-2.3$. A full catalog will be given in future work. 
Of the identified sources, 107 are also radio sources; however, only
2 had not previously been identified from the optical spectra.

556 (62$\%$) of the 894 distinct radio sources in the full field
have secure spectroscopic redshifts either from the literature or from our 
targeted spectroscopy of the sample. 
In the GOODS-N region (here defined as
the region that is well covered by the {\em HST\/} ACS observations
of Giavalisco et al.\ 2004),
367 (67\%) of 543 radio sources have spectroscopic
redshifts. These spectroscopic identifications primarily
come from UV/optical or NIR spectroscopy, but they also contain
the small number of sources with CO spectroscopic redshifts.

\subsection{Photometric Redshifts}
\label{secphotobs}

Photometric redshifts can extend the spectroscopically identified sample 
and provide a check on the spectroscopic redshifts.
Berta et al.\ (2011) compiled a multiwavelength catalog of sources
in the GOODS-N field and computed photometric
redshifts using the EAZY code of Brammer et al.\ (2008). 

In comparing the spectroscopic and photometric redshifts,
we only consider the spectroscopically identified radio sources 
in the spectroscopically well-covered area of the GOODS-N field
having a photometric redshift quality flag $Q_{z}<2$.
We also eliminate photometric redshifts $z\le0.06$, which are invariably
misidentifications of blue, higher redshift galaxies, and we restrict to galaxies
with {\em HST\/} ACS F850LP~$<25$ (Giavalisco et al.\ 2004).

\vskip 0.5cm
\begin{inlinefigure}
\centerline{\includegraphics[width=3.2in,angle=0]{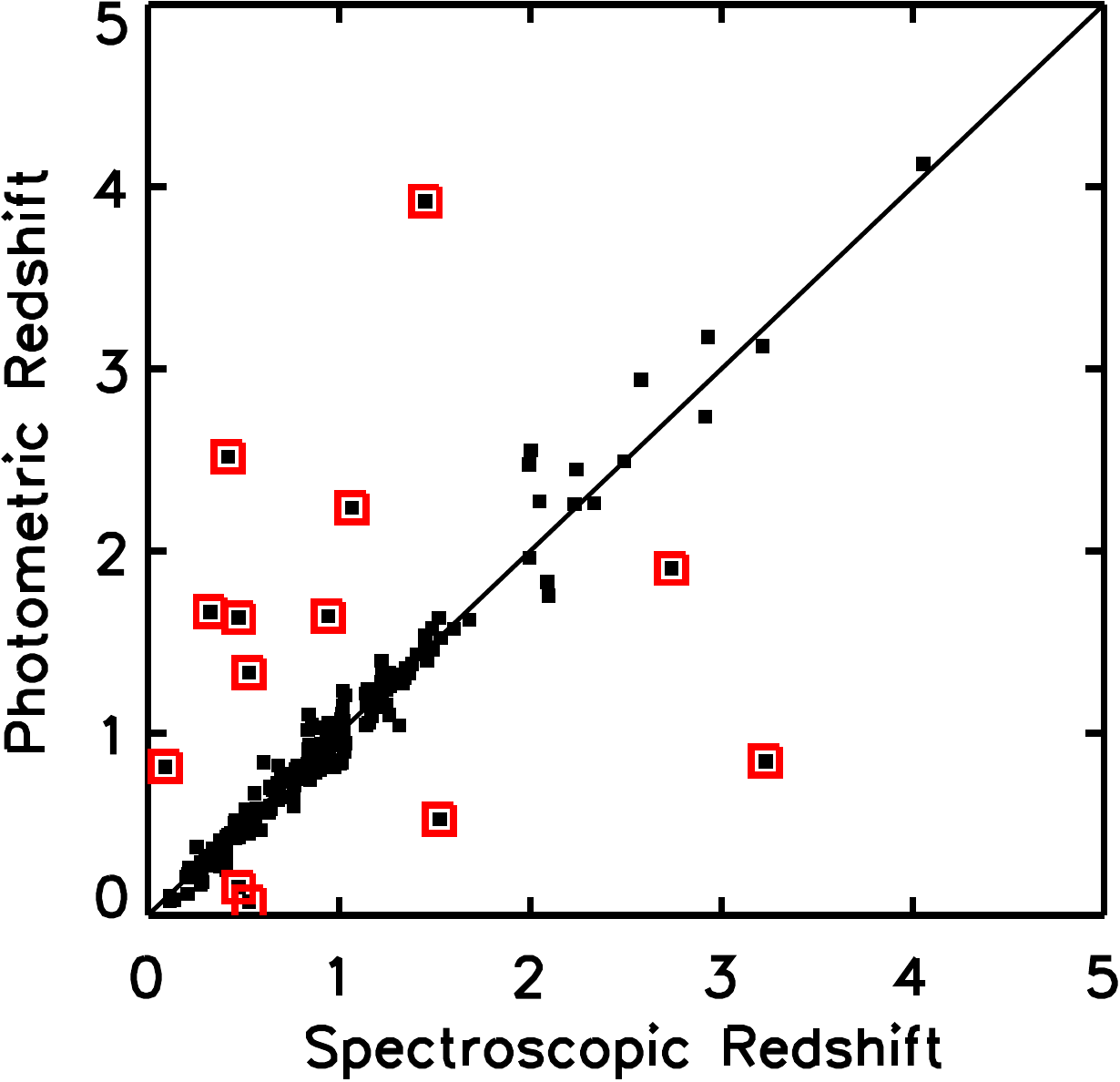}}
\caption{Comparison of spectroscopic redshifts with photometric redshifts 
from Berta et al.\ (2011). Only spectroscopically identified
radio sources lying in the spectroscopically well-covered area of the GOODS-N 
field and having F850LP~$<25$, a photometric redshift quality flag of $Q_{z}<2$, 
and $z>0.06$ are shown. Of the 303 objects, 13 have serious discrepancies. 
These are marked with the red squares.
\label{photspec}}
\end{inlinefigure}

We show a comparison of the photometric and spectroscopic redshifts
in Figure~\ref{photspec}. Only 13 (red squares)
of the 303 spectroscopic redshifts have 
strong disagreements with the corresponding photometric redshifts.
This number is a strong function of the quality flag.
For $Q_{z}=1$, we find 5 strongly discrepant sources out of 
243 radio sources, while for $Q_{z}=3$, this rises to 19 out of 323.
Thus, we adopt $Q_{z}<2$ to maximize the number of included
sources while not allowing too high of an error rate. We inspected
all 13 discrepant sources individually to confirm both the
galaxy identification and the spectral redshift measurement. We
concluded in all cases that the spectroscopic identification
was reasonable. In some cases the photometric redshift may
have been contaminated by blending of two distinct galaxies,
while in other cases strong emission lines in the spectrum
may have perturbed the photometric redshift estimate. However,
there were some cases where we could not find an obvious explanation 
for the discrepancy.

Rafferty et al.\ (2011) determined photometric redshifts over the full
field. Since these are based on more limited photometric information,
we do not use them in our subsequent analysis. However, we include
these in the photometric redshift column of Table~1 (marked with
an (R)) where no photometric redshift is available from Berta et
al.\ (2011).

\subsection{Millimetric Redshifts}
\label{secmillobs}

In Barger et al.\ (2012), we plotted the 1.4~GHz to 860~$\mu$m flux ratio 
versus $1+z$ for the 16 SMGs in our SMA sample with spectroscopic 
redshifts. We found that the Barger et al.\ (2000) Arp~220-based model 
agreed reasonably well with a power law fit over the observed spectroscopic 
redshift range. We therefore adopt this relation
(Equation~5 of Barger et al.\ 2000) to measure millimetric redshifts
(see also Carilli \& Yun 1999) for the SMGs in the SMA sample of Table~2. 
In Figure~\ref{figzmilli}, we compare these
millimetric redshifts with the spectroscopic redshifts for the sources in
the SMA sample with spectroscopic redshifts 
(see Table~2; black squares). The agreement is very good. 
For the SMGs without spectroscopic redshifts (blue diamonds), we
use the millimetric redshifts on both axes. We mark X-ray AGNs with red
small squares.

\vskip 0.5cm
\begin{inlinefigure}
\centerline{\includegraphics[width=3.2in,angle=0]{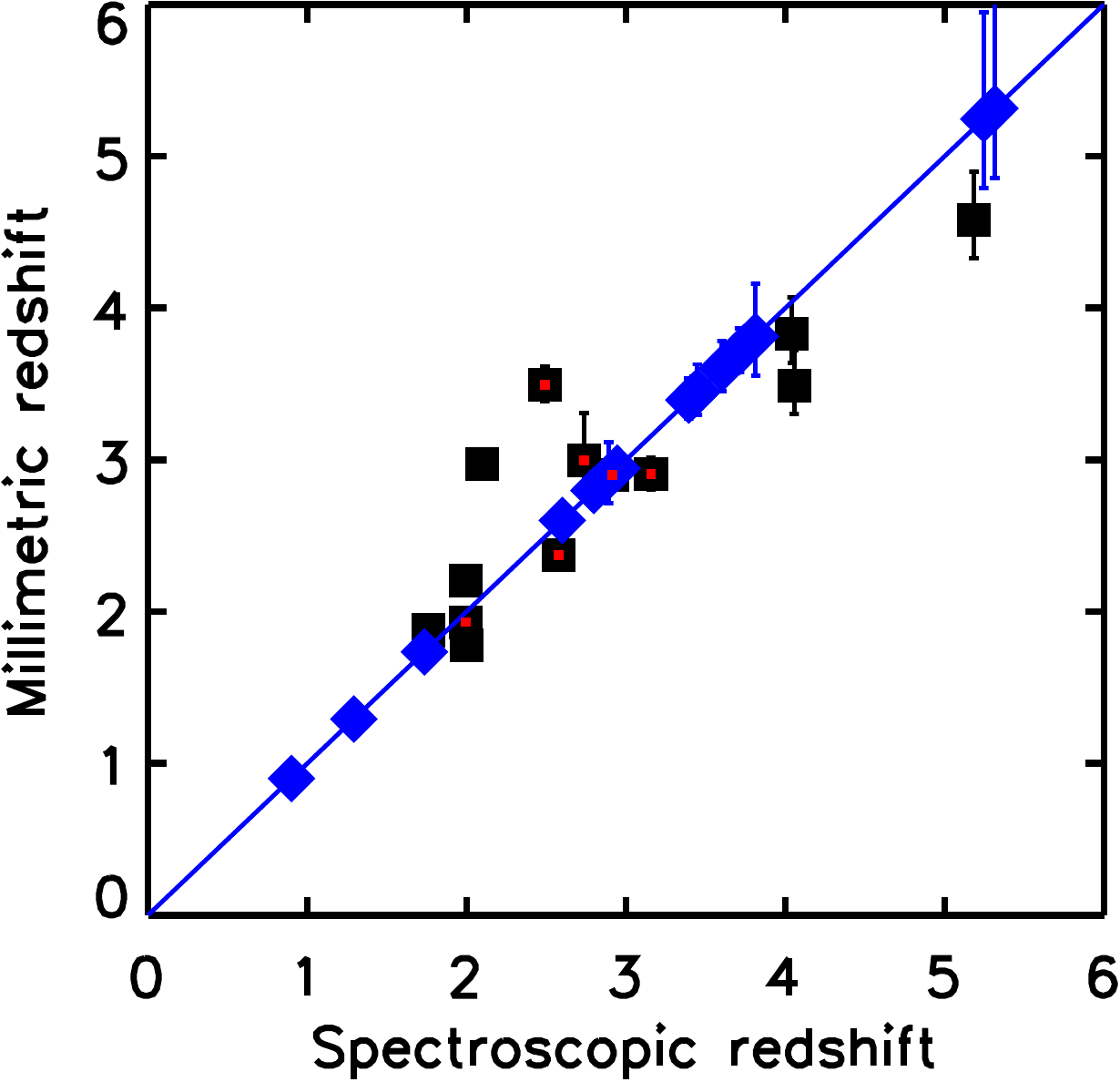}}
\caption{
Millimetric redshifts estimated from the 1.4~GHz to 860~$\mu$m 
flux ratio using the Barger et al.\ (2000) Arp~220-based model vs.
spectroscopic redshift.
The SMGs in the SMA sample with spectroscopic redshifts are denoted
by black squares.
Those without spectroscopic redshifts are
denoted by blue solid diamonds and are plotted at their millimetric 
redshifts on both axes. X-ray AGNs are marked with red small squares.
\label{figzmilli}
}
\end{inlinefigure}

We searched a $3''$ radius around each SMA position to find
X-ray counterparts in the 2~Ms 
{\em Chandra\/} catalog of Alexander et al.\ (2003).
Sources that contain AGNs are expected to follow the same relation as 
the non-AGNs, since both star-forming galaxies and radio-quiet AGNs obey 
the same tight empirical correlation between non-thermal radio
emission and thermal dust emission (e.g., Helou et al.\ 1985; Condon 1992).
This is seen to be true from
Figure~\ref{figzmilli}, where we mark the X-ray AGNs with red squares. Only
one of the five X-ray AGNs has only a millimetric redshift, which may be due 
to the fact that AGNs are easier to identify spectroscopically.

In Figure~\ref{fighist}(a), we show the spectroscopic (gray shading) and millimetric
(blue) redshift distributions for the SMA sample in histogram form. 
The millimetric redshifts predominantly fill in the $z\sim2.5-4$ range.
CDF15a and CDFN15b do not have radio counterparts, and are not
shown in  Figure~\ref{fighist}(a), but the lower limit on the millimetric redshift based on the upper
limit on the radio flux would place them at high redshift ($z>5$).
Unfortunately, it would be hard to model the selection effects for the SMA sample,
given the diverse reasons for the observations.

All but eight of the sources in the SCUBA-2 sample have either SMA identifications
or single radio sources in the SCUBA-2 beam. We show the spectroscopic
(gray shading) and millimetric (blue) redshift distributions in 
Figure~\ref{fighist}(b). The spectroscopic redshifts
range from $z=1$ to just above $z=5$, while the millimetric redshifts again
predominantly fill in the $z\sim2.5-4$ range. As in the SMA figure the four
sources with only radio upper limits, which are not shown in the figure,
are likely to lie at higher redshifts.

\vskip 0.5cm
\begin{inlinefigure}
\centerline{\includegraphics[width=3.2in,angle=0]{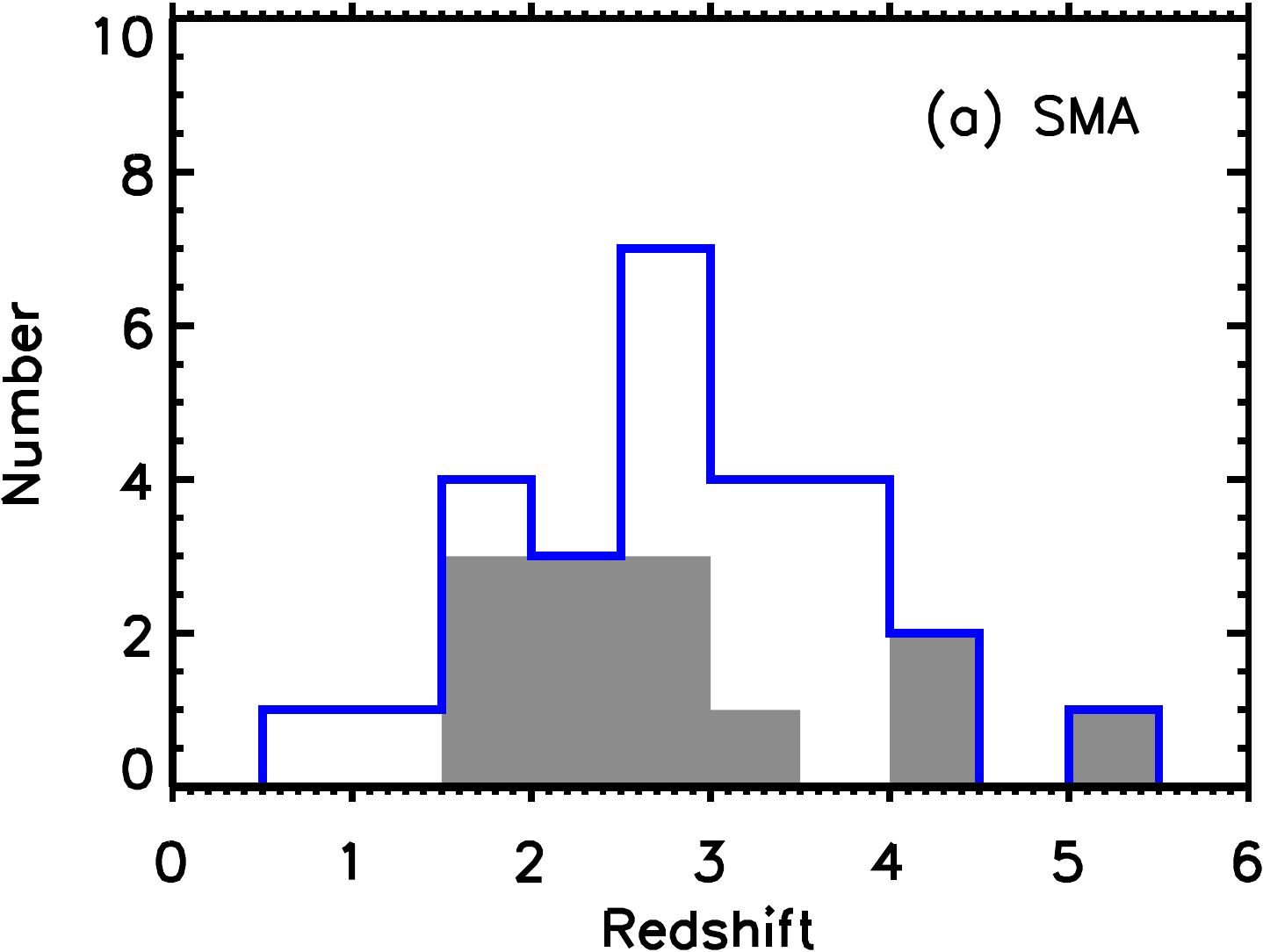}}
\vskip 0.2cm
\centerline{\includegraphics[width=3.2in,angle=0]{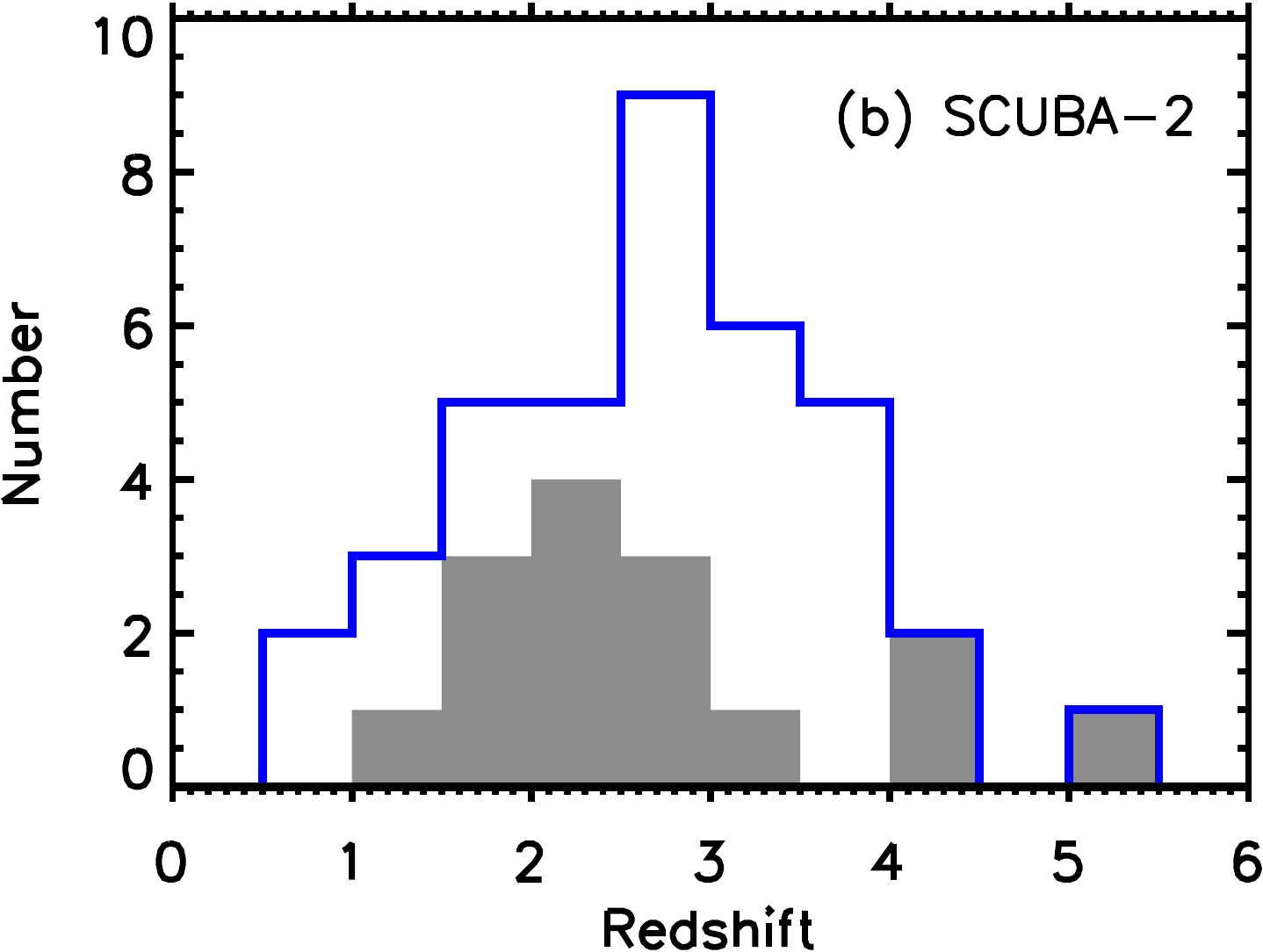}}
\caption{
Histograms of the spectroscopic (gray shading) and millimetric (blue) redshifts
for (a) the SMA sample and (b) the SCUBA-2 sample.
In (a) the 27 sources with radio counterparts in Table~2 are shown.
The 2 omitted sources without counterparts (CDN15a and CDFN15b)
are predicted to lie at $z>5$ by the millimetric estimate. 
In (b), only sources with SMA identifications
or single radio counterparts in the SCUBA-2 beam are shown (i.e., all but
8 of the 49 SCUBA-2 sources in Table~1).
\label{fighist}
}
\end{inlinefigure}

\section{Redshift Distribution of the Radio Sample}
\label{seczdist}

In Figure~\ref{radio_hist}(a), we show how the spectroscopic completeness 
(gray shading)
is very high for radio sources in the GOODS-N region (black histogram)
with bright optical or NIR counterparts, and how the
number of additional sources with secure photometric redshifts
(green shading) is small. In the GOODS-N field,
we have 367 spectroscopically identified radio sources, including
four with CO redshifts (cyan shading).
The photometric redshifts add only a further 30. 
The spectroscopic plus photometric redshift identifications
are highly complete to $K_{s}\sim21$, but nearly all have $K_{s}<22.5$.
In Figure~\ref{radio_hist}(b), we show similar results for
the radio sources in the full field (black histogram). 
We denote the spectroscopically identified 
sources with gray shading and the spectroscopically unidentified 
sources having clear submillimeter counterparts with blue shading. 
We can estimate millimetric redshifts for the latter sources, and 
CO redshifts may also be obtainable.

\vskip 0.5cm
\begin{inlinefigure}
\centerline{\includegraphics[width=3.2in,angle=0]{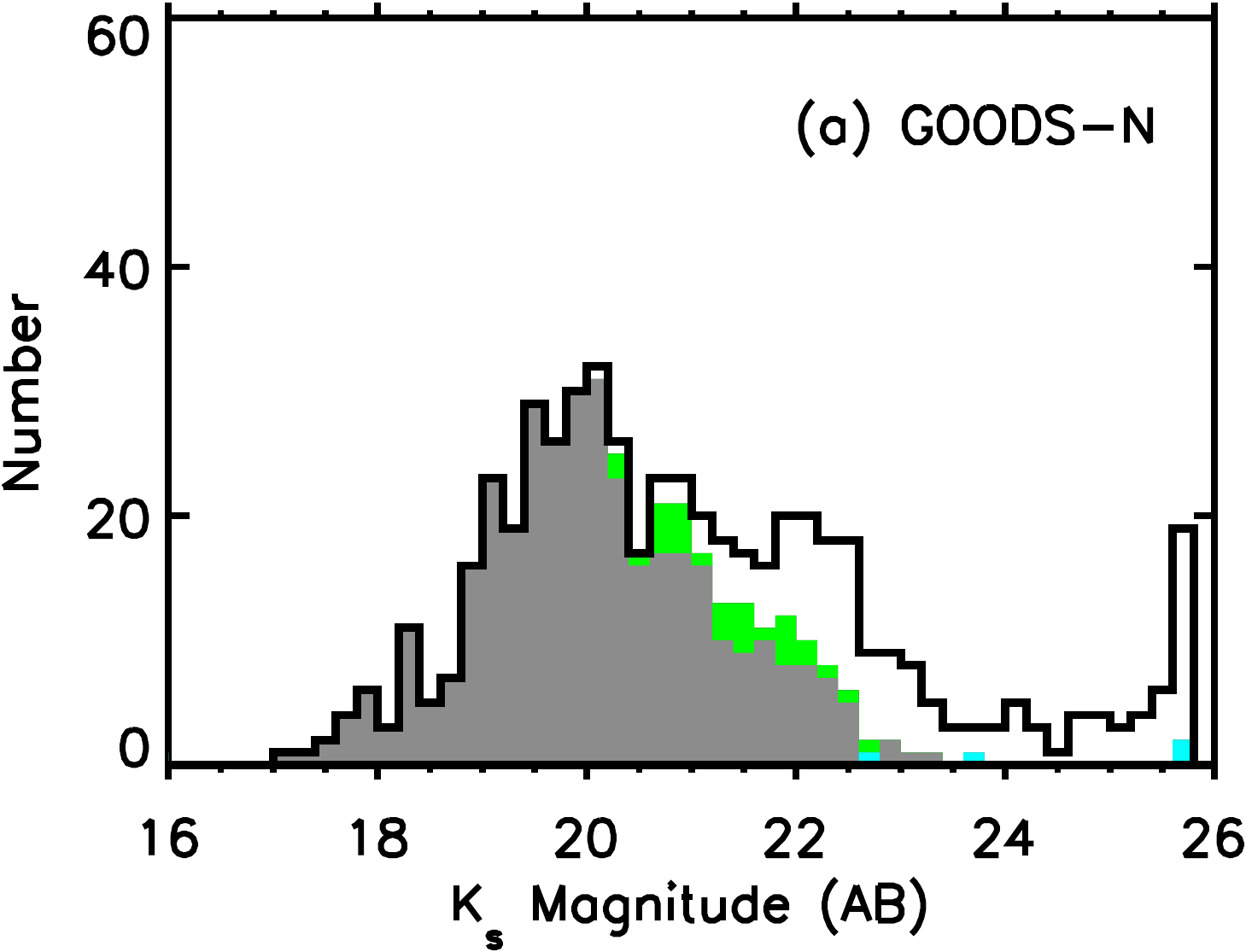}}
\vskip 0.2cm
\centerline{\includegraphics[width=3.2in,angle=0]{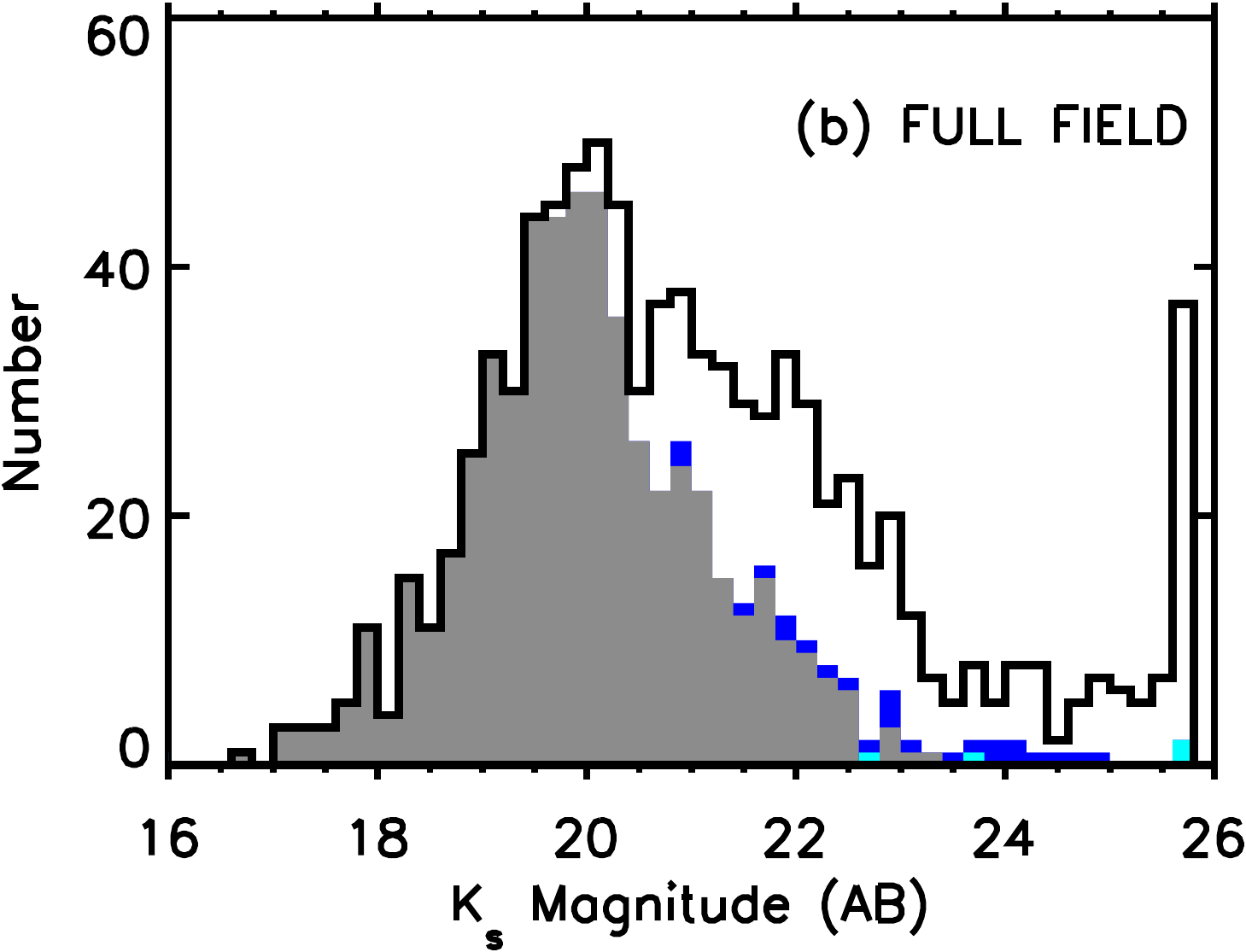}}
\caption{
Histogram of the $K_s$ magnitudes for the radio sources.
The step size is 0.2 mags. Sources with
$K_s$ fainter than 25.7 are all put into the faintest
bin.  In (a), we show the sources in the GOODS-N
region ({\em gray shading\/} --- spectroscopic
redshifts; {\em cyan shading\/} --- CO redshifts;
{\em green shading\/} --- additional sources with 
photometric redshifts; {\em no shading\/} --- sources without
redshifts). In (b), we show the sources in the full field
({\em gray shading\/} --- spectroscopic
redshifts; {\em cyan shading\/} --- CO redshifts; 
{\em blue shading\/} --- additional sources having clear submillimeter
counterparts for which we can measure millimetric redshifts;
{\em no shading\/} --- sources without redshifts).
\label{radio_hist}
}
\end{inlinefigure}

We see that there is an extended tail of radio sources with faint
NIR counterparts ($K_{s}>22.5$). This is the magnitude regime where 
many of the sources with millimetric redshifts lie and where all four
of the sources with CO redshifts lie (all with $z\gtrsim4$). 

In each figure, we lumped the
sources that are undetected in the $K_s$ image into the faintest bin. These
sources are also extremely optically faint: nearly all of the ones that lie
in the GOODS-N region are undetected in the {\em HST\/} ACS images. 
Unfortunately, such sources, which may lie at high redshifts, cannot be 
identified with either optical/NIR spectroscopy or photometry.

\section{The $K-\lowercase{z}$ Relation}
\label{seckz}

Because of the potential for estimating redshifts for radio galaxies in the 
NIR-faint (likely high-redshift) tail using the $K-z$ relation,
we now turn to examining that relation for our faint radio sample.
In Figure~\ref{radio_kmg}(a), we show $K_s$ magnitude versus
spectroscopic (black squares), CO (cyan squares), or photometric 
(green circles) redshift for 
the radio sources in the GOODS-N region.
By comparing with a $K_s$-selected field 
sample with spectroscopic redshifts in the GOODS-N region (Barger et al.\ 2008; 
purple contours show the surface density), we see that our radio sources
are nearly all located in the most $K_s$ luminous galaxies at all redshifts. 
Remarkably, the $K-z$ relation from Willott03 (red line; 
approximately converted from their $K$(Vega) to our $K_s$(AB) using a 
1.9~mag offset) accurately traces the $K_s$ luminous envelope of our sample
over a wide range in $K_s$ magnitude and redshift, indicating that some
faint radio sources lie in the same galaxy mass hosts as powerful radio
sources.

In Figure~\ref{radio_kmg}(b), we show $K_s$ magnitude versus 
spectroscopic (black squares), CO (cyan squares), or millimetric (blue diamonds) 
redshift for the radio sources in the full field. The millimetric redshifts fill in
the higher redshift part of the plot.
We again show the Willott03 relation (red line), as well as the
same relation with the $K_s$ magnitude made fainter by 2.5~mag (black line). 
Nearly all of the radio sources lie
in the band between the red and black lines.
At low redshifts (out to $z\sim1$), this could be
a consequence of the rapid evolution in the maximum
SFRs with increasing redshift, such that the radio
selection is always dominated by the highest redshift and highest
SFR galaxies (e.g., Condon 1989; Barger et al.\ 2007).

Eales et al.\ (1997) and Willott03 showed for the most powerful radio samples
that the $K-z$ relation has a weak dependence on radio flux selection, 
with lower radio flux sources having fainter $K$-band magnitudes for the same
redshift. More recently, Simpson et al.\ (2012) studied the $K-z$ relation for the
$S_{\rm 1.4~GHz} > 100~\mu$Jy sample in the Subaru/{\em XMM-Newton Deep Field},
which is about a factor of 1000 fainter than the faintest radio survey limit of Willott03,
and found that at $z\gtrsim1$, the sources 
were systematically fainter than the literature $K-z$ relations
(Willott03; Brookes et al.\ 2008; Bryant et al.\ 2009). 

Our $S_{\rm 1.4~GHz} > 11.5~\mu$Jy sample shows a clear dependence of the $K-z$ 
relation on the radio flux selection. We illustrate this in Figure~\ref{radio_kmg_byflux},
where we show $K_s$ magnitude versus spectroscopic or photometric (black squares),
CO (cyan squares), or millimetric (blue solid diamonds) redshift in four radio flux intervals for the
radio sources in the GOODS-N field. We mark X-ray AGNs with red squares and
SCUBA-2 sources with spectroscopic, photometric, or CO redshifts with blue
large open diamonds. In each radio interval, we adopt a $K-z$ relation 
having the Willott03 shape (see their Equation~2),
\begin{equation}
K_s = \Delta K_s + 4.53 \log_{10} z - 0.31 (\log_{10} z)^{2} \,.
\label{ks_eqn}
\end{equation}
We obtain least-squares fits to the data by adjusting the $K_s$ offset 
($\Delta K_s$). In each panel, we show our fits in black and the Willott03
relation ($\Delta K_s = 17.37$) in red.
Remarkably, the observed dependence on radio flux appears
to hold for all sources, whether they are star formation dominated or AGN
dominated. However, by $z\gtrsim3$, the SCUBA-2 sources (blue solid or open 
diamonds) appear fainter than expected, as was also observed
by Dunlop (2002) and Serjeant et al.\ (2003b) for SCUBA sources.

In Figure~\ref{kz_offset}, we plot the $\Delta K_s$ values found from our
fits versus the mean radio fluxes for each interval. These data points are
well fit by the functional form
\begin{equation}
\Delta K_s=21.73-0.84 \log_{10} f_{1.4~{\rm GHz}} \,,
\label{ks_off}
\end{equation}
which we show in the figure with the red line.
Here $f_{1.4~{\rm GHz}}$ is in $\mu$Jy.

\vskip 1.0cm
\begin{inlinefigure}
\centerline{\hskip 0.30cm \includegraphics[width=3.3in,angle=0]{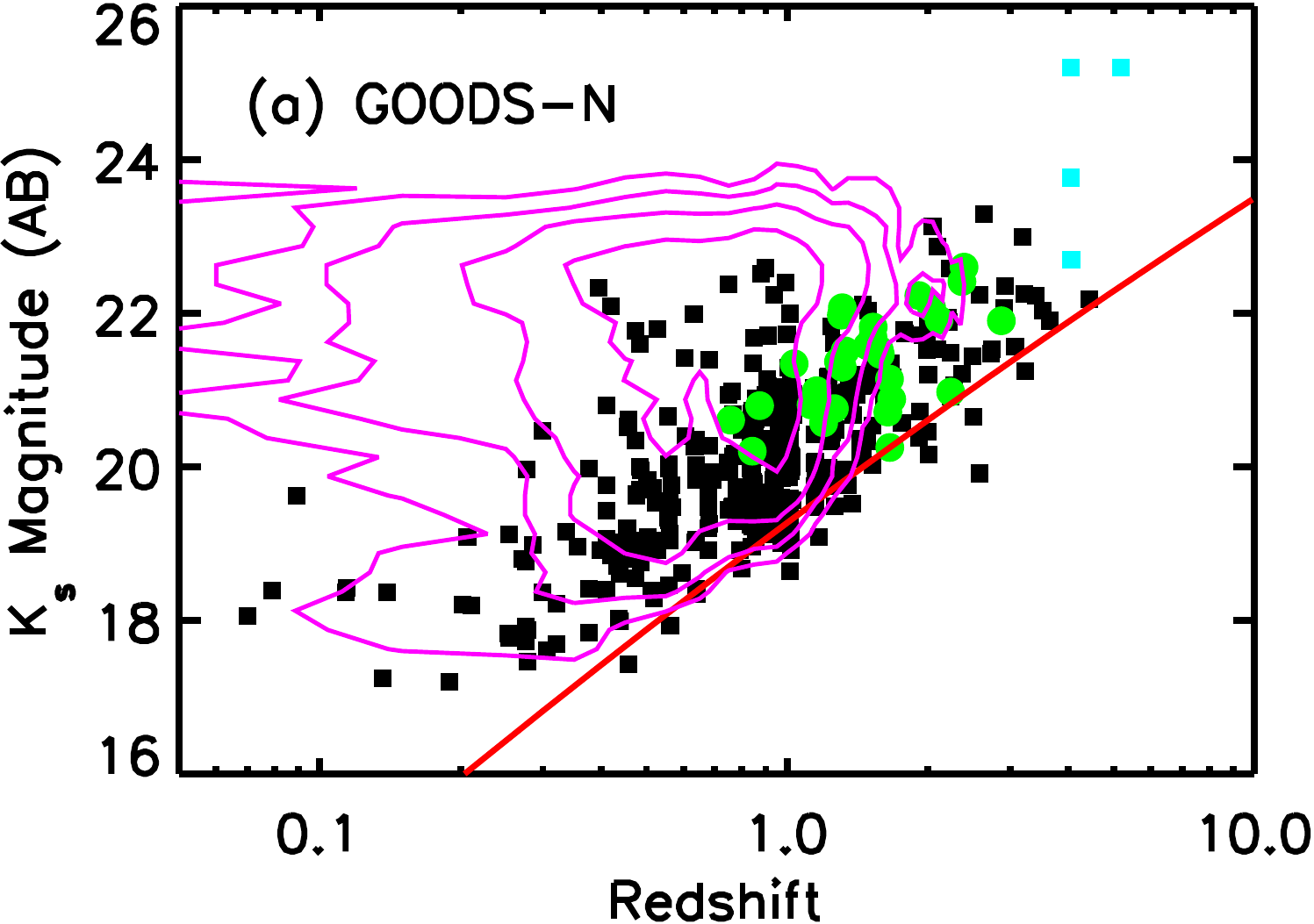}}
\vskip 0.2cm
\centerline{\includegraphics[width=3.2in,angle=0]{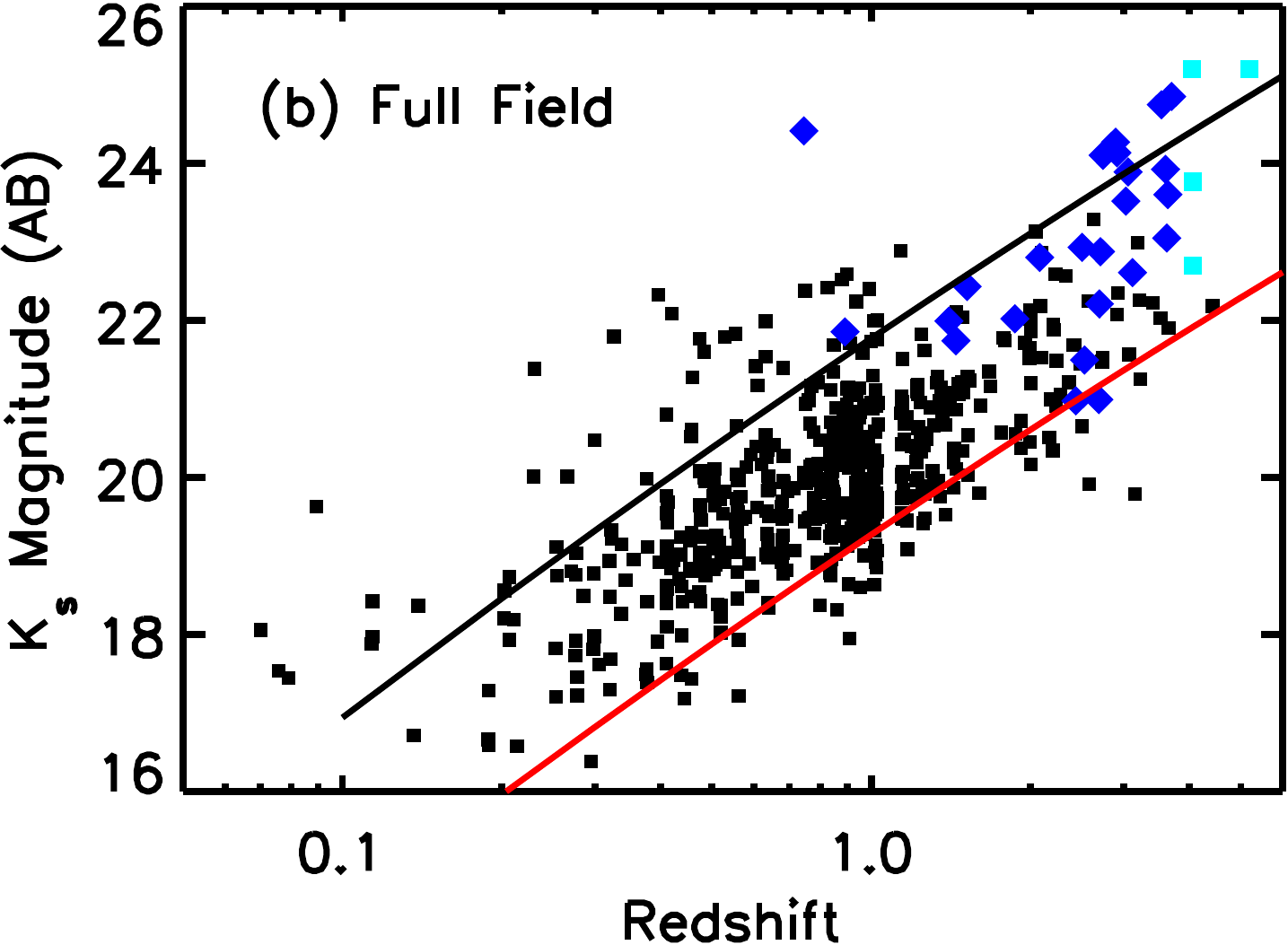}}
\caption{
$K_s$ magnitude vs. redshift for the radio sources. 
Sources undetected in $K_s$ are shown at a nominal magnitude of 25.2.  In (a), 
we show the radio sources in the GOODS-N region 
({\em black squares\/} --- spectroscopic redshifts; {\em cyan squares\/} --- CO redshifts;
{\em green circles\/} --- photometric redshifts).
The purple contours show the surface density of a $K_s$-selected
field sample with spectroscopic redshifts in the GOODS-N region
from Barger et al.\ (2008).
The contours rise by multiplicative factors of 2 from the lowest
contour with 1/40th of the peak surface density. 
In (b), we show the radio sources in the full field
({\em black squares\/} --- spectroscopic redshifts;
{\em cyan squares\/} --- CO redshifts;
{\em blue diamonds\/} --- millimetric redshifts).
The red line (also in (a)) shows the $K-z$ relation from Willott03.
The black line shows the same relation with the $K_s$ magnitude
made fainter by 2.5~mag.
 \label{radio_kmg}
}
\end{inlinefigure}

This relation also extrapolates approximately
to fit the highest radio flux samples of previous work. 
An exact comparison is difficult, however, because those
samples are chosen at different radio frequencies, and we
cannot precisely convert the NIR photometry.
Willott03 found a 0.55~mag difference in the mean $\Delta K$
for the 3CRR and 7C samples. Since the 7C sample is 20
times fainter, that would correspond
to a slope of 0.42 in Equation~\ref{ks_off}, suggesting that
brighter radio sources may follow a shallower relation.

From Equation~\ref{ks_off}, we can see that the observed 
$K_s$-band flux depends on the observed 1.4~GHz flux to the power 
of 0.34. This is a very weak dependence, which 
means that for a large range in radio flux, there is only a small 
range in host galaxy $K_s$-band flux.
Thus, for our sample, with $f_{1.4~\rm{GHz}}$ ranging from 
11.5 to 5276~$\mu$Jy, the range in $f_{K_s}$ is only 7.8. The corresponding
range in $K_s$ is 2.2~mag, consistent with the range
seen in Figure~\ref{radio_kmg}(a).

We may use the dependence of the $\Delta K_s$ values
on $f_{1.4~{\rm GHz}}$ to tighten the relation
between the $K_s$ magnitude and the redshift for the radio sources. We 
define a corrected $K_s$ magnitude, $K_{corr}$, as
\begin{equation}
K_{corr}\equiv K_s + 0.84 \log_{10} (f_{1.4~{\rm GHz}}/100~\mu {\rm Jy})
\label{k_corr}
\end{equation}
to move all of the sources to the track followed by 100~$\mu$Jy sources. 
In Figure~\ref{radio_kmg_correct}, we plot $K_{corr}$ versus redshift.

We made a fourth order polynomial fit to the data in Figure~\ref{radio_kmg_correct},
\begin{eqnarray}
K_{corr}&=&19.88 +3.20 \log_{10} z + 1.13(\log_{10} z)^{2} \nonumber \\ 
&+& 2.79(\log_{10} z)^{3}+2.58(\log_{10} z)^{4} \,.
 \label{kcorr_relation}
\end{eqnarray}
Above $K_{s}=22$, our spectroscopic and photometric identifications in the GOODS-N 
are extremely incomplete, with the spectroscopic identifications likely biased towards
star-forming galaxies and AGNs, so this turn up may not be representative
of the full radio population. However, we  can use Equation~\ref{kcorr_relation} 
to obtain rough redshift estimates for the radio sources. We show these in
Figure~\ref{figkz} plotted versus spectroscopic, photometric, millimetric, or CO 
redshift for the radio sources in the GOODS-N field with such information.

\begin{inlinefigure}
\centerline{\includegraphics[width=2.6in,angle=0]{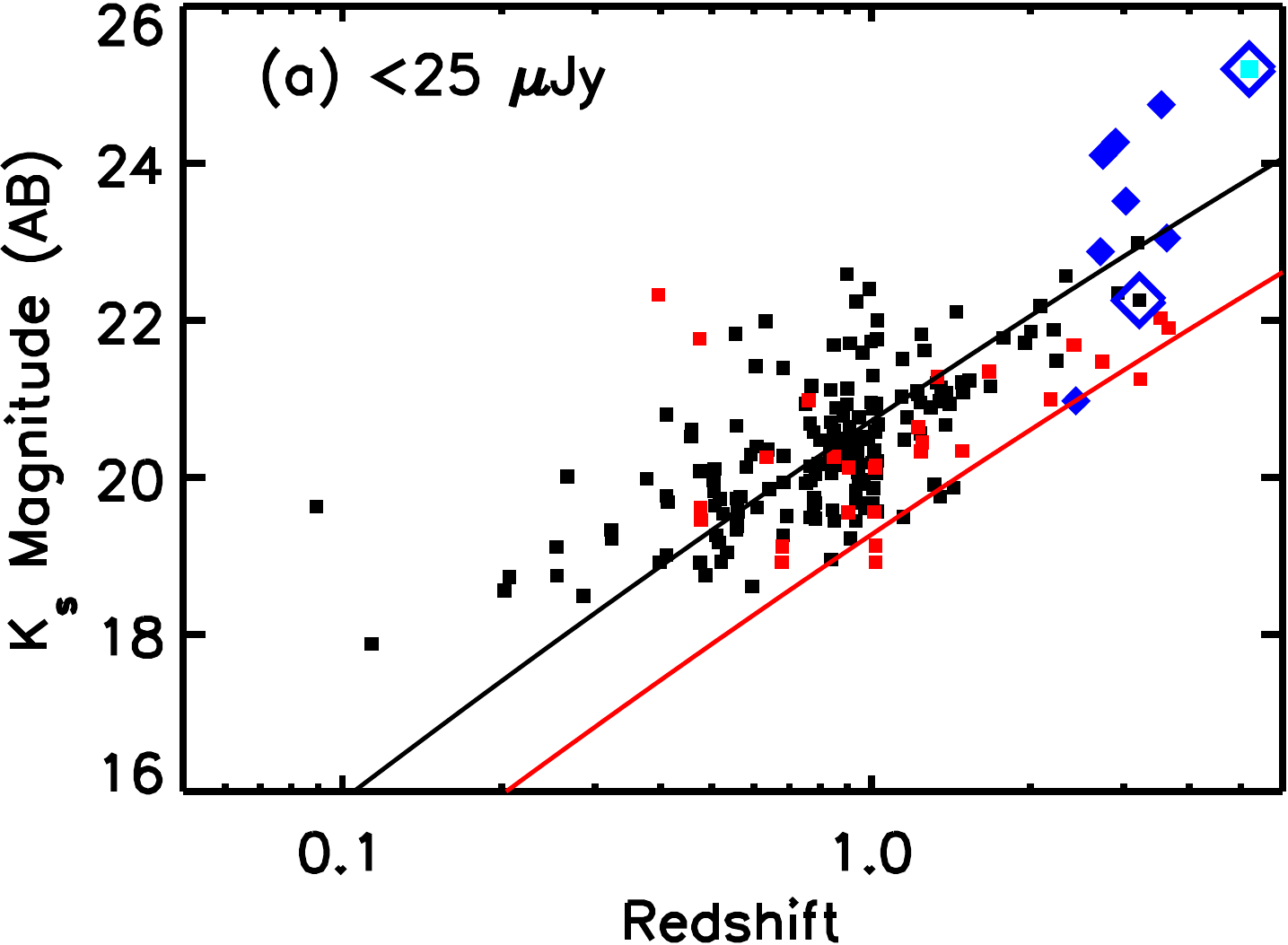}}
\centerline{\includegraphics[width=2.6in,angle=0]{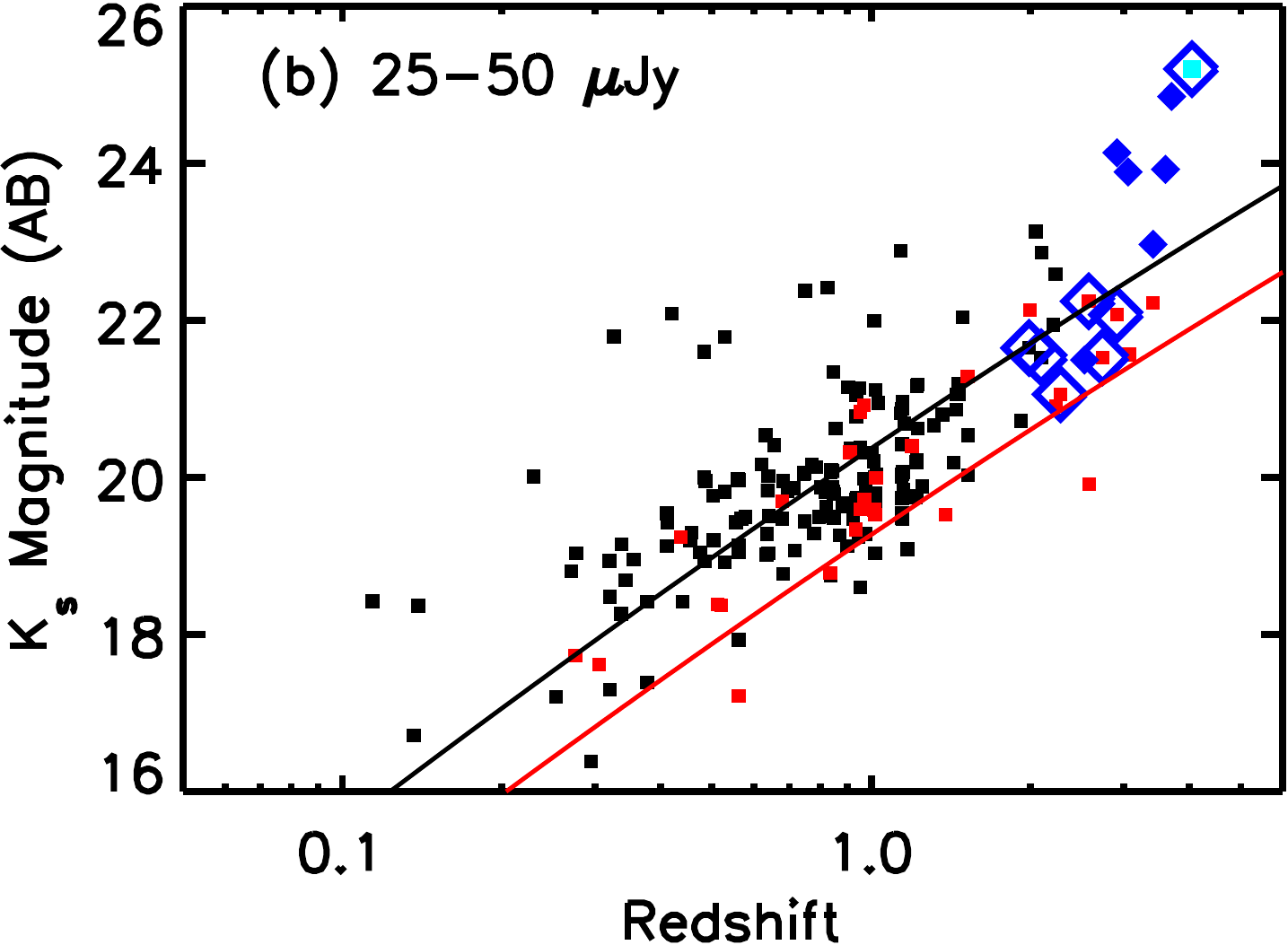}}
\centerline{\includegraphics[width=2.6in,angle=0]{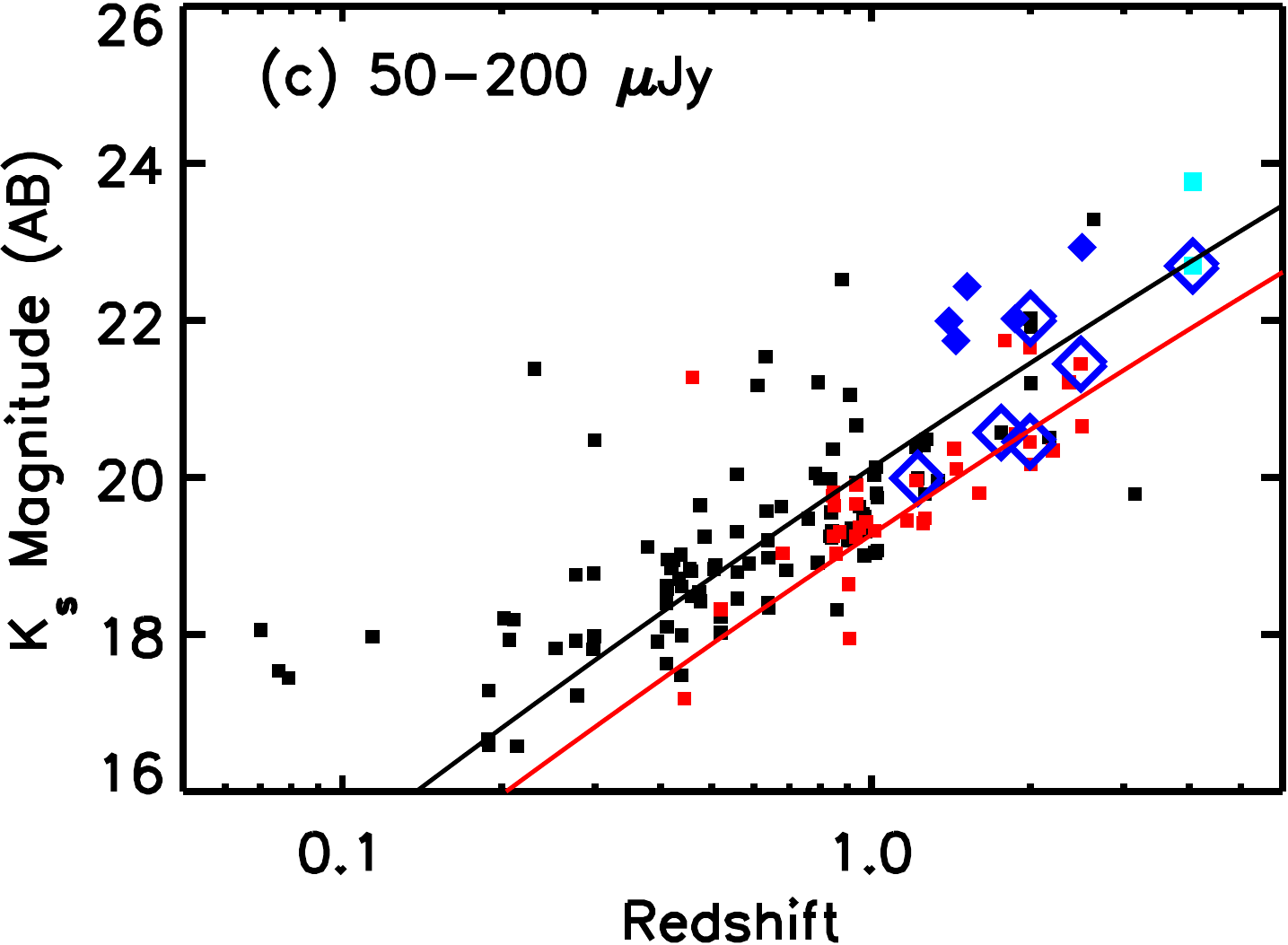}}
\centerline{\includegraphics[width=2.6in,angle=0]{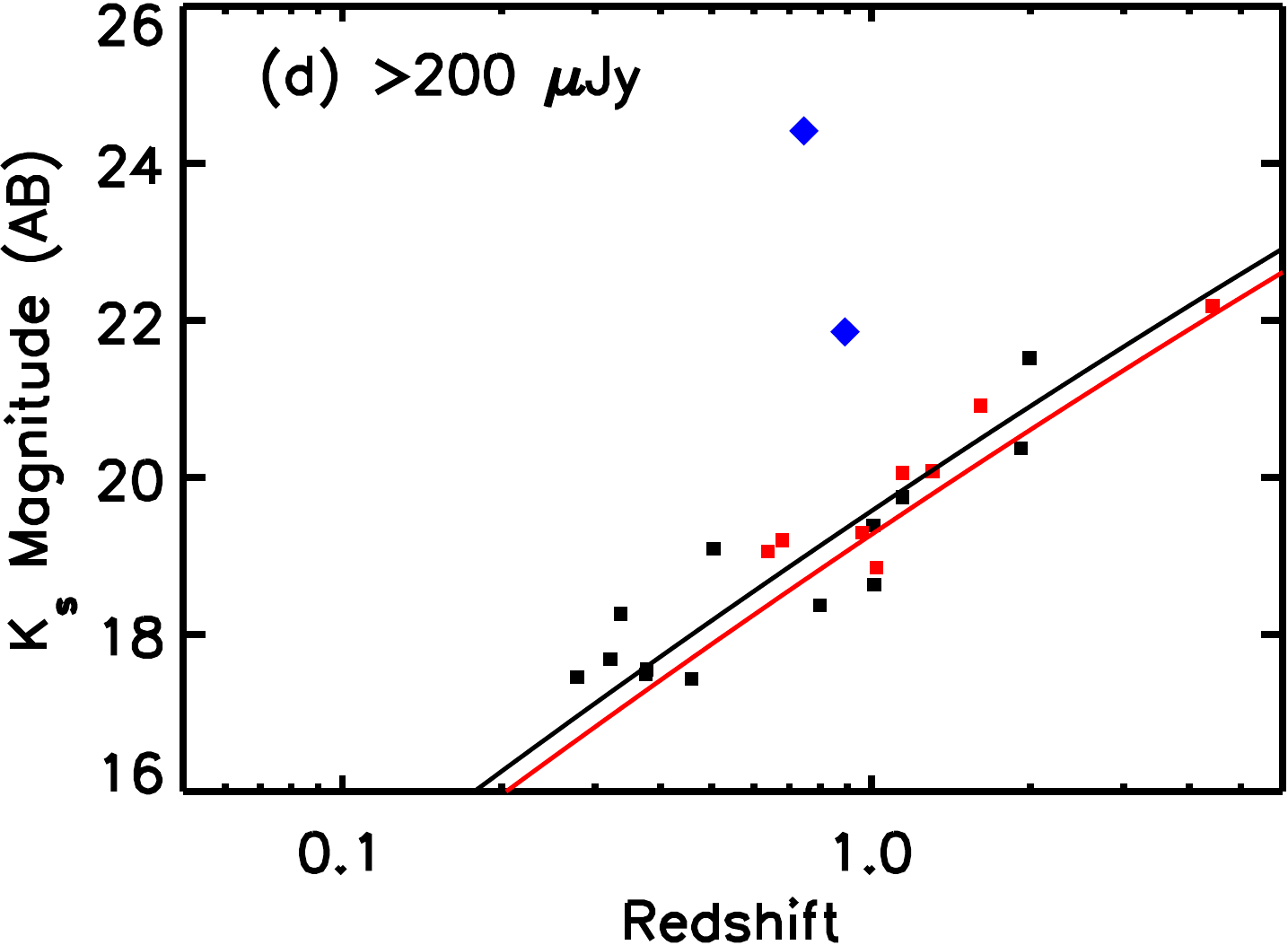}}
\caption{
$K_s$ magnitude vs. redshift in four 1.4~GHz flux intervals for radio
sources in the GOODS-N: (a) $<25~\mu$Jy,
(b) $25-50~\mu$Jy, (c) $50-200~\mu$Jy, and $>200~\mu$Jy.
Sources undetected in $K_s$ are shown at a 
nominal magnitude of 25.2.  We show
the radio sources in the full field ({\em black squares\/} ---
spectroscopic or photometric redshifts; {\em cyan squares\/} --- CO redshifts;
{\em blue solid diamonds\/} --- millimetric redshifts).
X-ray AGNs are marked with red squares.
Sources with spectroscopic, photometric, or CO redshifts
and SCUBA-2 detections are shown
surrounded by blue large open diamonds.
The red line shows the $K-z$ relation from Willott03. 
The black line shows the best-fit model using the shape
of the Willott03 $K-z$ relation, but allowing $\Delta K_s$
in Equation~\ref{ks_eqn} to vary.
 \label{radio_kmg_byflux}
}
\end{inlinefigure}

\begin{inlinefigure}
\centerline{\includegraphics[width=3.5in,angle=0]{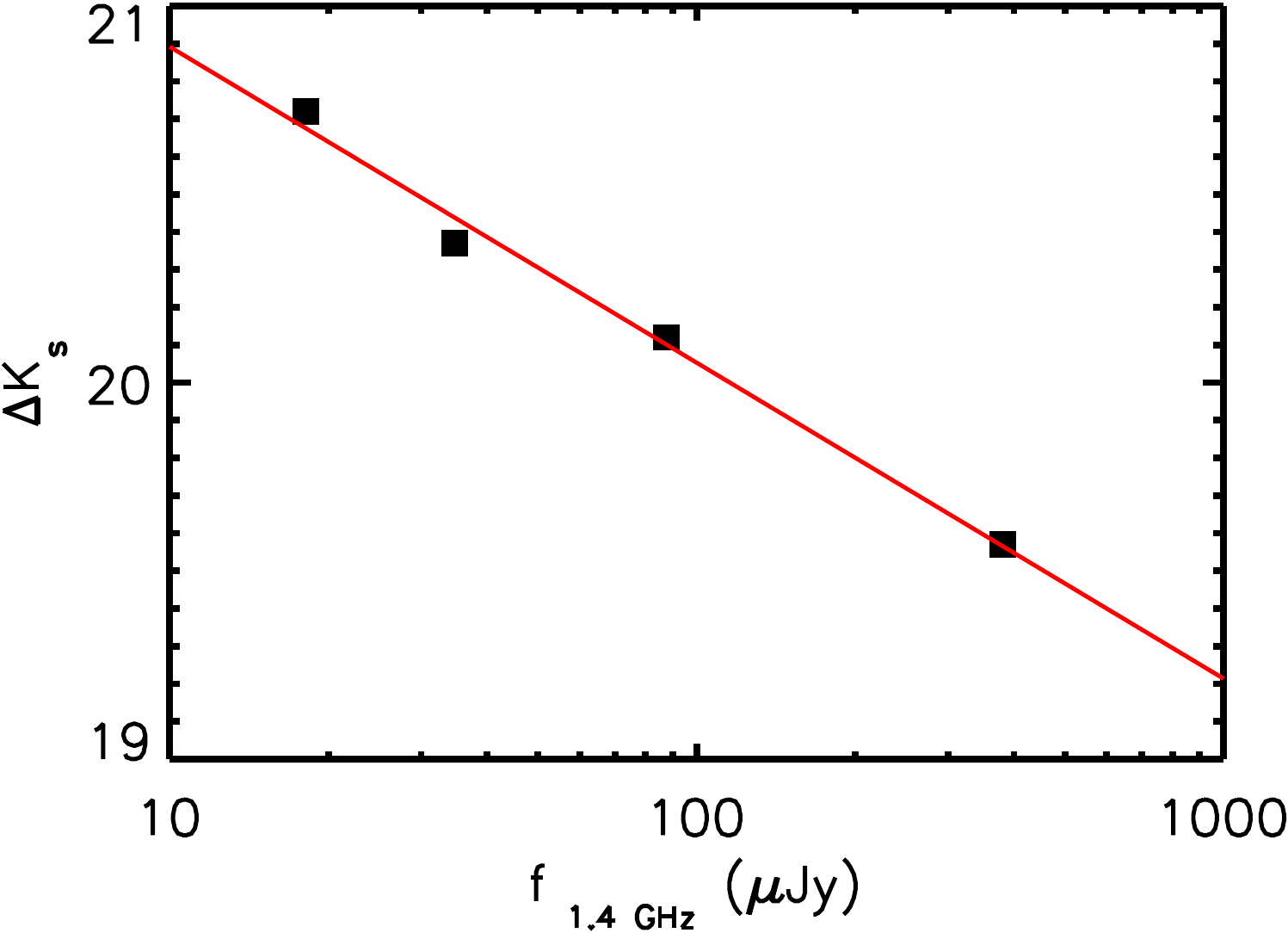}}
\caption{
Black squares show the $\Delta K_s$ values 
(20.72, 20.37, 20.12, 19.57)
determined in the radio flux intervals of Figure~\ref{radio_kmg_byflux} 
vs. the mean radio fluxes in the intervals (18.1, 34.7,
87.6, and 308.9~$\mu$Jy). The red line shows the least squares
fit of $\Delta K_s$ vs. $\log_{10}f_{1.4~{\rm GHz}}$
(Equation~\ref{ks_off}).
\label{kz_offset}
}
\end{inlinefigure}

\vskip 1.0cm
\begin{inlinefigure}
\centerline{\includegraphics[width=3.2in,angle=0]{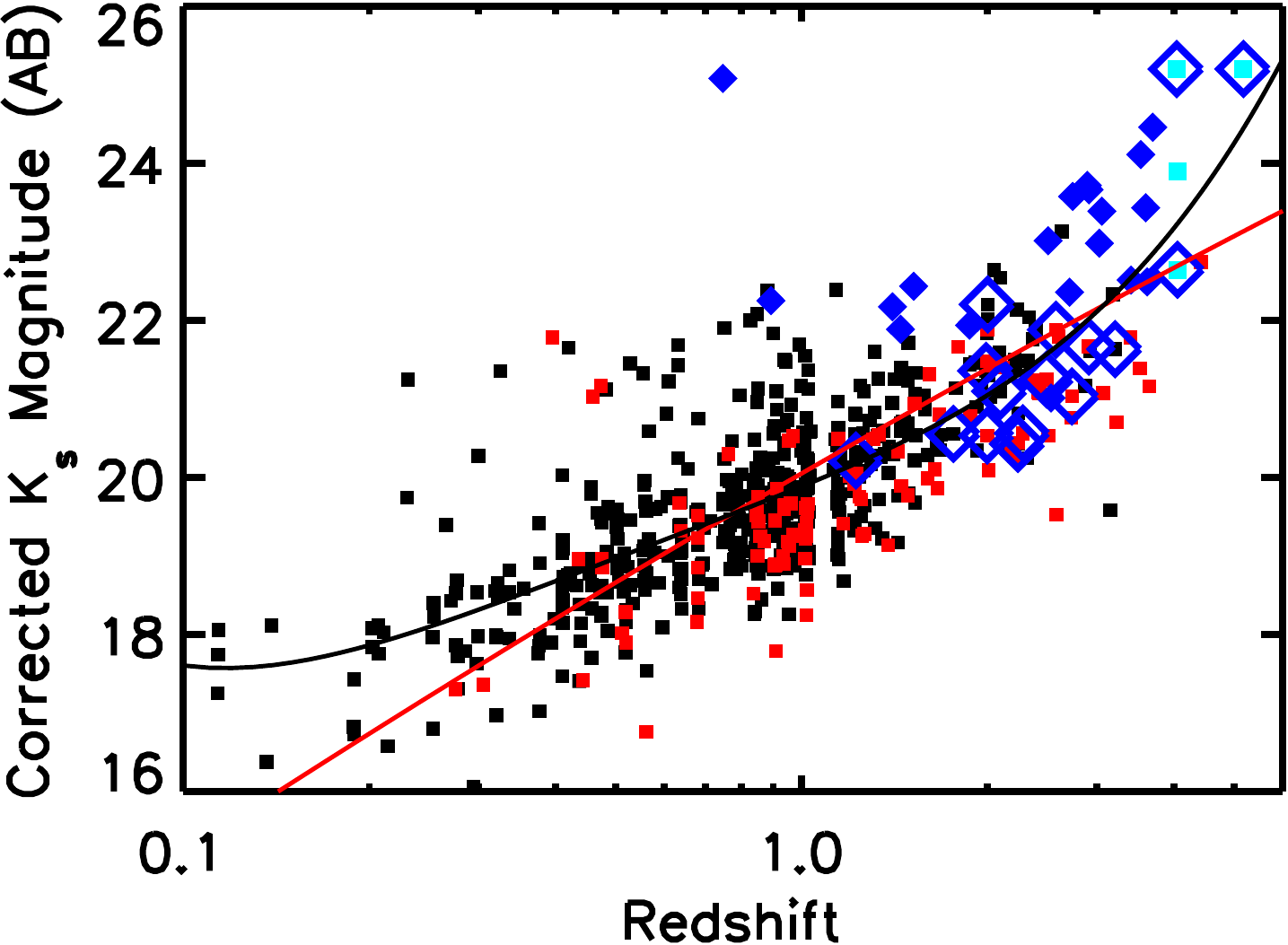}}
\caption{
$K_s$ magnitude corrected to $K_{corr}$ using Equation~\ref{k_corr}
for the radio sources in the GOODS-N vs. redshift 
({\em black squares\/} --- spectroscopic or
photometric redshifts; {\em cyan squares\/} --- CO redshifts; 
{\em blue solid diamonds\/} --- millimetric redshifts). 
X-ray AGNs are marked with red squares.
Sources with spectroscopic, CO, or photometric redshifts and
SCUBA-2 detections are shown surrounded by blue large open diamonds.
The curve from Equations~\ref{ks_eqn} and \ref{ks_off} for 
$f_{\rm 1.4~GHz}=100~\mu$Jy is shown in red.
The fourth order polynomial fit to the data from
Equation~\ref{kcorr_relation} is shown in black.
\label{radio_kmg_correct}
}
\end{inlinefigure}

\vskip 0.5cm
\begin{inlinefigure}
\includegraphics[width=3.2in,angle=0]{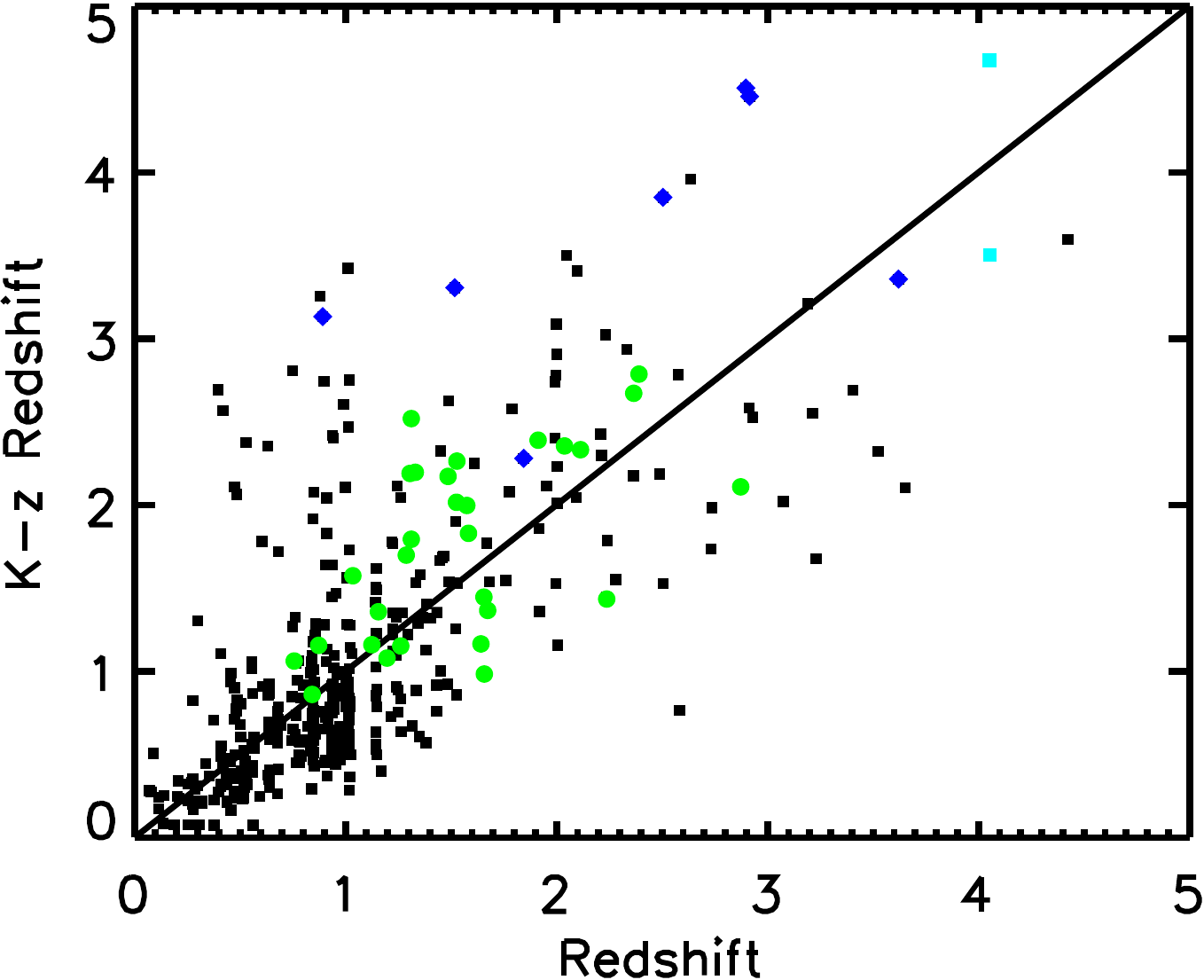}
\caption{
Redshift estimated from the $K-z$ relation (Equations~\ref{k_corr} and \ref{kcorr_relation})
vs. spectroscopic (black squares), CO (cyan squares),
photometric (green circles), or millimetric
(blue diamonds) redshift
for the radio sources in the GOODS-N field with such information.
\label{figkz}
}
\end{inlinefigure}

\section{Radio Power and Submillimeter Flux}
\label{secradiopower}

Locally, most radio sources more powerful than
$P_{1.4~{\rm GHz}}=10^{30}$~erg~s$^{-1}$~Hz$^{-1}$
are found to be associated with AGN activity 
(e.g., Condon 1989; Sadler et al.\ 2002; Best \& Heckman 2012).  
However, as we move to higher redshifts
where dusty galaxies with high SFRs become more common 
(e.g., Cowie et al.\ 2004a), it is possible 
that some fraction of the  
$P_{1.4~{\rm GHz}}>10^{30}$~erg~s$^{-1}$~Hz$^{-1}$
 radio sources are dominated by 
star formation. Indeed, as we discussed in the introduction, 
all $z\gtrsim4$ radio galaxies may have high SFRs 
($\gtrsim1000~M_{\sun}$~yr$^{-1}$) based on 
modeling of the $K-z$ relation (Rocca-Volmerange et al.\ 2004),
and such SFRs have been observed
(e.g., Dunlop et al.\ 1994; Ivison et al.\ 1998; Archibald et al.\ 2001; Reuland et al.\ 2004).
Thus, our radio sample should provide a powerful way of finding such sources.
However, separating the contributions to the radio emission
from star formation and AGN activity is not straightforward. 

We compute the rest-frame radio powers for the radio sources in the
full field assuming 
$S_\nu\propto \nu^\alpha$ and a radio spectral index of $\alpha=-0.8$
(Condon 1992; Ibar et al.\ 2010) using
\begin{equation}
P_{1.4~{\rm GHz}}=4\pi {d_L}^2 S_{1.4~{\rm GHz}} 10^{-29}
(1+z)^{\alpha - 1}~{\rm erg~s^{-1}~Hz^{-1}} \,.
\label{eqradio}
\end{equation}
Here $d_L$ is the luminosity distance (cm) and $S_{\rm 1.4~GHz}$
is the 1.4~GHz flux density ($\mu$Jy). 
In Figure~\ref{radio}, we show these radio powers versus redshift
for the sources with spectroscopic 
(including CO; black squares) or photometric (green circles)
redshifts. For the remaining sources (red plus signs), we use
the redshifts estimated from the $K_s-z$ relation
(Equations~\ref{k_corr} and \ref{kcorr_relation}).
We also plot the radio catalog limit of  11.5~$\mu$Jy ($5\sigma$) (blue dotted curve) 
and the radio powers of 
a luminous infrared galaxy (LIRG; $L_{\rm FIR}>10^{11}~L_\odot$) 
(dashed horizontal line) and an ultraluminous infrared 
galaxy (ULIRG; $L_{\rm FIR}>10^{12}~L_\odot$) (solid horizontal line), 
which we calculated by assuming that the FIR-radio correlation holds.
At $z\gtrsim3$, our radio observations are only sensitive to star-forming 
galaxies brighter than the ULIRG limit.

\vskip 0.5cm
\begin{inlinefigure}
\includegraphics[width=3.2in,angle=0]{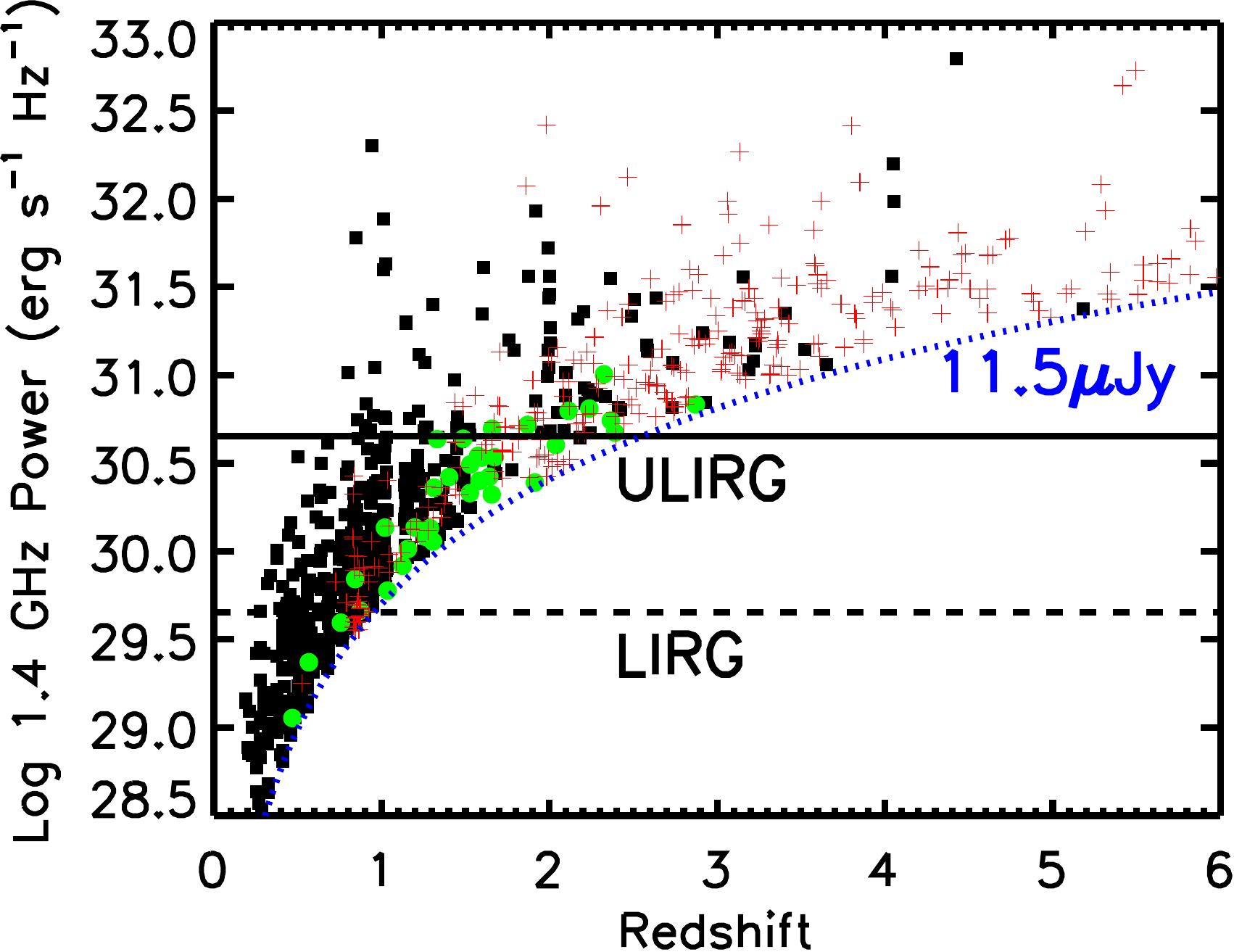}
\caption{
Radio power vs. redshift for the radio sources in the full field with spectroscopic
or CO redshifts (black squares) or photometric redshifts (green circles).
The remaining sources are shown at the redshifts that would be predicted from
the $K-z$ relation (red plus signs; Equations~\ref{k_corr} and \ref{kcorr_relation}).
The blue dotted curve shows the radio power corresponding to the 1.4~GHz
catalog limit of 11.5~$\mu$Jy  ($5\sigma$).
The dashed and solid horizontal lines show
the radio powers that correspond to the definitions of a LIRG 
and a ULIRG, respectively, which we calculated by assuming that the galaxies
are star formers and that the FIR-radio correlation holds.
\label{radio}
}
\end{inlinefigure}

\subsection{High Radio Power}
\label{subsechighradio}

Our primary interest in this paper is to determine whether there is a 
turn-down in the SFR distribution function at high redshifts. 
Thus, we now turn our attention 
to the high radio power sources in our sample, which we define as having
$P_{\rm 1.4~GHz}\ge10^{31}$~erg~s$^{-1}$~Hz$^{-1}$.

In Figures~\ref{radio_double}a and \ref{radio_double}b, we show blow ups 
of Figure~\ref{radio} to focus on the high radio power sources with secure redshifts.
We have 51 (50 are spectroscopic or CO, and 1 is photometric),
39 of which are at $z>1.5$.
Most of the spectroscopically unidentified sources are faint in $K_s$ and 
hence are likely to
lie at high redshifts based on the $K-z$ relation (see Figure~\ref{radio}).
For example, if we include our $K-z$ estimated redshifts, then the ratio of
secure to total $z>1.5$ sources would be 39/226, which would mean we have
secure identifications for $\lesssim1/5$ of the high-redshift, high radio
power sources. 

In Figure~\ref{radio_double}(a), we distinguish sources where the X-ray data show 
them to be X-ray AGNs (red squares) or X-ray quasars (red large squares). 
We use green squares to distinguish sources
that are likely to be elliptical galaxies,
as determined by their having rest-frame EW([OII]$\lambda3727)<10$~\AA. 
The latter distinction can only be made from the
optical spectra for galaxies at $z<1.5$.
We show with the blue dotted curve the radio limit of 11.5~$\mu$Jy ($5\sigma$).

In Figure~\ref{radio_double}(b), we mark radio
sources with 850~$\mu$m counterparts detected
above the $4\sigma$ level (blue circles). (If there is no SMA
measurement, then we only mark the source if there is a single radio 
counterpart within the SCUBA-2 beam.)
We show with the blue dashed curve the $850~\mu$m 
limit of 4~mJy ($4\sigma$) for the higher sensitivity region of the SCUBA-2 map 
(see Figure~\ref{area}) converted
to a radio power assuming an Arp~220 SED. 
Thus, over most of our SCUBA-2 map we would not expect to be able to detect
a source having a radio power much less than our high radio power 
source definition.

\vskip 0.5cm
\begin{inlinefigure}
\centerline{\includegraphics[width=3.2in,angle=0]{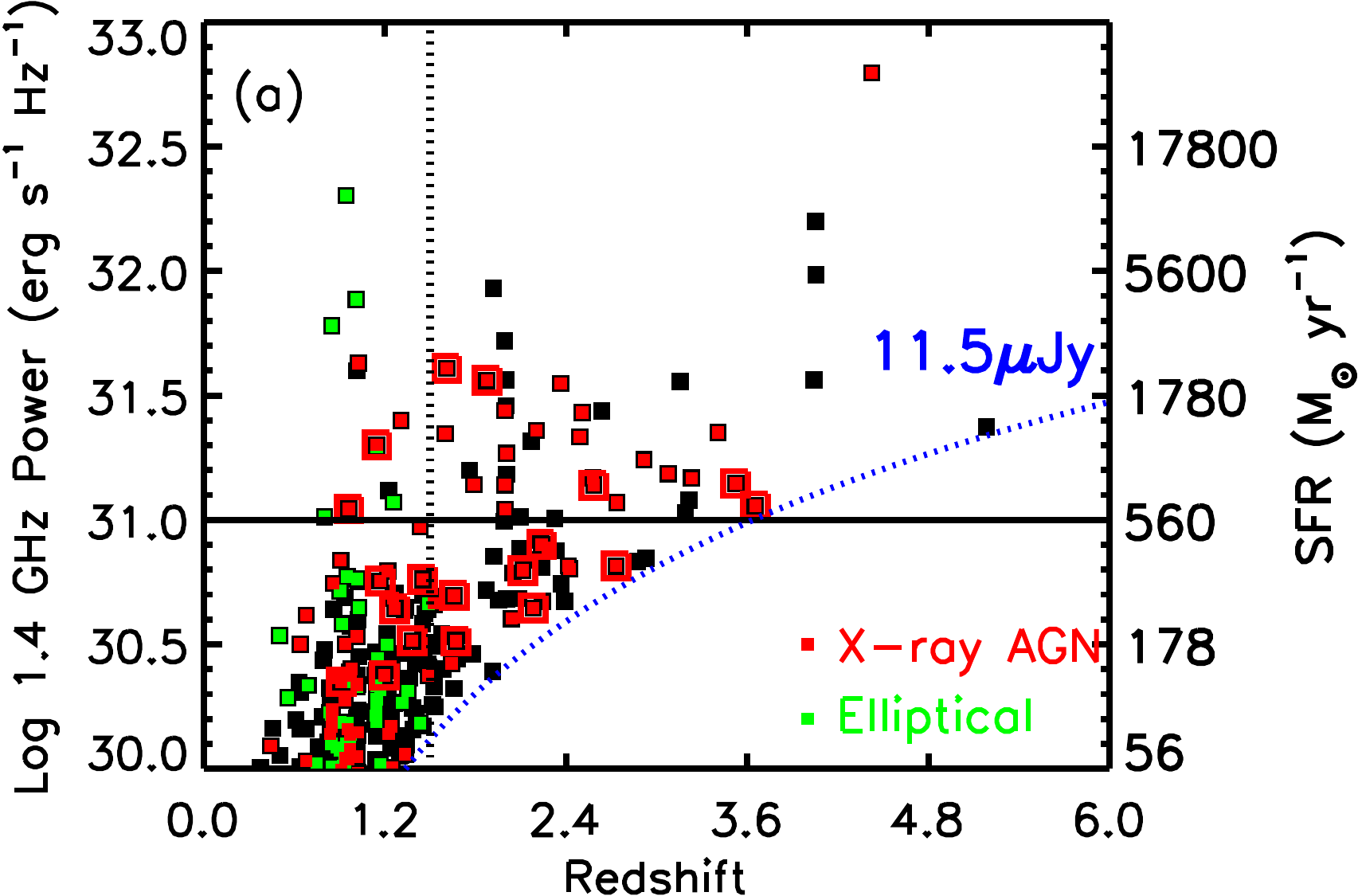}}
\centerline{\includegraphics[width=3.2in,angle=0]{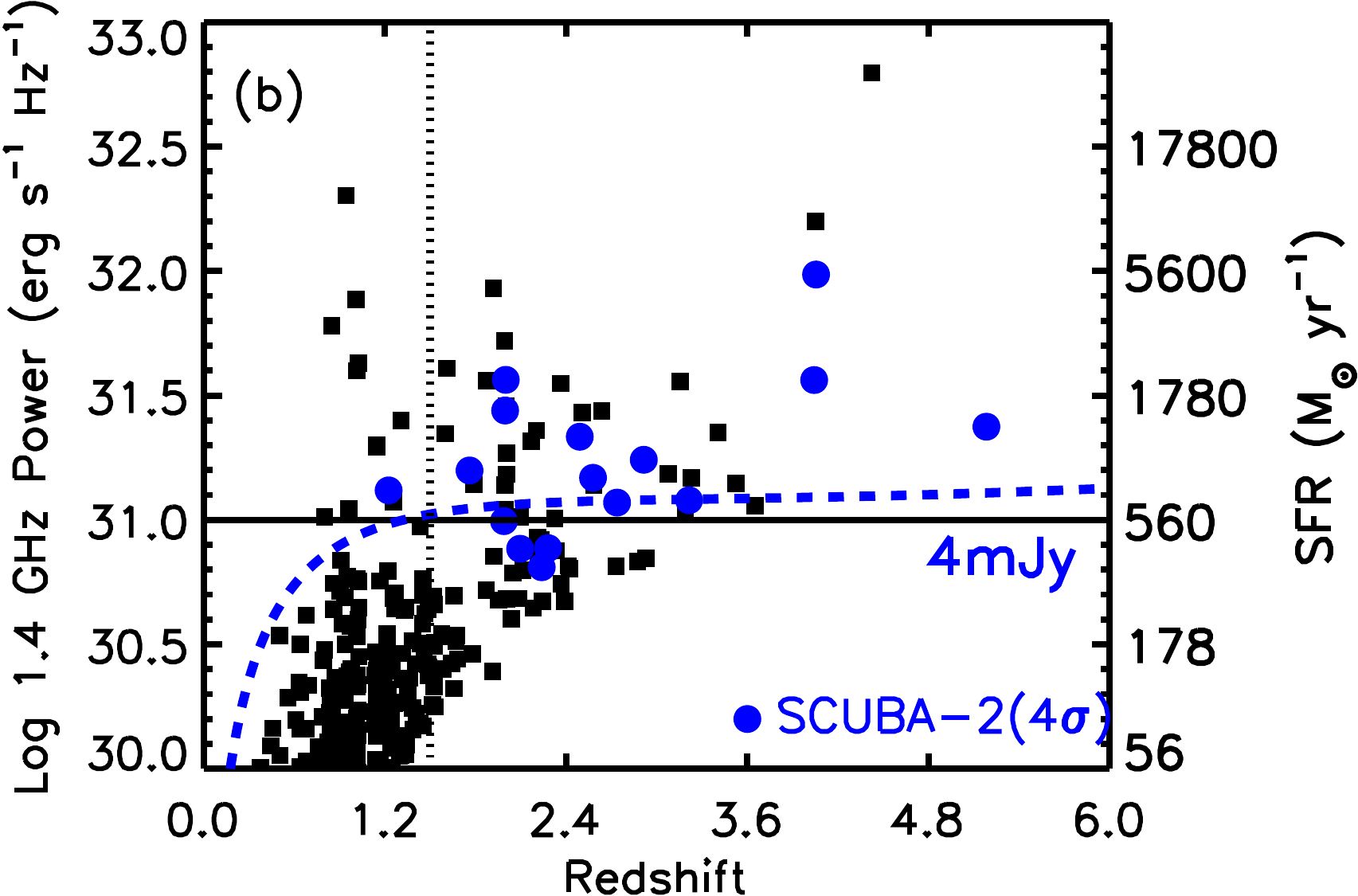}}
\caption{
Radio power vs. redshift for the 
$P_{\rm 1.4~GHz}>10^{30}$~erg~s$^{-1}$~Hz$^{-1}$ 
radio sources in the full field with spectroscopic, CO, or
photometric redshifts (black squares).
The solid horizontal line shows our high radio power dividing 
line of $P_{\rm 1.4~GHz}=10^{31}$~erg~s$^{-1}$~Hz$^{-1}$.
(a) Green squares show radio sources with measured rest-frame 
EW([OII]$\lambda3727)<10$~\AA; only $z<1.5$ sources can be classified
as elliptical galaxies this way. X-ray AGNs are marked with red squares, 
and X-ray quasars are marked with red large squares. 
The blue dotted curve shows the radio power corresponding to the 1.4~GHz
catalog limit of 11.5~$\mu$Jy  ($5\sigma$).
(b) Blue circles show single radio sources with 
$>4\sigma$ 850~$\mu$m counterparts. 
The blue dashed curve shows the 850~$\mu$m
limit of 4~mJy (4$\sigma$) for the higher sensitivity region 
of the SCUBA-2 map (see Figure~\ref{area})
converted to a radio power assuming an Arp~220 SED. 
The right-hand $y$-axes show the SFRs that would 
correspond to the radio powers if the sources are powered by star formation
(see Section~\ref{subsfr}).
\label{radio_double}
}
\end{inlinefigure}

Nearly all of the high radio power sources at $z<1.5$ are X-ray quasars or 
elliptical galaxies (see Figure~\ref{radio_double}(a)), both of which are likely to be 
AGN powered (Condon 1989; Sadler et al.\ 1989; Condon et al.\ 2013 and references
therein). Of the two remaining sources, one is a SCUBA-2 source, and the 
other is likely to be AGN powered, since it would be easily detectable with
SCUBA-2 if it were dominated by star formation
(see Figure~\ref{radio_double}(b)).

In contrast, at high redshifts ($z>1.5$), a substantial fraction of the high
radio power sources are detected at $>4\sigma$ in the submillimeter data.
In Section~\ref{subsfr}, we use the (albeit limited) available data to show 
that high radio power sources detected with SCUBA-2 are primarily extended
and dominated by star formation rather than spatially
compact and dominated by AGN activity. 
Thus, our detection at $850~\mu$m of 15 of the 41 high radio power sources 
at $z>1.5$ suggests that 37\% are star formers.

However, there are strong selection effects in the spectroscopic 
identifications of the radio sources at high redshifts, both in the targeting 
(e.g., by investing long integration times on the submillimeter detected galaxies
through multiple masks or by obtaining CO redshifts) and in the 
ease of making the identifications (e.g., by seeing strong emission line features).
Indeed, since ``red and dead'' galaxies would be hard to identify spectroscopically
at high redshifts and hence do not appear on Figure~\ref{radio_double}, 
we may expect that our star-forming fraction is overestimated.

We test the impact of our spectroscopic incompleteness on our estimated 
star-forming fraction using the $K_s>21$ high radio power sources.
At $K_s\le21$, the combined spectroscopic and photometric redshifts provide an
essentially complete identification of the radio sample in the GOODS-N region
(see Figure~\ref{radio_hist}(a)).
It is only above this magnitude where the identifications become substantially 
incomplete.

In Figure~\ref{kmg_flux}, we show $850~\mu$m signal-to-noise 
ratio versus $K_s$ magnitude for the $K_s>21$ high radio power sample in
the region of the SCUBA-2 image where the rms 850~$\mu$m noise 
is $<1$~mJy.
Based on the $K-z$ relation, the unidentified $K_s>21$ 
sources are estimated to lie at high redshifts.
Consistent with this are the high redshifts of the sources with spectroscopic, 
CO, or photometric identifications, which we mark on the plot using 
red ($z=1-2$), pink ($z=2-4$), and cyan squares ($z>4$).
Thus, we can roughly compare with our 
previous estimate made for the secure $z>1.5$ sources.
We find that 29 of the 179 (16\%) sources in the figure are detected above 
the $3\sigma$ level at $850~\mu$m, and 23 of 179 (13\%)  are detected 
above the $4\sigma$ level.
This result is insensitive to choosing a fainter $K_s$ magnitude threshold.

\vskip 0.5cm
\begin{inlinefigure}
\centerline{\includegraphics[width=3.2in,angle=0]{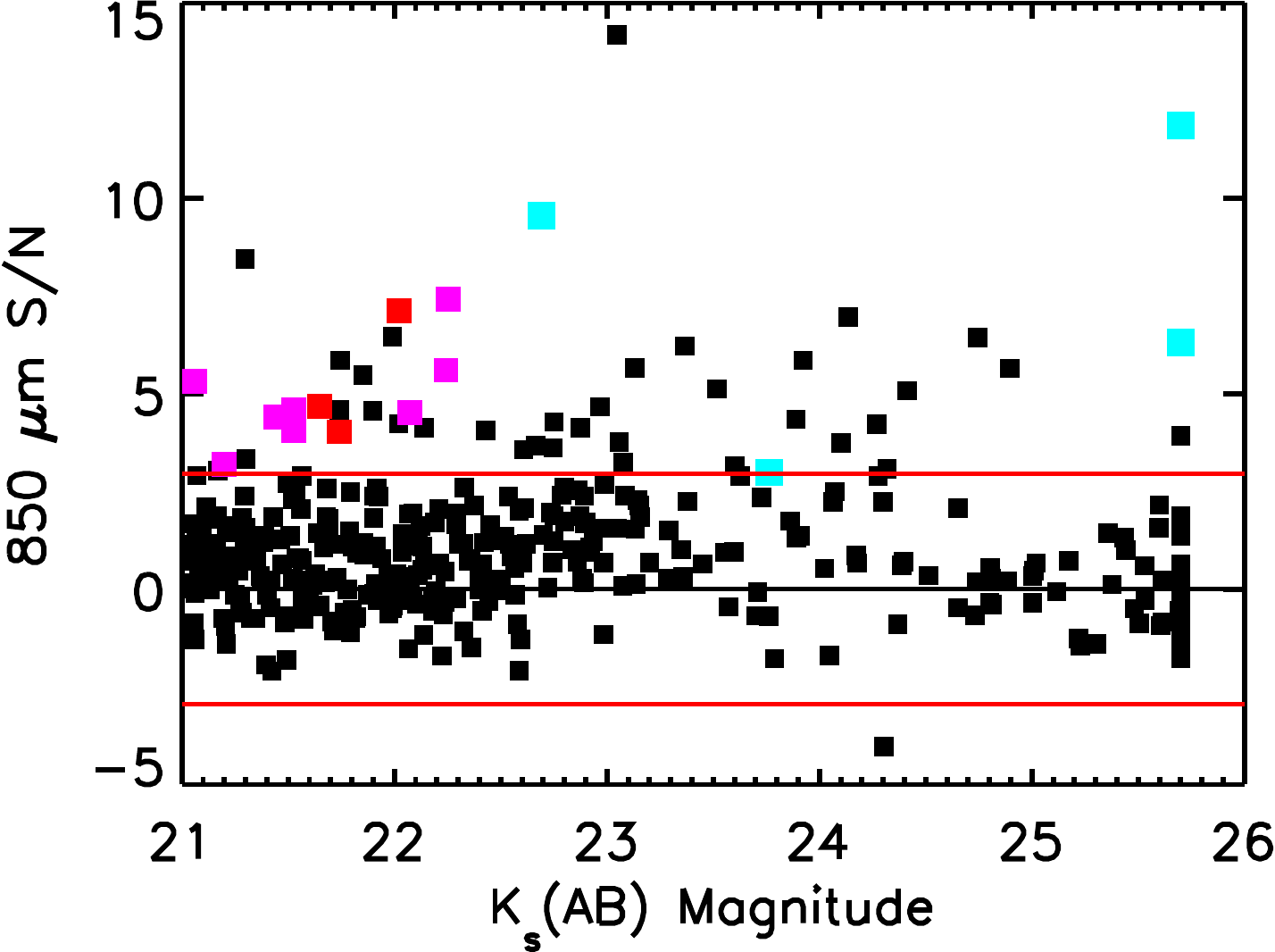}}
\caption{
850~$\mu$m signal-to-noise ratio vs. $K_s$ magnitude for
the $K_s>21$ radio sources in the region of the SCUBA-2 image
where the rms 850~$\mu$m noise is less than 1~mJy.
The red horizontal lines mark the 3$\sigma$ positive and negative noise levels.
The red (pink) squares show galaxies with spectroscopic 
or photometric redshifts $z=1-2$ ($z=2-4$). 
The cyan squares show galaxies with CO redshifts (these are all at $z>4$;
see Figure~\ref{radio_kmg}). 
\label{kmg_flux}
}
\end{inlinefigure}

\subsection{Radio Power Based Star Formation Rates}

If, as we shall argue in Section~\ref{subsfr}, the submillimeter detected  
radio galaxies are primarily star formation dominated, then we can calculate the 
SFRs for the individual galaxies from their radio powers. However, in order to do so,
we need to assume that the FIR-radio correlation is roughly invariant
to $z\sim6$. Barger et al.\ (2012) showed this to be true at $z=2-4.2$ for luminous 
SMGs, but here we assume that the FIR-radio correlation extends to
our highest sensitivity $850~\mu$m flux threshold of 2~mJy ($4\sigma$), which 
is about a factor of 2 lower than the fluxes of the 5 SMGs in the Barger et al.\ (2012) 
clean SMA sample (see their Section~5) with well-determined SEDs and measured FIR 
luminosities (hereafter, the 5 SMGs).

We convert radio power to SFR using the FIR-radio correlation
(Helou et al.\ 1985; Condon et al.\ 1991), parameterized by the quantity $q$,
\begin{equation}
q = \log \left(\frac{L_{\rm FIR(8-1000~\mu {\rm m})}}{3.75\times 10^{12}~{\rm erg~s^{-1}}} \right) - \log \left(\frac{P_{\rm 1.4~GHz}}{\rm erg~s^{-1}~Hz^{-1}} \right) \,,
\end{equation}
and the Kennicutt (1998) relation between $L_{{\rm FIR}(8-1000~\mu {\rm m})}$
and SFR. This gives
\begin{equation}
\log {\rm SFR} (M_\odot~{\rm yr}^{-1}) = \log {P_{\rm 1.4~GHz}~({\rm erg~Hz^{-1}})} - A \,.
\label{sfr_power}
\end{equation}
To determine the normalization constant, 
$A$, we use the $\langle q\rangle=2.51$ value obtained by Barger et al.\ (2012) 
from the 5 SMGs. We find $A=28.25$, a value which
is almost identical to that determined by Bell (2003) ($A=28.26$) 
and about a  factor of two lower than that determined by 
Condon (1992) based on the Milky Way. 
Barger et al.\ (2012) followed Cowie et al.\ (2011) in using an intermediate
normalization of $A=28.1$, which is a factor of 1.4 higher than the present value.
However, here, in order to be consistent with our submillimeter determinations of the 
SFRs, we stay with $A=28.25$.
The factor of two range between the Bell and Condon determinations is probably a
reasonable measure of the systematic uncertainty in the SFR-radio power relation.

The SFRs that we obtain from Equation~\ref{sfr_power}  are for a -1.35 
power-law Salpeter (1955) initial mass function (IMF) extending 
from $0.1-100~M_\odot$.
(This assumption is built into our adopted value of the normalization constant $A$
through our use of the Kennicutt (1998) relation, which is calculated for that IMF.)
The Salpeter IMF only differs significantly from the current
best IMFs of Kroupa (2001) and Chabrier (2003) below $1~M_\odot$.
One can roughly convert the Salpeter IMF SFRs
into Chabrier IMF SFRs by dividing by 1.39 and the Salpeter IMF SFRs
into Kroupa IMF SFRs by dividing by 1.31.

On the right-hand $y$-axes of Figure~\ref{radio_double},
we show the SFR scale corresponding to the radio power scale for
the radio star formers.
Our high radio power source definition corresponds to a
SFR of $800~M_{\sun} ~{\rm yr}^{-1}$ for a star formation
dominated radio source. The submillimeter detected sources are seen to
have SFRs up to $\sim6,000~M_{\sun} ~{\rm yr}^{-1}$.

At very high redshifts, the relation between SFR and radio power,
and presumably also the FIR-radio correlation,
must begin to break down, particularly for less luminous galaxies,
because the Compton cooling of the relativistic electrons on the
Cosmic Microwave Background (CMB), which increases rapidly
with increasing redshift, will begin to dominate over synchrotron
losses (e.g., Condon 1992). This will decrease the radio power
for a given SFR. The cross-over point will occur when the energy 
density in the CMB becomes comparable to the magnetic field energy
density in the galaxy. 

We emphasize that such additional sources of cooling would cause us to 
underestimate the SFRs based on the observed radio power. 
However, for the ULIRGs of the present
sample, where the magnetic field and relativistic energy density
are expected to be extremely high, this breakdown of the FIR-radio
correlation may not occur over the $z<6$ redshift range that
we are considering.

\subsection{Submillimeter Flux Based Star Formation Rates}

For sources with spectroscopic or photometric redshifts, we can compute the 
SFR directly from the observed frame $850~\mu$m flux using the
Kennicutt (1998) relation between $L_{{\rm FIR}(8-1000~\mu {\rm m})}$
luminosity and SFR, if we assume a spectral shape, such as Arp~220.
%
%
As is well known, this relation is almost redshift independent
for sources above $z=1.5$ (Blain \& Longair 1993; see blue dashed curve 
in Figure~\ref{radio_double}(b)). For an Arp~220 shape obtained from
the fits of Klaas et al.\ (1997) over the $8-1000~\mu$m range, the relation is
${\rm SFR}_{850~\mu{\rm m}} = 180 \times S_{850~\mu{\rm m}}$, where
${\rm SFR}_{850~\mu{\rm m}}$  is in $M_{\sun}$~yr$^{-1}$, and 
$S_{850~\mu{\rm m}}$ is in mJy.

\vskip 0.5cm
\begin{inlinefigure}
\centerline{\includegraphics[width=3.2in,angle=0]{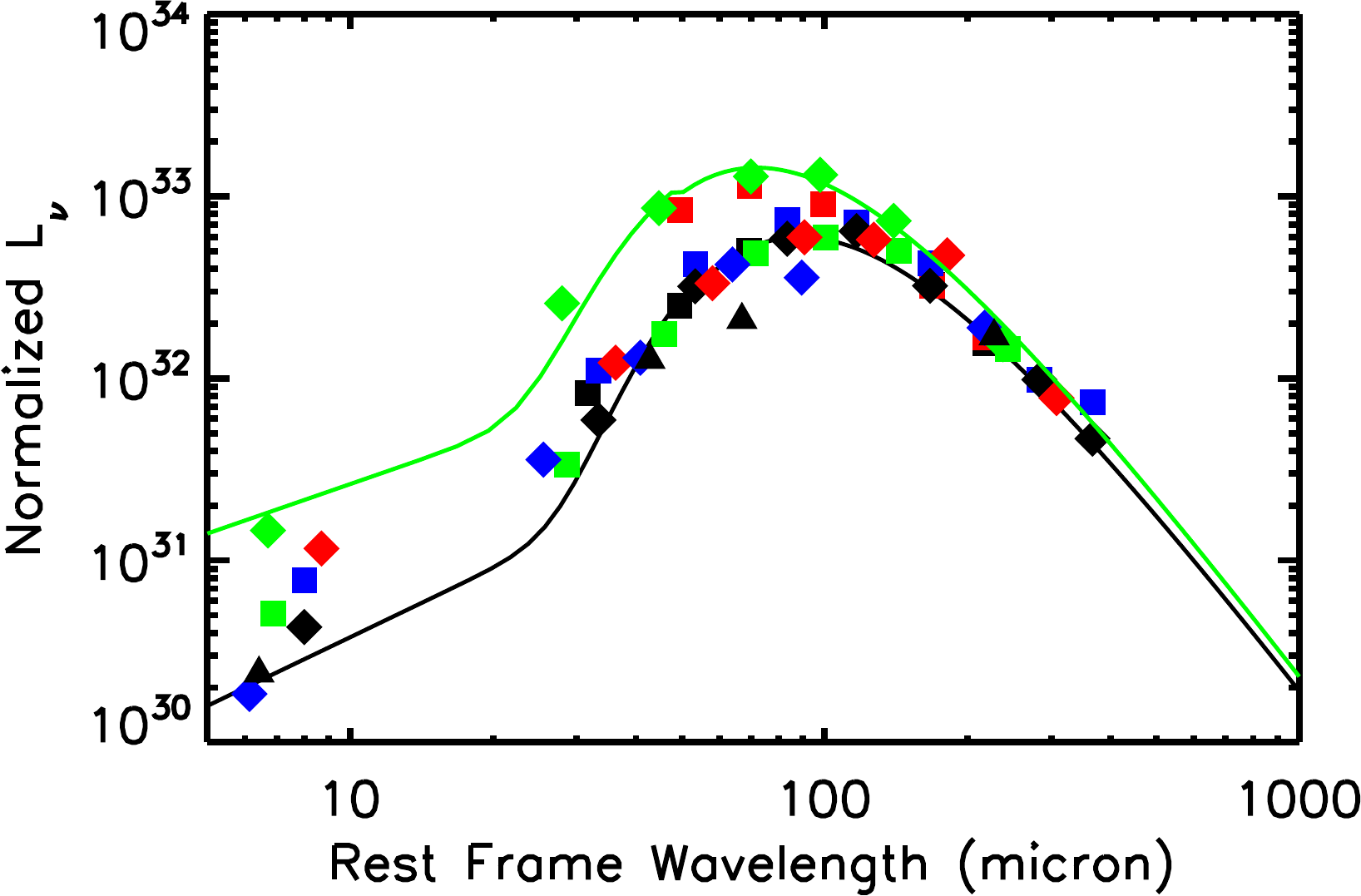}}
\caption{SEDs for the 9 isolated SMGs with spectroscopic or CO redshifts 
between $z=1.5$ and 5 and substantial coverage from the {\em Herschel\/} satellite. 
We show rest-frame $L_{\nu}$ divided by observed-frame SMA flux. 
The SEDs are very similar for 
seven of the sources but higher for the remaining two.
The sources shown are CDFN3 (black
squares), CDFN11 (red squares), CDFN13 (blue squares), CDFN14 (green squares),
CDFN16 (black diamonds), CDFN22 (red diamonds), CDFN27 (blue diamonds),
CDFN29 (green diamonds), and CDFN37 (black triangles). The two solid
curves show the combined gray body and power law fits to the SEDs
of CDFN3 (black) and CDFN29 (green).
\label{show_fir_850}
}
\end{inlinefigure}

We may directly measure the conversion for the 9 SMGs
that are spatially isolated based on the SMA and 24~$\mu$m 
images, have spectroscopic redshifts greater than 1.5, and are covered
by the {\em Herschel\/} data. (The 5 SMGs
of Barger et al.\ (2012) are a subset of this sample.) We first compiled
the fluxes in the 24, 70, 100, 160, 250, 350, 500, 860, and 1100~$\mu$m
bands from Magnelli et al.\ (2012, 2013), Barger et al.\ (2012),
and Perera et al.\ (2008).
In Figure~\ref{show_fir_850}, we show the rest-frame $L_{\nu}$ 
divided by the observed-frame SMA flux for these 9 sources.
Hereafter, we will say 850~$\mu$m everywhere instead of alternating 
between 850~$\mu$m (SCUBA-2) and 860~$\mu$m (SMA), since, within 
the uncertainties, the differences are not important.
 
For each source we fitted both a gray body and, below a rest-frame 
wavelength of 50~$\mu$m, a power law (see, e.g., Casey 2012
for a discussion of the fitting procedures). We show two sample fits
in Figure~\ref{show_fir_850}. In Figure~\ref{sma_conversion}, we
show the $L_{{\rm FIR}(8-1000~\mu{\rm m})}$ to observed-frame 
850~$\mu$m flux ratios that we determined from the fits.
Converting these luminosities to SFRs using the Kennicutt (1998) 
formula, we find a mean conversion at $z>1.5$ of
\begin{eqnarray}
{\rm SFR}_{850~\mu{\rm m}} &=&  200\times S_{850~\mu{\rm m}} \,,
\label{sfr_850}
\end{eqnarray}
with a multiplicative range over the individual values of just over 2 in 
each direction about the mean.  (Seven of the sources have similar conversions,
while two of the sources have higher conversions.) We adopt this value,
which is quite close to that inferred from the Arp~220 shape, as our conversion factor.
We will assume that the multiplicative range of 2 is the systematic error in the 
individual SFRs determined from the submillimeter fluxes based on the
variations in their SEDs.

\vskip 0.5cm
\vskip 0.1cm
\begin{inlinefigure}
\centerline{\includegraphics[width=3.2in,angle=0]{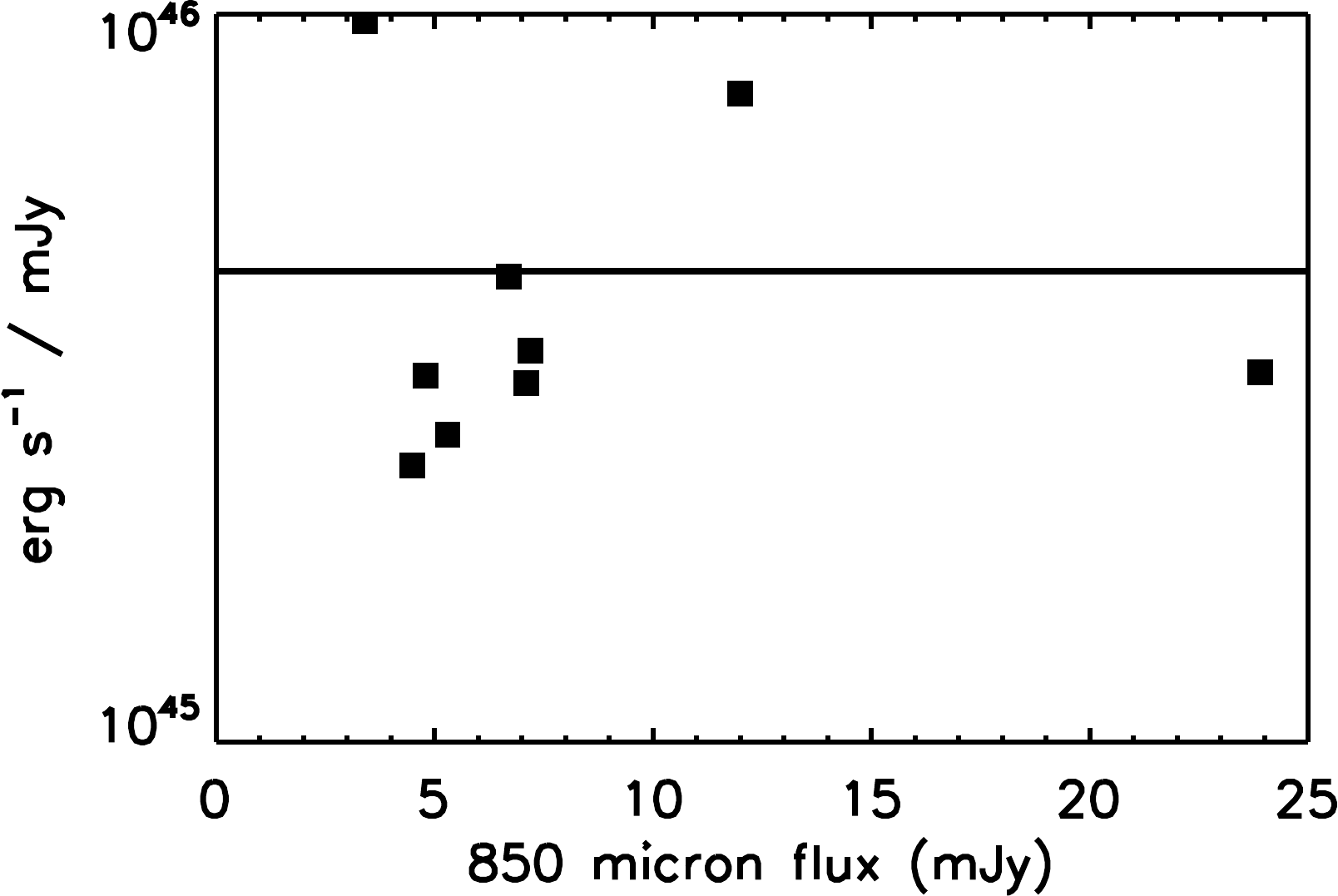}}
\caption{
Ratio of measured $L_{{\rm FIR}(8-1000~\mu {\rm m})}$
to observed-frame SMA flux vs. observed-frame SMA flux
for the 9 isolated SMGs with 
spectroscopic or CO redshifts above 1.5 and substantial 
coverage from the {\em Herschel\/} satellite. The solid line
shows the mean conversion of $4.44\times10^{45}$~erg~s$^{-1}$~mJy$^{-1}$.
\label{sma_conversion}
}
\end{inlinefigure}

We also computed the mean fluxes for the isolated sample 
based on direct measurements in the GOODS-{\em Herschel\/}
images (DR1 release; Elbaz et al.\ 2011). 
We measured the fluxes using a matched filter equal to the
point spread function of the image centered at the radio position. 
We compared the SFR 
conversion for the fainter sources at 850$~\mu$m ($2-5$~mJy) 
with that for the brighter ($>5$~mJy) sources.
For both samples, we computed the mean luminosity normalized
to the observed-frame 850~$\mu$m flux at the mean
rest-frame wavelength of the sample.
We only included sources
lying within regions of the 100~$\mu$m image where the
exposure time was greater than 25\% of the maximum
exposure time. We computed the background
correction and the 68$\%$ confidence limits by constructing
100 equivalent samples with the same redshift distribution
but randomized positions.

In Figure~\ref{sed_flux}(a), we show the results for the spectroscopic 
sample only, and in Figure~\ref{sed_flux}(b), we show the results for 
a much larger sample that uses millimetric redshifts when 
we do not have spectroscopic redshifts. The $2-5$~mJy sample
conversion is 2\% higher than the $>5$~mJy sample conversion
if we use the spectroscopic
sample, and it is 25\% lower than the $>5$~mJy sample 
conversion if we use the spectroscopic plus millimetric redshift
sample.
However, in both cases, the SEDs are consistent within
the errors. (Note that because the noise is due to confusion,
the errors are correlated between bands.) We therefore conclude
that the SFR conversion is not strongly dependent on flux
over the observed flux range. A similar test shows no evolution
in the SFR conversion as a function of redshift.

\vskip 0.5cm
\vskip 0.1cm
\begin{inlinefigure}
\centerline{\includegraphics[width=3.2in,angle=0]{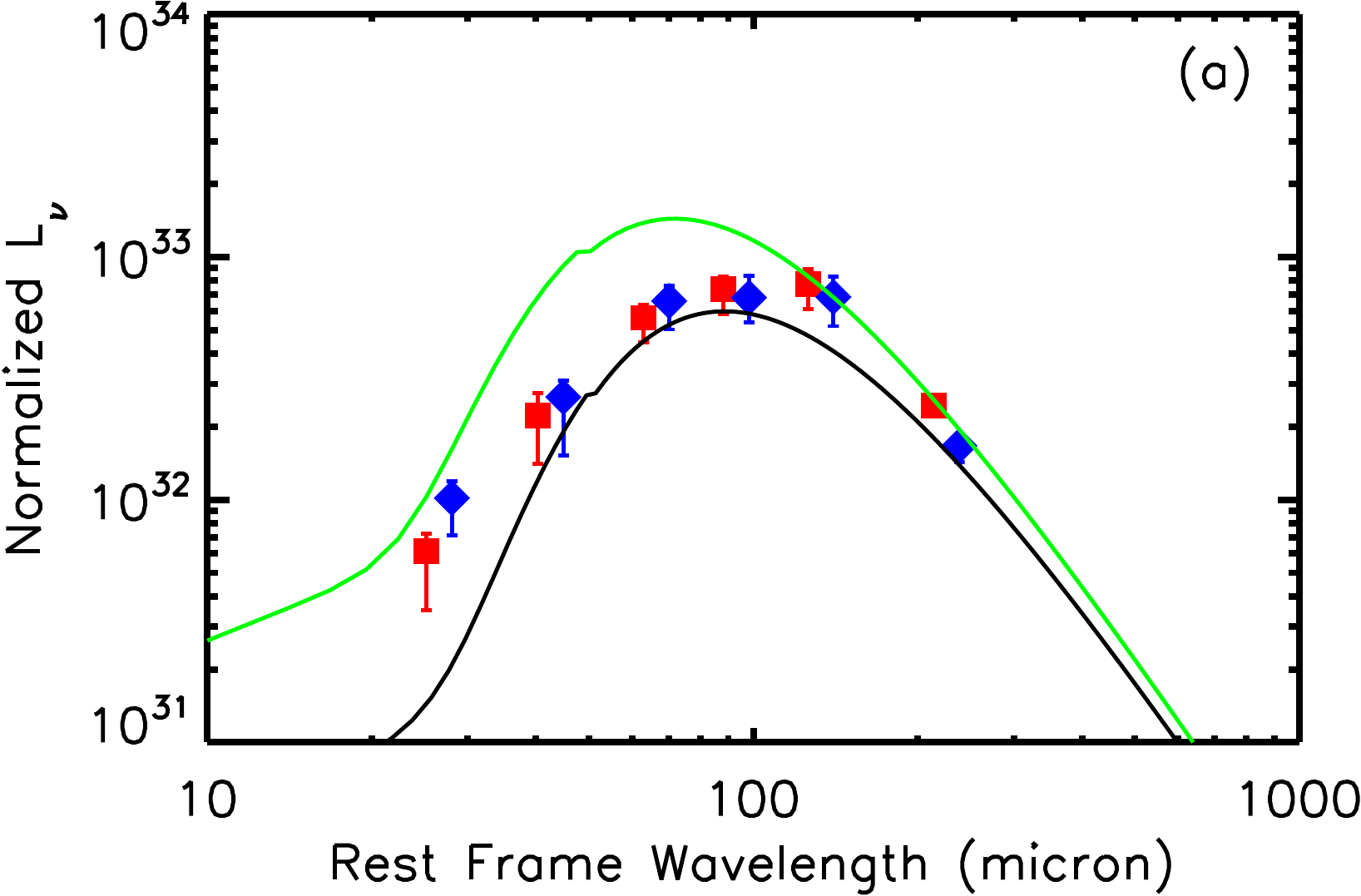}}
\centerline{\includegraphics[width=3.2in,angle=0]{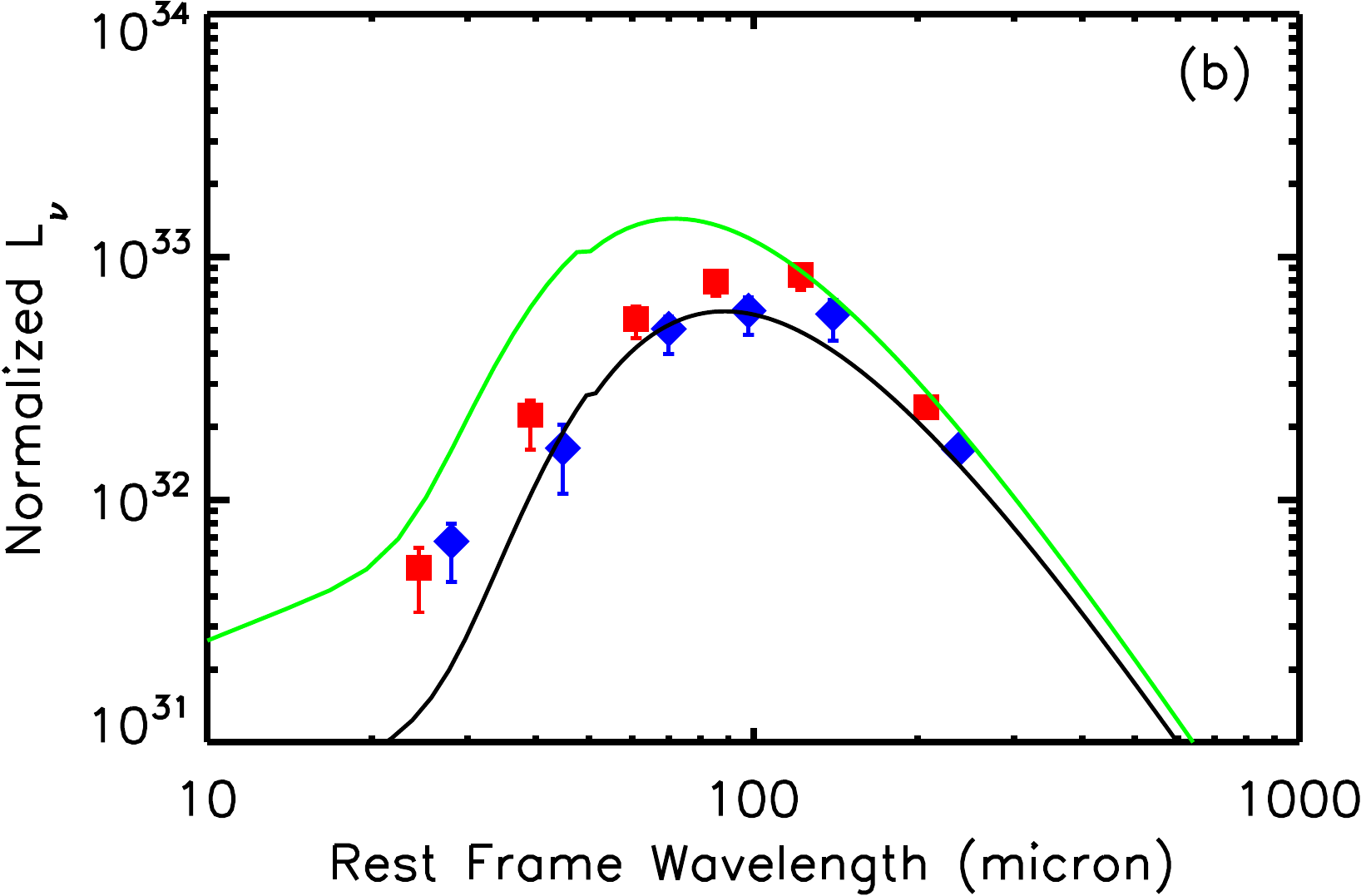}}
\caption{Mean value of $L_\nu$ divided by observed-frame 850~$\mu$m flux
in each of the {\em Herschel\/} bands from 100 to 500~$\mu$m and in the
SCUBA-2 850~$\mu$m band vs. the mean rest-frame wavelength of the
sample. The blue diamonds (red squares) show the values for isolated sources
with radio source identifications and 850$~\mu$m fluxes between 2 and 5~mJy
($>5$~mJy). (a) Sources with spectroscopic redshifts.
(b) Sources with spectroscopic redshifts or with millimetric 
redshifts when there is no spectroscopic identification. The latter includes 
12 sources in the high flux range and 16 sources in the low flux range.
The two solid curves show the combined gray body and power law fits to the 
SEDs of CDFN3 (black) and CDFN29 (green) from Figure~\ref{show_fir_850}, 
which give the range of the individual fits.
\label{sed_flux}
}
\end{inlinefigure}

We can only obtain independent estimates of the SFRs from the radio power
and the submillimeter flux where we have spectroscopic, CO, or photometric
redshifts. Where we only have millimetric redshifts, the SFRs
obtained from the radio power will agree with those from the
submillmeter fluxes by construction, since we are using a consistent
assumption about the FIR SED.

In Figure~\ref{sfr_comparison}, we compare the SFRs derived from the 
two methods for sources with spectroscopic, CO, or photometric redshifts $z>1.5$.
The SFRs derived are in good agreement, though the submillimeter derived SFRs are about
$5\%$ higher,  on average, than the radio derived SFRs. This is well
within the uncertainties in the absolute calibration of the SFRs, and we conclude
that either method will produce similar results.

\vskip 0.5cm
\begin{inlinefigure}
\centerline{\includegraphics[width=3.2in,angle=0]{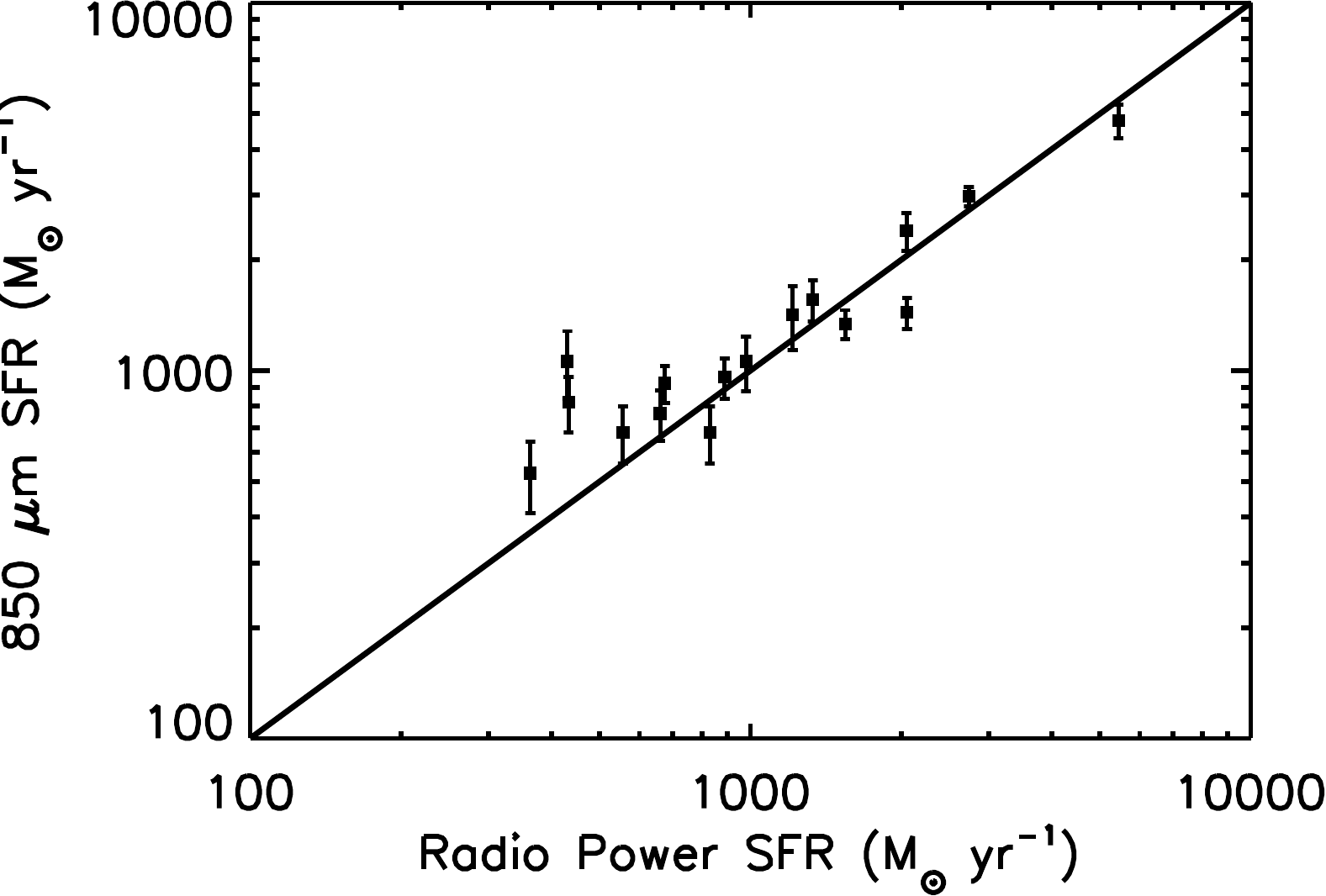}}
\caption{
Comparison of the SFRs derived from the radio power with those
derived from the 850~$\mu$m flux for the sources with 
spectroscopic, photometric, or CO redshifts $z>1.5$. 
The submillimeter derived 
SFRs are about 5$\%$ higher, on average, than the radio derived SFRs. 
\label{sfr_comparison}
}
\end{inlinefigure}

\subsection{Star Formers in the High Radio Power Population}
\label{subsfr}

At high radio powers, the submillimeter detected sources are clearly quite 
distinct from other radio sources.
In Figure~\ref{radio_flux}, we show $850~\mu$m flux versus
radio power for the radio sources with spectroscopic,
photometric, or CO redshifts  $z>1$ in the region of
the SCUBA-2 field where the rms 850~$\mu$m noise
is less than 1.5~mJy (black squares). 
We mark radio sources with 850~$\mu$m counterparts 
detected above the $4\sigma$ level with blue solid circles.
(If there is no SMA observation, then we only mark the
source if there is a single radio counterpart within the SCUBA-2 beam.)
These SMGs begin
to enter at a radio power of $\approx~5\times10^{30}$~erg~s$^{-1}$~Hz$^{-1}$,
as would be expected for sources with an Arp~220 SED obeying the
local FIR-radio correlation (cyan curve), given our $4\sigma$ $850~\mu$m
flux limit of 2~mJy. (This radio power corresponds to a
SFR of  $\sim 400~M_{\sun}$~yr$^{-1}$, or an $850~\mu$m
flux of 2.5~mJy, assuming the sources are powered by star formation.)
We mark X-ray AGNs with red squares and X-ray quasars with red large
squares.  None of the submillimeter detected sources are X-ray quasars.

Above this radio power, we see  a bifurcation, with some sources being undetected
in the submillimeter, even at very high radio luminosities,
while others follow the FIR-radio track of the cyan curve. We shall refer
to the two tracks as submillimeter-bright and submillimeter-blank
radio sources.

Based on the small number of sources with high-resolution
radio observations in the field, the submillimeter-bright sources
appear to be predominantly extended and star formation dominated.
The three submillimeter-bright sources in our radio sample with high-resolution 
1.4~GHz observations from either the Multi-Element Radio Linked Interferometer
Network (MERLIN)+VLA (Chapman et al.\ 2004b)
or the Very Long Baseline Interferometer (Momjian et al.\ 2010)
have all been confirmed as being extended 
(CDFN9/GOODS~850-3/GN6, CDFN7/GOODS~850-36, 
and CDFN3/GN20).
(Note that CDFN3/GN20 lies outside the area shown in Figure~\ref{radio_flux},
but it lies smoothly on the submillimeter-bright track.)

Two submillimeter-blank sources in our radio sample at $z>1$ were
classified as AGNs by Guidetti et al.\ (2013) using high-resolution 
5~GHz observations with e-MERLIN combined with existing 1.4~GHz
MERLIN+VLA observations obtained by Muxlow et al.\ (2005).
We show these enclosed in black large squares in Figure~\ref{radio_flux}.
(A further 3 of the submillimeter-blank sources shown
in Figure~\ref{radio_flux} were observed by Guidetti et al.\ but were
not clearly classified.)

We shall assume in the following that the submillimeter-bright
sources are star formation dominated, though the number of sources
used to come to this conclusion is small.

\vskip 0.5cm
\begin{inlinefigure}
\centerline{\includegraphics[width=3.2in,angle=0]{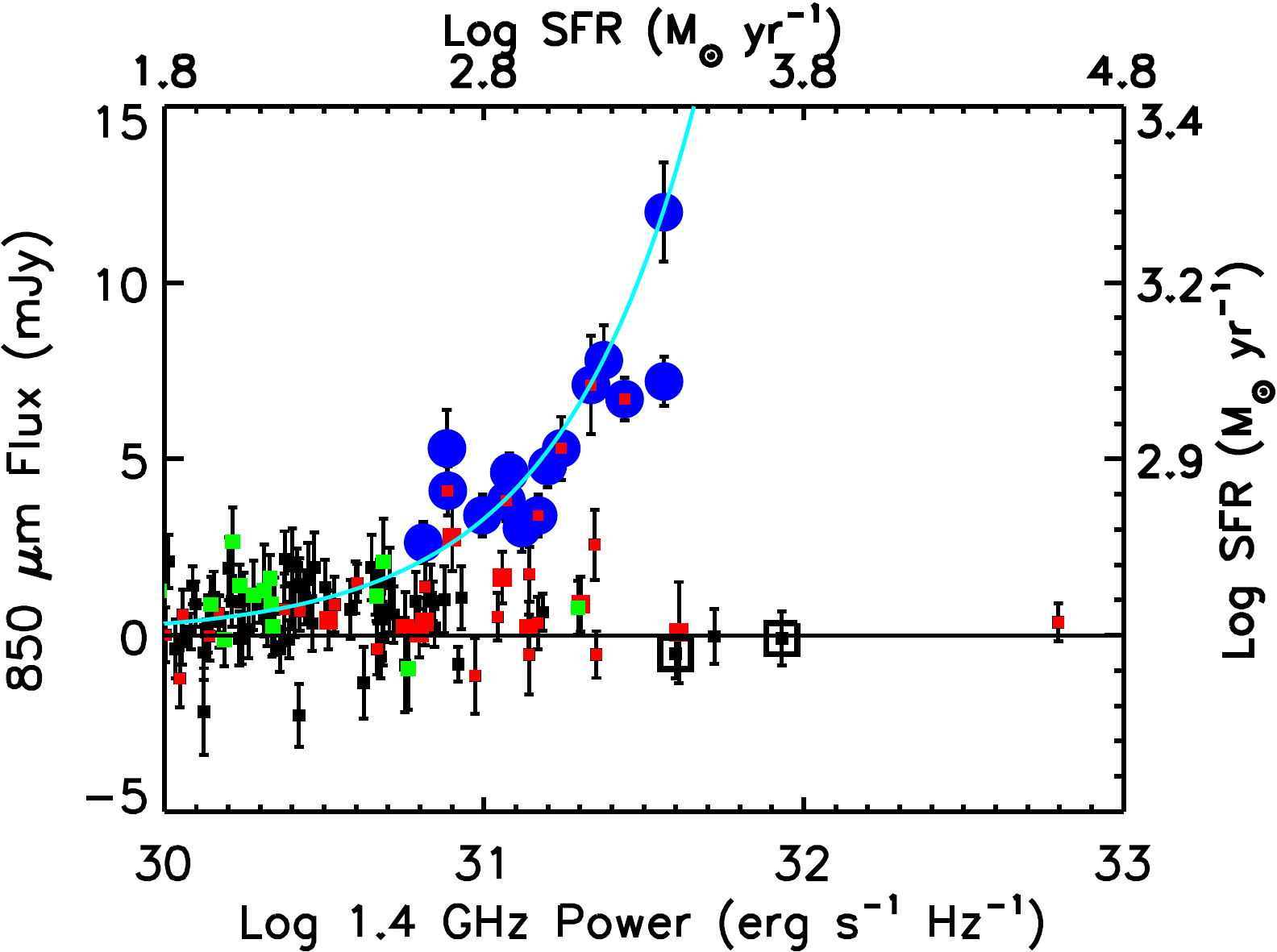}}
\caption{
850~$\mu$m flux vs. radio power for the radio sources with 
spectroscopic, photometric, or CO redshifts $z>1$ 
in the region of the SCUBA-2 image where the
rms 850~$\mu$m error is $<1.5$~mJy (black squares). 
X-ray AGNs are marked with red squares, and
X-ray quasars are marked with red large squares.
Green squares show radio sources with measured rest-frame
EW([OII]$\lambda3727)<10$~\AA; only $z<1.5$ sources can be classified
as elliptical galaxies this way.
Blue solid circles show single radio sources with well-determined 
$>4\sigma$ 850~$\mu$m counterparts. The sources enclosed in 
black large squares are sources classified as AGNs by Guidetti et al.\ (2013). 
Error bars for all the symbols are $\pm1\sigma$. 
The cyan curve shows the submillimeter flux expected for an Arp~220 SED
based on Equations~\ref{sfr_power} and \ref{sfr_850}.
The top axis (right-hand axis) shows the SFR that would correspond to the 
radio power (submillimeter flux), if the source is powered by star formation.
\label{radio_flux}
}
\end{inlinefigure}

\section{Star Formation Rate Distribution Function}
\label{secsfh}

A major goal of this paper is to search for evidence of a 
turn-down in the SFR distribution function, which would indicate a 
characteristic maximum SFR in galaxies. Here, we
use the SCUBA-2 sample of Table~1 to explore 
the shape of the SFR distribution function at high redshifts.

Of the 49 SCUBA-2 sources in Table~1, 24 have SMA 
observations that directly determine the radio counterparts. Three of these 
SCUBA-2 sources have multiple SMA/radio counterparts, giving
a total of 27 SMA detected sources. These correspond to all but two of
the sources in the SMA sample of Table~2; i.e.,
GOODS~850-13a and GOODS~850-13c are not included in the 
SCUBA-2 selection, because they lie below the detection threshold.
There are a further 18 SCUBA-2 sources for which there is only a single
radio source within the SCUBA-2 beam, which we take to be
the counterpart. The remaining 7 SCUBA-2 sources either
have multiple radio sources within the beam (this is the case for three sources)
or no radio counterpart (this is the case for four sources, including
the single SCUBA-2 source/SMA pair CDFN15a and CDFN15b
where both SMA counterparts are undetected in the radio). 
Some of the latter category could be spurious when they are close to the 
$4\sigma$ threshold, but if they are real, as is clearly the case for
CDFN15, then they are the most plausible extremely high-redshift galaxy candidates.

\vskip 0.5cm
\begin{inlinefigure}
\centerline{\includegraphics[width=3.2in,angle=0]{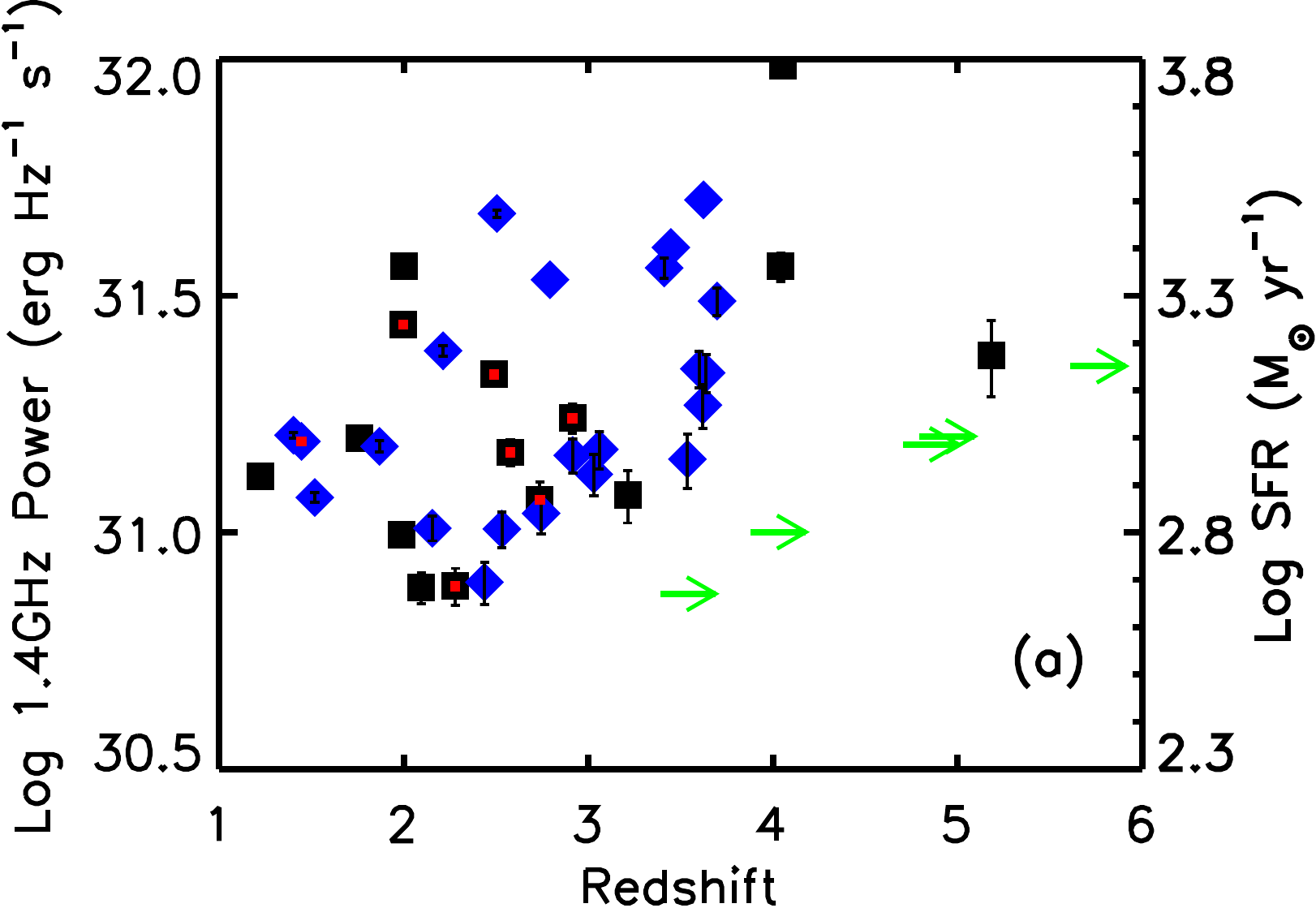}}
\vskip 0.2cm
\centerline{\includegraphics[width=3.2in,angle=0]{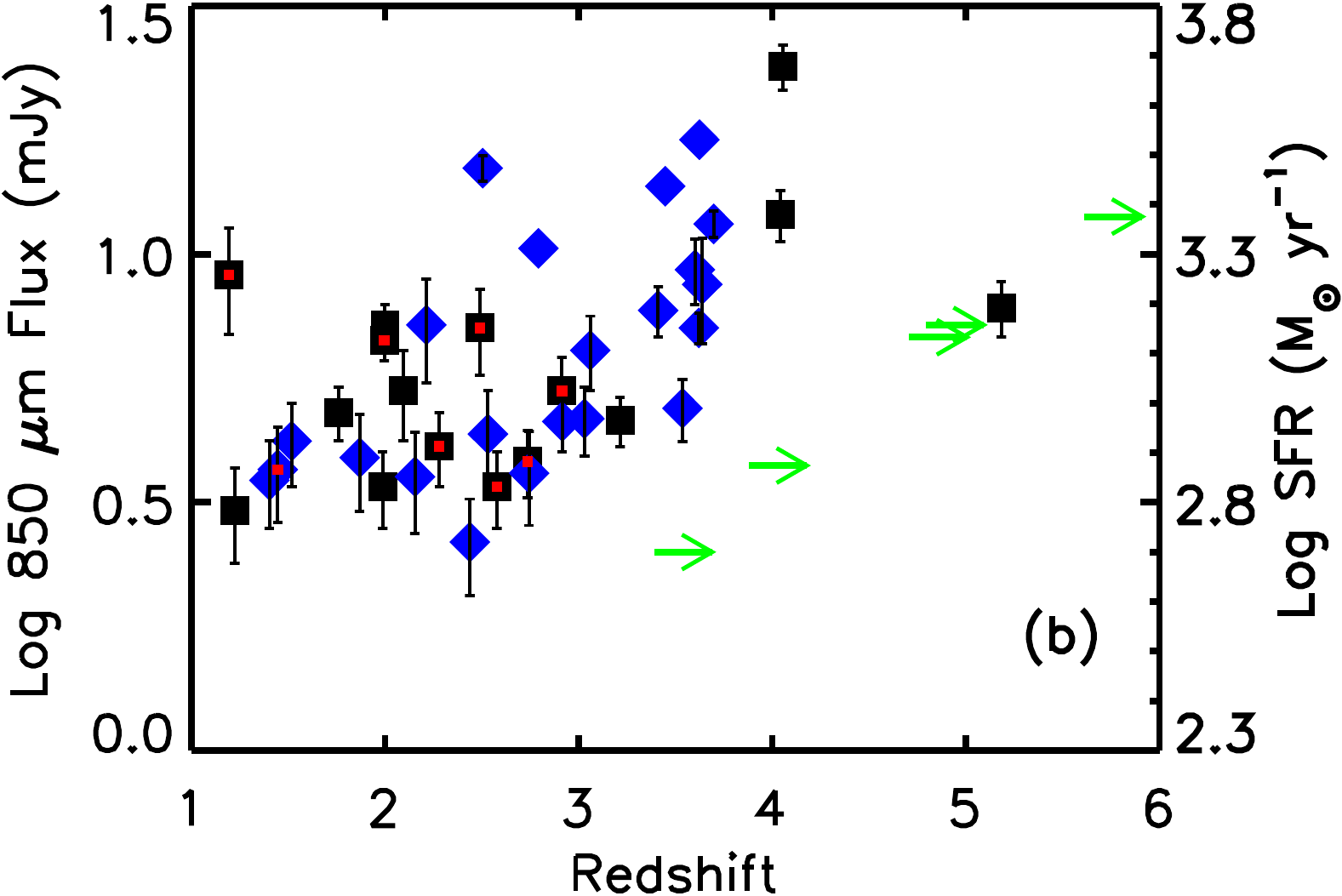}}
\caption{
(a) Radio power vs. redshift for the SCUBA-2 sample 
with single radio counterparts at $z>1$ (black squares - spectroscopic, 
photometric, or CO redshifts; blue diamonds - millimetric redshifts),
as well as the five without radio counterparts (green right-pointing 
arrows; we computed the minimum millimetric redshifts for these by 
assuming a 1.4~GHz flux of $10~\mu$Jy).
X-ray AGNs are marked with red squares.
None of the sources are X-ray quasars.
The right-hand axis shows the SFRs calculated from the radio powers
using Equation~\ref{sfr_power},
assuming the sources are powered by star formation.
(b) 850~$\mu$m flux vs. redshift for the same sample and using the
same symbols as in (a). In this panel,
the right-hand axis shows the SFRs calculated from the submillimeter 
fluxes using Equation~\ref{sfr_850},
assuming the sources are powered by star formation. 
This axis is only valid for sources at $z>1.5$.
\label{rad_power}
}
\end{inlinefigure}

In the following, we restrict our analysis to the SCUBA-2 SMGs with SMA/radio 
detections or single radio counterparts, giving a total sample of 45 galaxies.
(Note, however, that with some reasonable assumptions, we also present 
results that include the five sources without radio counterparts.)
Where possible, we use the spectroscopic, photometric, or CO redshifts.
As summarized in Table~1, 19 of the 45 sources have such redshifts,
14 of which lie at $z>1.5$. For the remaining 26 sources, we use
the millimetric redshifts from Table~1, 22 of which lie at $z>1.5$.

In Figure~\ref{rad_power}(a), we show radio power (left-hand $y$-axis) and 
the SFR calculated from the radio power using Equation~\ref{sfr_power}
(right-hand $y$-axis) versus redshift for the SMGs at $z>1$.
In Figure~\ref{rad_power}(b), we show submillimeter flux  (left-hand $y$-axis)
and the SFR calculated from the submillimeter flux using Equation~\ref{sfr_850} 
(right-hand $y$-axis) versus redshift for the same sample.
We denote sources with spectroscopic, photometric, or CO redshifts  
with black squares, and we denote sources with millimetric redshifts with blue 
diamonds. We mark X-ray AGNs with red squares. None of the sources
are X-ray quasars. We show the five sources without radio 
counterparts as green right-pointing arrows. We computed the minimum millimetric
redshifts for these by assuming a 1.4~GHz flux of 10~$\mu$Jy.

In both panels, the SFRs range from $400~M_{\sun}$~yr$^{-1}$ to 
$\sim6000~M_{\sun}$~yr$^{-1}$. For homogeneity, we decided to calculate 
the SFRs from the submillimeter fluxes in our subsequent analysis, but our results 
are not significantly changed if we instead compute the SFRs from the radio 
powers. 

For each source, we determined the area over which a $4\sigma$ detection
would have been made in the SCUBA-2 image. 
We then used this to determine the accessible volume in the redshift interval
z1 to z2. Since the conversion from $850~\mu$m flux to SFR
is nearly redshift invariant, this is just the comoving volume
between z1 and z2 that corresponds to the area for that source.
We then formed the SFR per unit volume per $\log$ SFR in the redshift interval
by summing the inverse volumes and dividing by the bin size width. We used
bins stepped by 0.5 in $\log$ SFR.
 
In Figure~\ref{star_formers}, we show the number density of sources per unit
comoving volume per unit 
$\log$ SFR versus $\log$~SFR for the $z=1.5-6$ SCUBA-2 sources with SMA
detections or single radio counterparts (black squares). Here and subsequently,
we only use the SMGs with SFRs $>500~M_\sun$~yr$^{-1}$
corresponding to $850~\mu$m fluxes $\gtrsim3$~mJy,
where we have substantial area coverage (see Figure~\ref{area}; this
only eliminates two SMGs).
The green diamonds show the same 
but assuming that the five SCUBA-2 sources without radio 
counterparts also lie in this redshift interval. Because there is no redshift 
dependence in the SFR conversion (see Equation~\ref{sfr_850}),
the submillimeter fluxes of these  sources place them in the appropriate
SFR bin. We have not included the three SCUBA-2 sources that have multiple radio 
sources within the SCUBA-2 beam, but  if they also are at
$z=1.5-6$, then they contain just under 10\% of the total
submillimeter flux, or, equivalently, of the total SFR.
Thus, the overall normalization should not be 
increased by more than this amount with their inclusion.

The red solid line shows the shape that would be required
to produce the same amount of star formation
in each logarithmic SFR interval.  The two lowest
SFR bins fall on this relation; however, above
$\log {\rm SFR} \sim 3.3$, the measured volume
density begins to drop below this relation. This drop is highly statistically 
significant, since a constant amount of star formation
in each logarithmic SFR interval would imply that we would have
23 objects above $\log$ SFR\,$\sim 3.3$ in the field, whereas we see 
only four. Over the range of the two lowest data points 
($500-2000~M_{\sun}$~yr$^{-1}$), the total SFR density is
$0.016~M_{\sun}$~yr$^{-1}$~Mpc$^{-3}$, while the contribution
from sources with SFRs above $2000~M_{\sun}$~yr$^{-1}$
is only $0.004~M_{\sun}$~yr$^{-1}$~Mpc$^{-3}$. Thus,
we appear to have a characteristic maximum SFR of 
$\sim2000~M_{\sun}$~yr$^{-1}$.

\vskip 0.5cm
\begin{inlinefigure}
\centerline{\includegraphics[width=3.2in,angle=0]{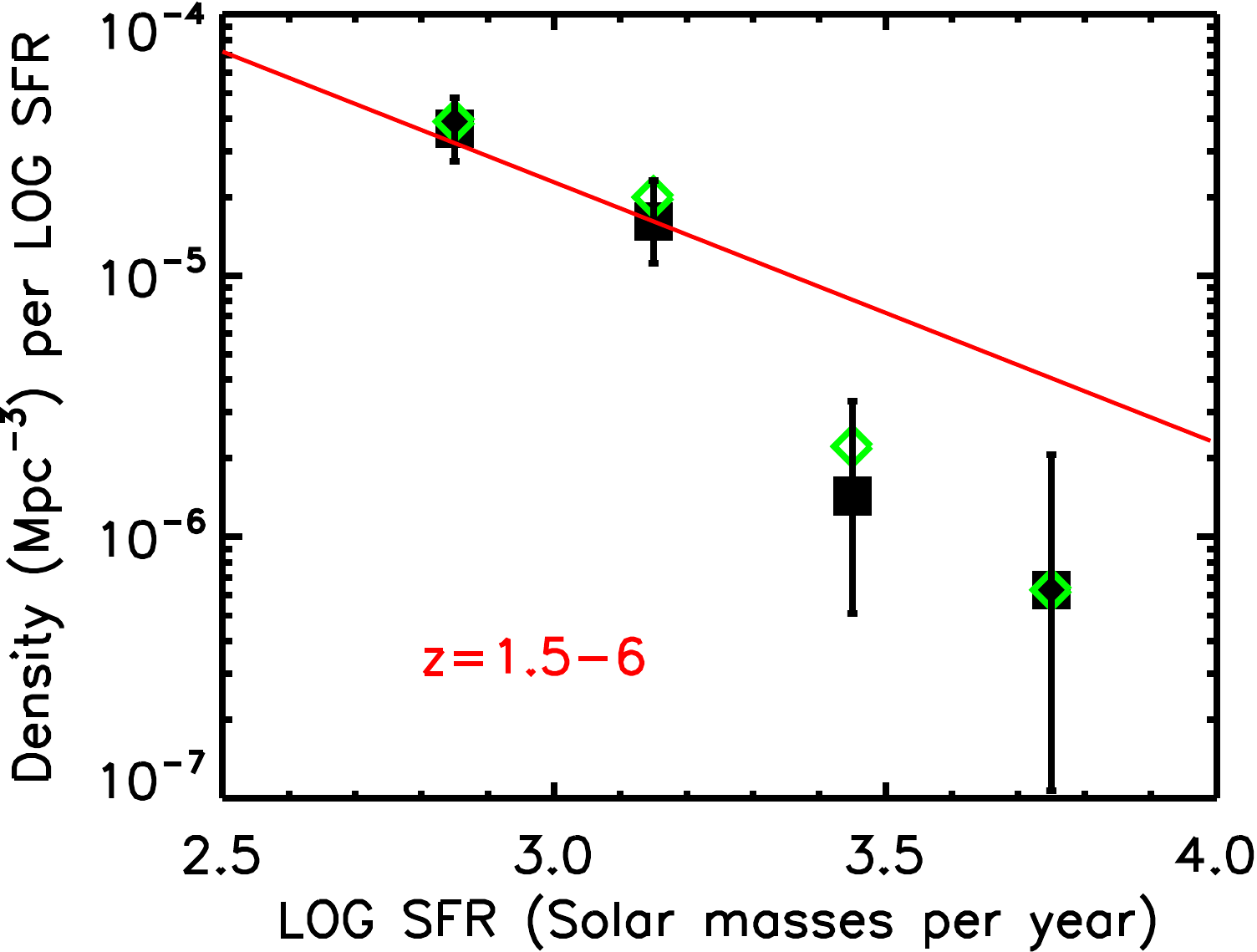}}
\caption{
Number density per unit comoving volume
per unit $\log$ SFR vs. $\log$~SFR for the $>4\sigma$ SCUBA-2
sources at $z=1.5-6$ with SFRs $>500~M_\odot$~yr$^{-1}$.
Black squares show the sources with SMA detections or single radio counterparts.
The error bars are 68$\%$ confidence ranges based on the number of 
sources in each bin.
The green diamonds show the results if the five SMGs
without radio counterparts are also assumed to lie in this redshift interval. 
The red solid line shows the shape of the SFR distribution
function that would produce equal amounts of star formation in each 
$\log$ SFR interval. 
\label{star_formers}
}
\end{inlinefigure}

It is unlikely that this result could be affected by gravitational lensing
of the submillimeter/1.4~GHz sources. While the bright end sources
in ultra-wide fields surveys are dominated by lensed sources (Negrello
et al.\ 2010), there is only a low probability of seeing a significantly lensed
source in a field of the present size (e.g., Takahashi et al.\ 2011). 
We searched around the
brightest SMGs for neighboring bright foreground galaxies
that could be plausible lensers and found only two. One
of these is HDF850.1 (Hughes et al.\ 1998) or GOODS 850-1, 
which has a nearby elliptical galaxy 
at  $z=1.224$ from Barger et al.\ (2008). Walter et al.\ (2012), using their
new redshift and position for the SMG from the IRAM Plateau de Bure 
Interferometer, derived only a modest possible amplification factor of $\sim1.4$.

We can compare the contributions that we found from the very massively
star-forming galaxies in the SCUBA-2 sample to the contributions 
from rest-frame UV selected samples. 
In Figure~\ref{star_formers_plus}, we plot volume density
versus $\log$~SFR for the SCUBA-2 galaxies from Figure~\ref{star_formers}
and for Lyman Break Galaxy (LBGs) from
the extinction-corrected UV luminosity functions of van der Burg et al.\ (2010)
(red triangles for $z=4.8$, green diamonds for $z=3.8$, 
and blue squares for $z=3.1$) and Reddy \& Steidel (2009) (blue curve 
for $z\sim3$ and cyan curve for $z\sim2$). We converted their luminosity
functions to the units of Figure~\ref{star_formers_plus} using the Kennicutt (1998) 
conversion of 1600~\AA\  luminosity to SFR for a Salpeter IMF.  

van der Burg et al.\ (2010) adopted luminosity-dependent 
dust correction factors from Bouwens et al.\ (2009).
Reddy \& Steidel (2009) also used luminosity-dependent dust corrections, 
but theirs were significantly smaller.
Indeed, van der Burg et al.\ directly compared their 
extinction-corrected SFR densities with those Reddy \& Steidel in their Figure~14
and found them to be quite different, illustrating the level of uncertainty in the 
extinction corrections.

While the distribution of SMG SFRs appears to extend smoothly
from the distribution of LBG SFRs, the LBG SFRs determined from the 
extinction-corrected UV selected samples are not as high as those of the
SMGs but instead cut off at $\sim300~M_\odot$~yr$^{-1}$.
Thus, either the SMGs are completely omitted from the UV selected samples, or
the extinction corrections applied to some UV sources are substantially 
underestimated (see discussion in Bouwens et al.\ 2009). 
Even if catastrophically wrong extinction corrections are applied to some 
UV sources, causing lower SFRs to be assigned to sources that 
genuinely have high SFRs, the UV distributions in Figure~\ref{star_formers_plus} 
would remain the same. The reason is that the volume density of SMGs is 
much smaller than that of LBGs, which means the number of sources that 
would be affected would be too small to make a difference.

\vskip 0.5cm
\begin{inlinefigure}
\centerline{\includegraphics[width=3.2in,angle=0]{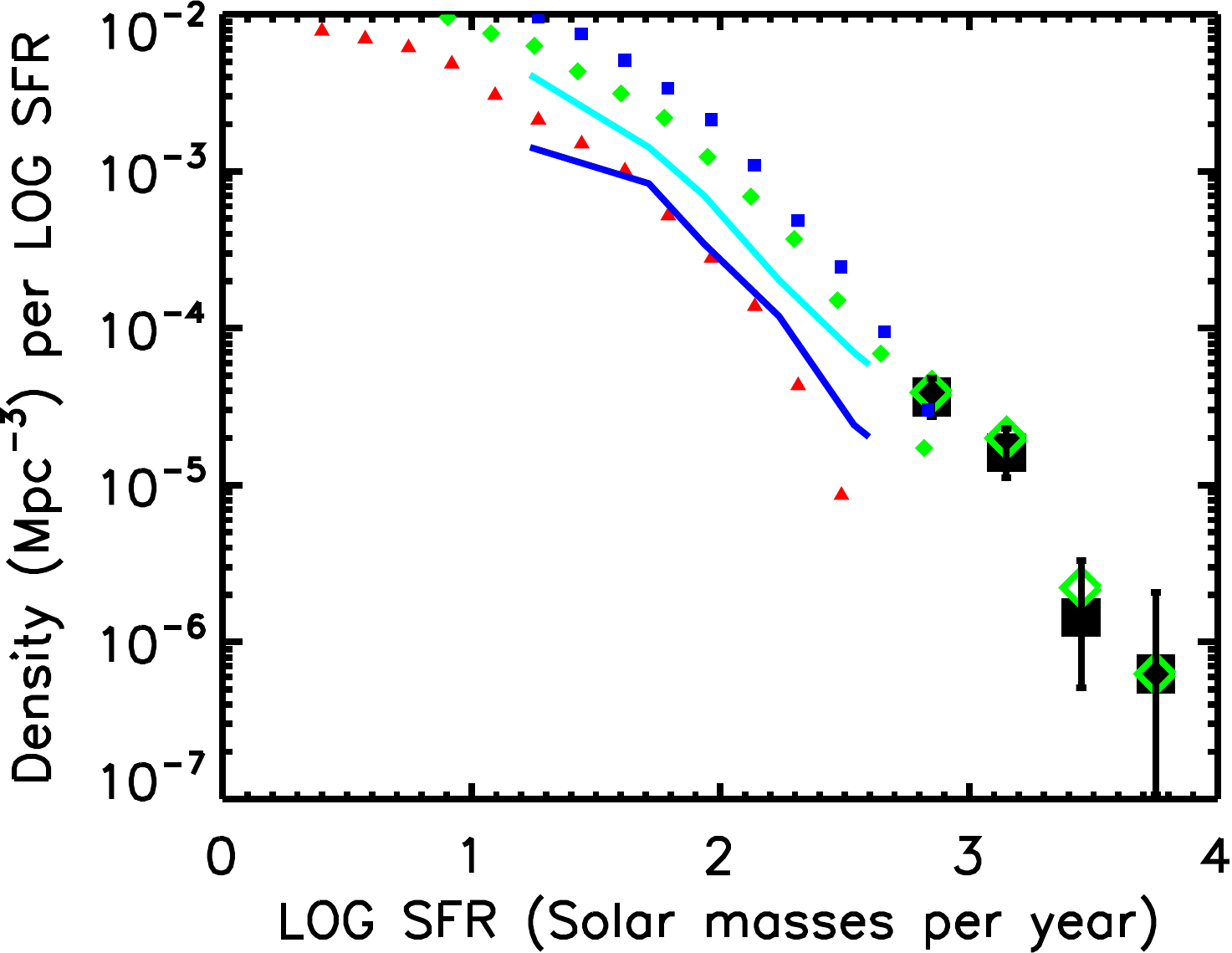}}
\caption{
Number density per unit comoving volume
per unit $\log$ SFR vs. $\log$~SFR for the $>4\sigma$ SCUBA-2
sources at $z=1.5-6$ with SFRs $>500~M_\odot$~yr$^{-1}$.
Black squares show the sources with SMA detections or single radio counterparts.
The error bars are 68$\%$ confidence ranges based on the number of 
sources in each bin.
The green diamonds show the results if the five SMGs
without radio counterparts are also assumed to lie in this redshift interval. 
For comparison, the small symbols and curves show extinction-corrected UV
results from van den Burg et al.\ (2010) 
(red triangles - $z=4.8$; green diamonds - $z=3.8$; blue squares - $z=3.1$)
and Reddy \& Steidel (2009) (blue curve - $z\sim3$; cyan curve - $z\sim2$),
assuming the Kennicutt (1998) conversion of UV luminosity to SFR for a 
Salpeter IMF.
\label{star_formers_plus}
}
\end{inlinefigure}

Since the LBGs' brightest submillimeter fluxes are only $\sim0.2-0.3$~mJy 
based on stacking analyses
(e.g., Peacock et al.\ 2000; Chapman et al.\ 2000; Webb et al.\ 2003),
with the present submillimeter sensitivities, which are set by the 
blank field confusion limit, the SMG SFRs do not overlap with LBG SFRs.
Thus, there is a gap between the two populations where the SFR distribution
function is poorly determined. 

In Figure~\ref{sfr_history}, we plot the SFR density per unit
comoving volume for the SCUBA-2 sample with SFRs $>500~M_\sun$~yr$^{-1}$
(black squares) versus redshift and compare it with the compilation by 
Hopkins \& Beacom (2006; black solid curve) 
for extinction-corrected UV selected samples over the same redshift range.
We compare these results with the SFR density history
of Barger et al.\ (2012)---who used a substantially smaller
SCUBA selected and SMA confirmed sample in the GOODS-N---after
reducing their points by a factor of 1.4 (blue open squares)
to adjust them to the SFR calibration of the present paper.
Note that Casey et al.\ (2013) constructed the most recent SFR density history
using both 450~$\mu$m and 850~$\mu$m selected 
SCUBA-2 samples in the COSMOS field, which they compared with
Barger et al.\ (2012) at 850~$\mu$m, Chapman et al.\ (2005) at 850~$\mu$m 
using SCUBA, Wardlow et al.\ (2011) at 870~$\mu$m using LABOCA, 
Roseboom et al.\ (2012) at 1.2~mm using MAMBO,
and Casey et al.\ (2012a,b) at $250-500~\mu$m using {\em Herschel}-SPIRE
(see their Figure~14).

\vskip 0.5cm
\begin{inlinefigure}
\centerline{\includegraphics[width=3.2in,angle=0]{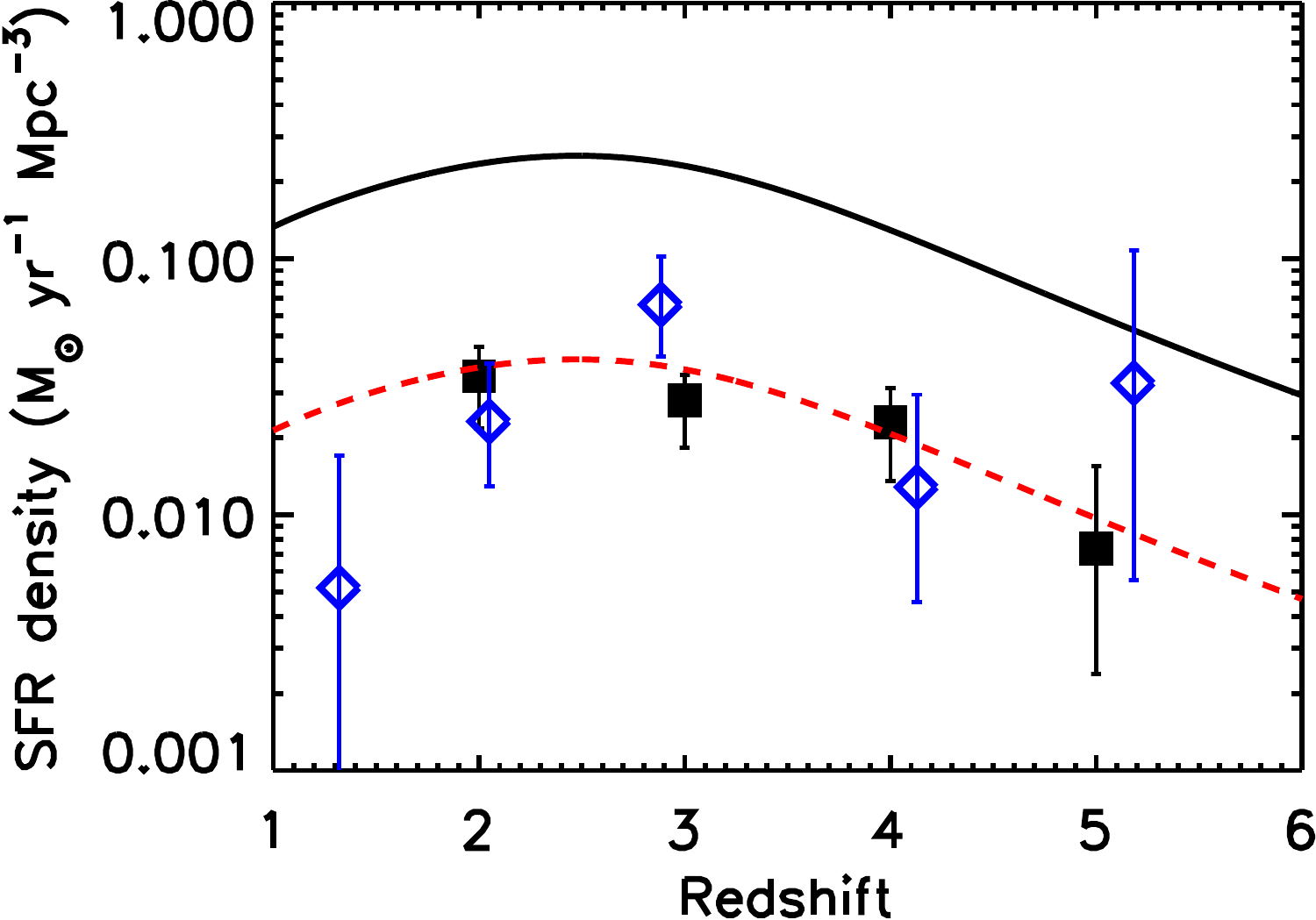}}
\caption{
SFR density per unit comoving volume vs. redshift for the SCUBA-2
sample with SFRs $>500~M_\sun$~yr$^{-1}$.
The black squares show the computations at $z=1.5-2.5$, $2.5-3.5$, $3.5-4.5$, 
and $4.5-5.5$ and are plotted at the mean redshift of each bin.
Sources without radio counterparts are placed
at their minimum millimetric redshifts, and we have renormalized
the points by a multiplicative factor of 1.1 to allow for the sources
with multiple radio counterparts. The error bars
are 68$\%$ confidence ranges based on the number of sources in the bin.
The black solid curve shows the SFR density history computed by 
Hopkins \& Beacom (2006) based on
UV selected samples, and the red dashed curve shows this
multiplied by 0.16 to match roughly the SFR density history of
the current sample. The blue open diamonds show
the SFR density history computed by Barger et al.\ (2012) based
on the smaller SCUBA sample in the GOODS-N field. We have
reduced these points by a factor of 1.4 to correspond to the present
SFR calibration.
\label{sfr_history}
}
\end{inlinefigure}

With our relatively large sample, we see a smoother evolution than
we saw in Barger et al.\ (2012), and one which closely matches the shape 
of the Hopkins \& Beacom (2006) SFR density history. The massive 
star-forming galaxies studied in this paper contain about 16\% of the SFR 
density seen in the UV selected population 
(the red dashed curve shows a scaled down version of Hopkins \& Beacom), 
though the systematic uncertainties in the SFR determinations and in the UV 
extinction corrections could easily change this by multiplicative factors of two. 
As we saw from Figure~\ref{star_formers_plus}, the contributions are
essentially disjoint, and the SMG contributions should be added to the 
extinction-corrected UV contributions.

The fraction of the total star formation in galaxies with SFRs $>500~M_\sun$~yr$^{-1}$
is substantially higher than what was found by Rodighiero et al.\ (2011)
using a {\em Herschel\/}-PACS selection, which suggests that the longer
wavelength samples of the present paper are more effective
at finding these galaxies. Indeed, only 13 of the 27 SMA detected sources 
lying in the GOODS-{\em Herschel\/} region
(which have exact positions and confirmed 850~$\mu$m fluxes)
are detected above the $4\sigma$ threshold in the PACS
100~$\mu$m image.

The combined extinction-corrected UV and submillimeter data 
in Figure~\ref{sfr_history} show that the shape of the SFR density
does not change much in the redshift range $z=1.5-5.5$, though there is
about a factor of 3 increase in the absolute normalization at $z=2$ 
relative to that at $z=5$.

\section{Discussion}
\label{secdisc}

In this discussion, our goal is to fit together the various pieces of information 
that we have presented in this paper to form a comprehensive picture.
From Figure~\ref{star_formers}, we saw a turn-down in the SFR
distribution function for SFRs above $2000~M_\sun$~yr$^{-1}$.
However, even $2000~M_\sun$~yr$^{-1}$ is an extraordinarily high
SFR, and we aim to show that this rate cannot be sustained for any long
period ($\gg 10^8$~yr) in individual galaxies
without producing too many ultra-massive galaxies overall.

Under the simple assumption that all the
SCUBA-2 galaxies have a fixed time period of star formation, $\tau$, that does
not depend on luminosity or redshift, then each
galaxy forms a mass $M=\tau~\times$~SFR. We obtain the number density
by integrating the SFR distribution function (Figure~\ref{star_formers}) over 
cosmic time and
dividing by $\tau$. To make the integration,
we assume that the shape of the SFR distribution function is fixed with redshift, 
but we allow the normalization to vary with redshift in order to match the
shape of the red dashed curve in Figure~\ref{sfr_history}.

\vskip 0.5cm
\begin{inlinefigure}
\centerline{\includegraphics[width=3.4in,angle=0]{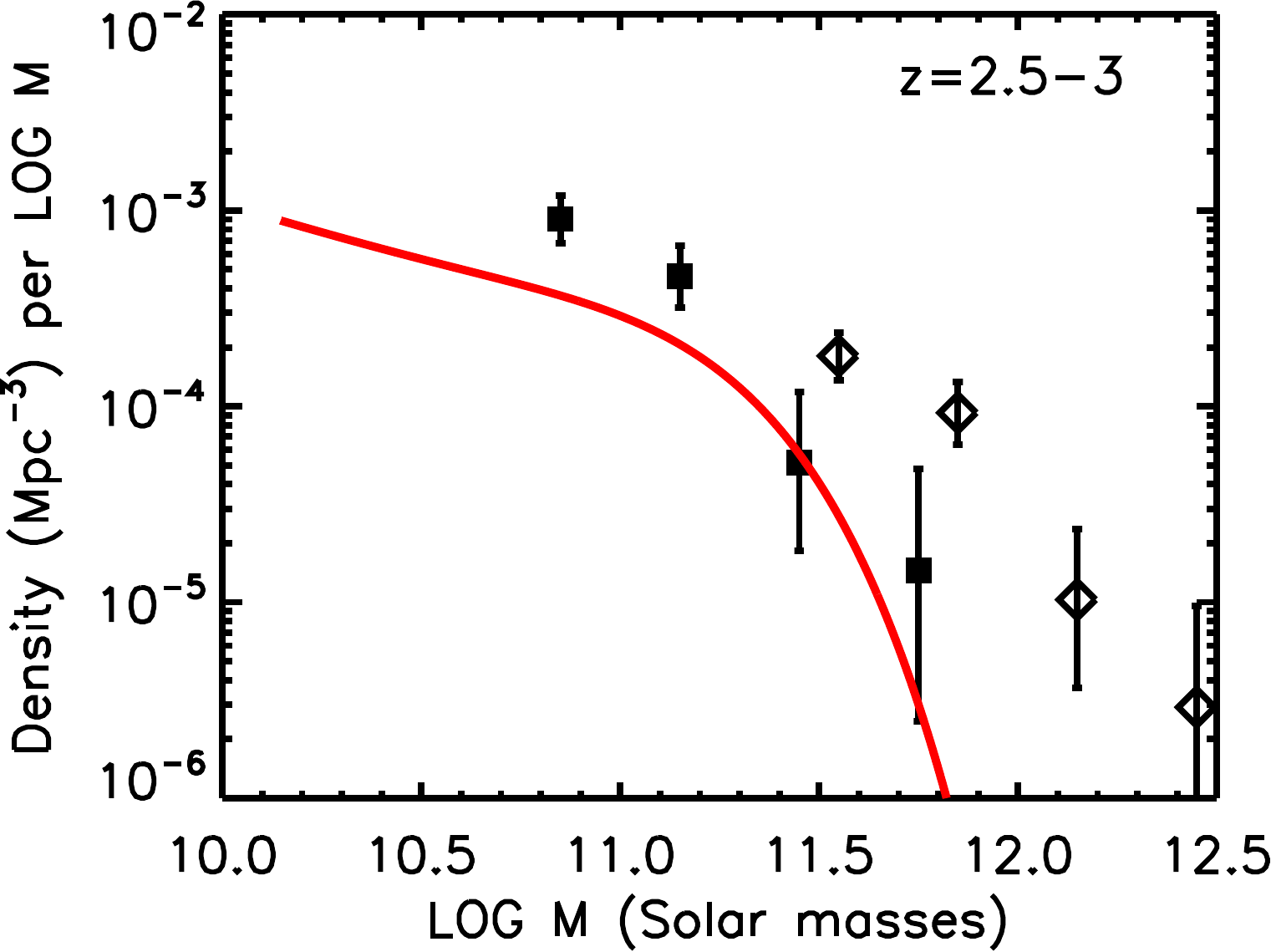}}
\caption{
Mass density per unit comoving volume per unit $\log$~M vs.
$\log$~M for $z=2.5-3$, computed from the SFR distribution function
of Figure~\ref{star_formers} assuming a fixed shape with redshift
but allowing the normalization to vary.
The open diamonds (black squares) show the values calculated 
if each SFR episode lasts for $5\times10^{8}$~yr 
($10^{8}$~yr). The red curve shows the mass distribution function
of Ilbert et al.\ (2013) in this redshift interval corrected to a Salpeter IMF.
\label{mass_den}
}
\end{inlinefigure}

In Figure~\ref{mass_den}, we plot the mass distribution function at 
$z=2.5-3$ that we predict using two values of $\tau$,
$10^8$~yr  (squares) and $5\times10^8$~yr (diamonds).
We compare these with the mass distribution function determined by
Ilbert et al.\ (2013) using the UltraVISTA DR1 data release for the same 
redshift interval, after correcting theirs to a Salpeter IMF (red curve).
Longer star-forming lifetimes than $10^8$~yr greatly over-predict the 
high mass end of the mass distribution. This is also consistent
with the measured gas reservoirs in the SMGs, which are only
large enough to sustain such high SFRs for a limited length of time. 
For example, for GN20, Hodge et al.\ (2012, 2013a) give a CO based mass of
$1.3\times10^{11}~M_\sun$ and a dynamical
mass of $5.4\times10^{11}~M_\sun$, which could only
sustain GN20's high SFR for a short period. 

Thus, there are many generations of high SFR galaxies contributing
to the SFR distribution function through the redshift
interval that are continuously switching on, forming a
massive galaxy, and then switching off. Correspondingly, there
will be a wide range of ages in the most massive galaxies at any redshift.
However, at later redshifts, the distribution of ages will be centered around 
a more mature age.

The $K-z$ relation shows that nearly
all high-redshift radio sources lie in very  massive galaxies.
The most powerful radio sources lie in galaxies with stellar
masses  $>10^{12}~M_\sun$, which must have formed at
high redshifts ($z\gg5$) (Willot03; Rocca-Volmerange et al.\ 2004). 
However, the mass dependence on radio power is extremely weak.
We find that even sources $1000-10000$ times less 
powerful in the radio than the sources considered in Willott03 must lie 
in galaxies with masses in excess of $10^{11}~M_\sun$.
Such galaxies are moderately rare at all redshifts and lie on the exponential 
tail of the mass distribution function (see Figure~\ref{mass_den}).

However, approaching this from the other direction, we also find that a substantial
fraction of high-mass galaxies contain powerful radio sources. We
show the mass versus redshift relation for galaxies in the GOODS-N
region in Figure~\ref{radio_mass}. 
We calculated the masses following Cowie \& Barger (2008) and using
a Salpeter IMF. We mark the galaxies with spectroscopic redshifts
with black squares and the galaxies with photometric
redshifts with green circles. We enclose the galaxies with high radio powers
($P_{\rm 1.4~GHz}\ge10^{31}$~erg~s$^{-1}$~Hz$^{-1}$) in red diamonds.

\vskip 0.5cm
\begin{inlinefigure}
\centerline{\includegraphics[width=3.4in,angle=180]{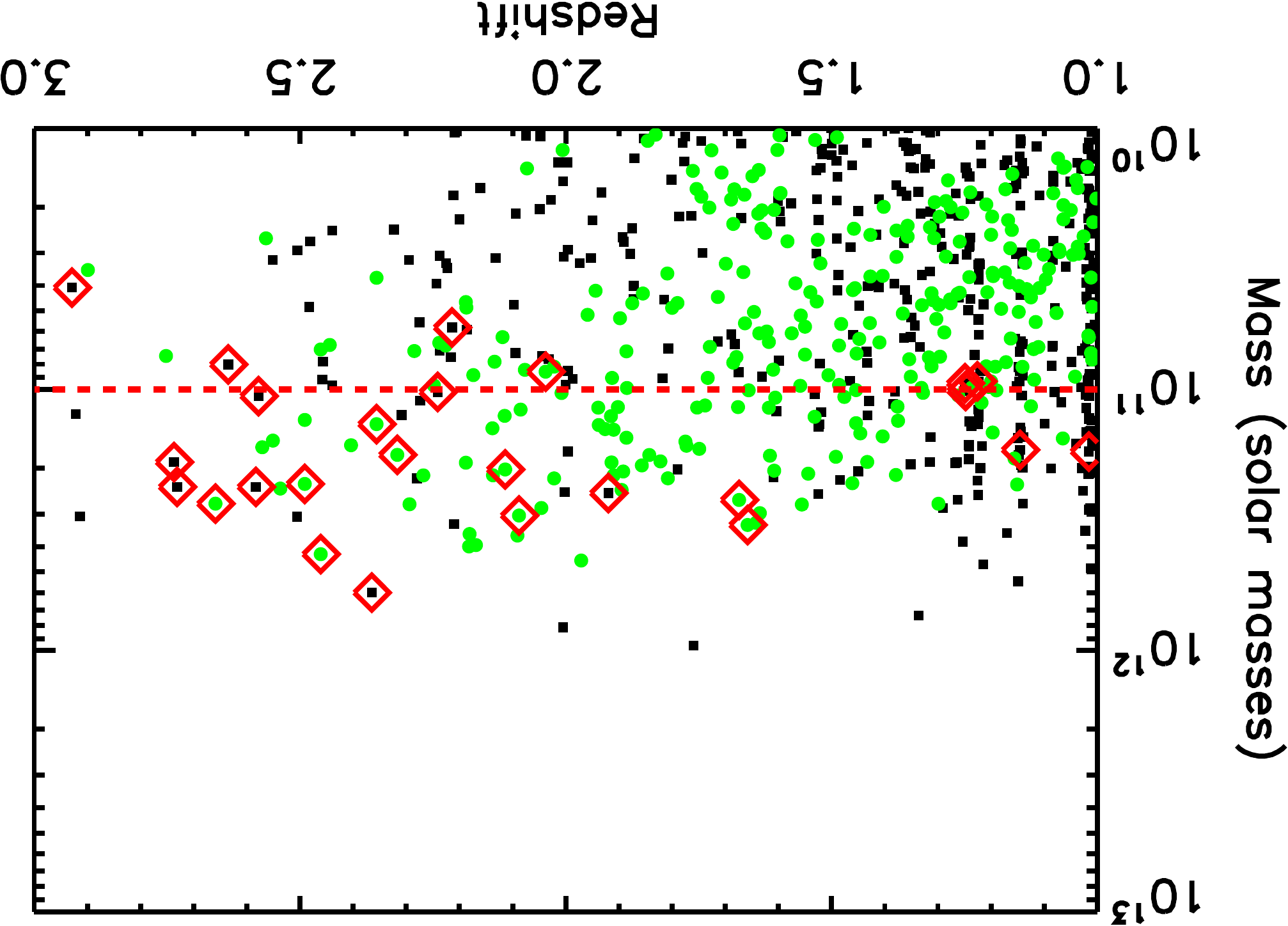}}
\caption{
Mass vs. redshift in the GOODS-N region. Black
squares denote spectroscopic redshifts, and green circles denote
photometric redshifts. The red diamonds mark high radio power 
sources ($P_{\rm 1.4~GHz}\ge10^{31}$~erg~s$^{-1}$~Hz$^{-1}$). 
The red dashed line shows the $10^{11}~M_\sun$ limit.
\label{radio_mass}
}
\end{inlinefigure}

As expected from the $K-z$ relation,
nearly all the high radio power sources have galaxy masses $>10^{11}~M_\sun$.
In the other direction, above $z=2$, $34\%\pm11\%$ of the galaxies satisfying 
this mass limit contain high radio power sources. Thus,
the integrated lifetime of the powerful radio period
must be long. At $z=2$, the age of the universe is $3.2\times10^{9}$~yr,
and in order to have one-third of the massive galaxies be powerful
radio sources, this phase (or phases) must have a lifetime in excess 
of $10^{9}$~yr.

The relative timescales of the star-forming period ($10^{8}$~yr or less)
relative to that of the radio-powerful AGN period(s)
(in excess of $10^{9}$~yr)
imply that, at least at $z\sim2$, $\sim10\%$ of the 
high radio power sources will be star formers. 
This is roughly consistent with the fraction that we determined in
Section~\ref{subsechighradio} ($13-16$\%).
Thus, after the $>10^{11}~M_\odot$ galaxies form in the short initial starburst,
they spend a substantial fraction of their subsequent lifetime as high radio
power AGNs.

\section{Summary}
\label{secsum}

In this paper, we presented an integrated SCUBA-2, SMA, and 1.4~GHz study of
a 400~arcmin$^2$ area surrounding the GOODS-N field.  Using the SCUBA-2
data, we constructed an 850~$\mu$m catalog of 49 sources to 2~mJy ($4\sigma$).
We looked for counterparts to these sources in the ultradeep (11.5~$\mu$Jy
at $5\sigma$) radio data. In cases where there were multiple radio counterparts, 
we often were able to use new and existing SMA data to determine the correct 
counterparts (the correct radio counterparts to only three SMGs remain uncertain).  
Only five SMGs have no 
radio counterparts, making them extremely high redshift galaxy candidates.

We either obtained ourselves or located in the literature extensive spectroscopic 
redshifts for the radio sources in the field. In the GOODS-N proper, the redshift 
identifications are highly complete to $K_s=21$ after including a small number
of photometric redshifts. 
For the SMGs without spectroscopic, CO, or photometric redshifts,
we used an Arp~220 based model from Barger et al.\ (2000) to 
measure millimetric redshifts from the 1.4~GHz to 860~$\mu$m flux ratios.
The millimetric redshifts predominantly fill in the $z\sim2.5-4$ range.

We found an extended tail of radio sources
with faint optical/NIR counterparts, the faintest of which are undetected even in the 
{\em HST\/} ACS images.  These sources are not identifiable with optical/NIR
spectroscopy or photometry and may lie at high redshifts.  Indeed, we found that
there is a strong correlation between $K_s$ magnitude and redshift in the radio
sample (the $K-z$ relation), making it possible to use the $K_s$ magnitudes as a 
crude redshift estimator for the radio sources.

We computed rest-frame radio power for the radio sources with spectroscopic, CO, or
photometric redshifts.  At $z\gtrsim3$, even these ultradeep observations are only 
sensitive to sources brighter than the ULIRG limit calculated assuming the FIR-radio
correlation.

We are particularly interested in the high radio power 
($P_{\rm 1.4~GHz}\ge10^{31}$~erg~s$^{-1}$~Hz$^{-1}$)
sources at high redshifts, as those at $z<1.5$ mostly appear to be AGN powered.
At $z>1.5$, a substantial fraction (37\%) of the spectroscopically identified
high radio power sources are detected in
the submillimeter, suggesting that they are massive star formers. However, it is difficult to 
determine the true fraction of high radio power star formers, because there are strong 
selection effects in the spectroscopic identifications of the radio sources at high redshifts.
Based on the $K-z$ relation, the unidentified radio sources at $K_s>21$ should
lie at high redshifts. Using the 850~$\mu$m signal-to-noise ratio for the high radio
power sources at these magnitudes, we found a likely star-forming fraction of $13-16$\%. 

We computed SFRs for the individual sources from the 1.4~GHz power, assuming
that the FIR-radio correlation is roughly invariant to $z\sim6$ for SMGs down to
our 850~$\mu$m flux threshold of 2~mJy, and from the submillimeter fluxes, assuming
using an SFR conversion computed from the average SEDs of isolated galaxies
in the sample. (The SFR conversion is quite close to that which would be computed from Arp 220.)  
We found that the SFRs derived from the two methods are in good agreement for
the sources with spectroscopic, CO, or photometric redshifts $z>1.5$, though the
submillimeter derived SFRs are about 5\% higher, on average, than the radio
derived SFRs. This is well within the uncertainties in the absolute calibration of
the SFRs.

We found that at high radio powers, the submillimeter detected sources are quite
distinct from the other radio sources. A small number of these had
high-resolution radio observations that showed them to be predominantly extended or
star formation dominated, so we assumed that they were all star formation
dominated.

We found that the SFRs of the SMGs ranged from $400~M_\odot$~yr$^{-1}$ to
$\sim6000~M_\odot$~yr$^{-1}$. We constructed the SFR distribution function
for the SMGs at $z>1.5$ with
SFRs~$>500~M_\odot$~yr$^{-1}$ and found a characteristic maximum SFR of
$\sim2000~M_\odot$~yr$^{-1}$.  It should be emphasized that while we only
have spectroscopic redshifts for about 40$\%$ of the sources, and the
remaining sources have relatively uncertain redshifts primarily based on the
radio to submillimeter flux, the results are very insensitive to even fairly large
errors in the redshifts. Because the conversion from 850~$\mu$m flux depends
extremely weakly on redshift at $z>1.5$, the result would only change if an
SMG was moved outside this redshift interval by the redshift uncertainty.

We compared our submm results with extinction-corrected
UV selected samples and saw that the LBGs do not have as high of SFRs as the SMGs
but instead cut off at $\sim300~M_\odot$~yr$^{-1}$. Thus, the two samples are
essentially disjoint.

We constructed the SFR density history for the SMG sample and compared it with 
the extinction-corrected UV selected SFR density history
compilation of Hopkins \& Beacom (2006) over the
redshift range $z=1-6$. The shapes closely match, with the SMG SFR density
history being about 16\% of the extinction-corrected UV selected SFR density history.
However, since the samples are disjoint, the SMG contributions and the
extinction-corrected UV contributions should be added for a fuller accounting 
of the overall star formation history.

Finally, we discussed how the above information could be put together to form a 
comprehensive picture. We concluded that nearly all high radio power
sources have galaxy masses $>10^{11}~M_\odot$ and that in order to avoid 
over-predicting the high galaxy mass end, there must be many generations
of high SFR galaxies that are continuously switching on, forming a massive galaxy in
a period of $<10^8$~yr, and then switching off. However, the powerful radio
period lasts much longer ($>10^9$~yr), making the high radio power sources without
submillimeter counterparts the most common type of
high radio power source, even at high redshift.

\acknowledgements

We thank the anonymous referee for a thoughtful report.
We gratefully acknowledge support from
the University of Wisconsin Research Committee with funds 
granted by the Wisconsin Alumni Research Foundation and the 
David and Lucile Packard Foundation (A.~J.~B.),
NSF grants AST-1313150 (A.~J.~B.), AST-0709356 (L.~L.~C., C.-C.~C.), 
and AST-1313309 (L.~L.~C.), and 
National Science Council of Taiwan grant
102-2119-M-001-007-MY3 (W.-H.~W.).
C.~M.~C. was generously supported by a Hubble Fellowship provided 
by Space Telescope Science Institute, grant HST-HF-51268.01-A.
We acknowledge the cultural significance that the summit of 
Mauna Kea has to the indigenous Hawaiian community.


\newpage
\clearpage
\begin{deluxetable*}{lccrrrcccccrrrrr}
\renewcommand\baselinestretch{1.0}
\tablewidth{0pt}
\tablecaption{SCUBA-2 Sample}
\scriptsize
\tablehead{\# & R.A. & Decl. &  Flux  & Error & S/N & SMA & Flux & R.A. & Decl. & $K_{s}$ & z$_{spec}$ & z$_{phot}$ & z$_{milli}$\\ 
& J2000.0 & J2000.0 & 850~$\mu$m & 850~$\mu$m & & 860~$\mu$m & 1.4~GHz & J2000.0 & J2000.0 & & & & \\ 
& ($^{\rm h}~^{\rm m}~^{\rm s}$) & ($^\circ~'~"$) & (mJy) & (mJy) & & (mJy) & ($\mu$Jy) & ($^{\rm h}~^{\rm m}~^{\rm s}$) & ($^\circ~'~"$) & (AB) & & & \\
(1) & (2) & (3) & (4) & (5) & (6) & (7) & (8) & (9) & (10) & (11) & (12) & (13) & (14)}
\startdata
1    &   12   35   55.53    &   62   22  36.56  &   16.95      &   3.45   
 &    4.92     &   17.0(1.9)        &  58.5(7.6)  &        12       35 55.88  & 
      62       22 39.0  &  22.12  &  \nodata  &     &  3.624 \cr 
2    &   12   35   51.51    &   62   21  46.07  &   16.90      &   3.16   
 &    5.35     &   13.7(2.8)        &  51.9(7.3)  &        12       35 51.37  & 
      62       21 47.2  &  22.11  &  \nodata  &     &  3.447 \cr 
3  &   12   37   11.92    &   62   22  09.31  &   16.41      & 
 2.26     &    7.27     &   23.9(2.5)        &  87.50(15.)  & 
      12       37 11.89  &        62       22 12.7  &  22.69  &  4.055  &  4.125
 &  3.487 \cr 
4  &   12   37   30.53    &   62   12  59.30  &   12.21    
 &   0.77     &   15.81     &   14.9(0.9)        &  126.0(4.9)  & 
      12       37 30.80  &        62       12 58.7  &  22.92  &  \nodata  & 
2.627  &  2.504 \cr 
5    &   12   35   58.96    &   62  06  07.16  &   11.90      &   2.83   
 &    4.21     &                    &  $<$7.5  &    &    &     &    &     &    \cr 
6    &   12   35   46.74    &   62   20  10.76  &   11.13      &   2.74   
 &    4.06     &                    &  53.40(12.)  &        12       35 47.06
 &        62       20 9.70  &  18.97  &  0.543  &  0.568   &        \cr 
         &                      &              &              &            & 
           &                    &  85.7(6.7)  &        12       35 46.64  & 
      62       20 13.3  &  22.75  &  \nodata  &     &        \cr 
7    &   12   35   48.70    &   62   19  05.41  &   10.27      &   2.43   
 &    4.23     &                    &  71.3(6.4)  &        12       35 48.84  & 
      62       19 4.91  &  22.85  &    &     &  2.792 \cr 
8    &   12   36   26.98    &   62  06  06.27  &    9.46      &   2.29   
 &    4.13     &   11.5(0.7)        &  34.20(2.9)  &        12       36 27.21
 &        62 06 5.59  &  24.84  &  \nodata  &     &  3.697 \cr 
9    &   12   36   28.16    &   62   21  38.46  &    9.09      &   2.20   
 &    4.13     &                    &  35.60(3.2)  &        12       36 28.67
 &        62       21 39.7  &  20.39  &  1.194  &  1.191   &  3.160 \cr 
10   &   12   35   50.72    &   62   10  41.78  &    8.69      &   2.09   
 &    4.16     &                    &  25.00(3.0)  &        12       35 50.35
 &        62       10 41.9  &  23.60  &    &     &  3.638 \cr 
11 &   12   36   33.41    &   62   14  08.10  &    8.24    
 &   0.63     &   13.06     &   12.0(1.4)        &  33.20(5.6)  & 
      12       36 33.38  &        62       14 8.40  &  26.59  &  4.042  & 
\nodata  &  3.373 \cr 
12   &   12   36   31.86    &   62   17  12.97  &    7.84      &   0.71   
 &   10.99     &   7.1(0.5)         &  21.60(2.5)  &        12       36 31.94
 &        62       17 14.5  &  23.04  &  \nodata  &  3.757  &  3.620 \cr 
13  &   12   37   11.29    &   62   13  29.89  &    7.74      & 
 0.65     &   11.89     &   6.7(0.6)         &  123.8(5.4)  & 
      12       37 11.34  &        62       13 30.9  &  20.45  &  1.995  & 
1.801  &  1.893 \cr 
14 &   12   37  07.11  &   62   14  09.08  &    7.62      & 
 0.57     &   13.27     &   7.1(1.4)         &  58.40(11.)  & 
      12       37 7.210  &        62       14 7.91  &  21.44  &  2.490  &  2.451
 &  2.488 \cr 
15   &   12   36   34.70    &   62   19  21.37  &    7.13      &   0.98   
 &    7.29     &   (a) 7.2(1.7)     & 63.0(2.4)   &  12  36 34.92 &  62 19 23.6 &  21.74   &    &     &  2.568  \cr 
         &                      &              &              &            & 
           &   (b) 6.8(1.7)     & 60.0(2.4)   &  12 36 34.63  &62 19 26.0    &  19.30   & 0.871   & 0.839    & 2.560   \cr 
16  &   12   36   18.24    &   62   15  48.98  &    7.12      & 
 0.77     &    9.25     &   7.2(0.7)         &  163.6(4.9)  & 
      12       36 18.35  &        62       15 50.4  &  22.02  &  2.000  & 
2.508  &  1.333 \cr 
17 &   12   36   51.83    &   62   12  25.79  &    6.04    
 &   0.55     &   11.01     &   7.8(1.0)         &  12.50(2.4)  & 
      12       36 52.03  &        62       12 25.9  &  99.00  &  5.183  & 
\nodata  &  4.277 \cr 
18   &   12   37   41.20    &   62   12  21.21  &    5.74      &   0.99   
 &    5.77     &   7.1(1.8)         &  27.40(3.1)  &        12       37 41.18
 &        62       12 21.0  &  23.36  &  \nodata  &  3.029    &  3.396 \cr 
19   &   12   36   44.15    &   62   19  37.19  &    5.66      &   0.97   
 &    5.82     &   6.4(1.1)         &  25.40(2.9)  &        12       36 44.08
 &        62       19 38.7  &  23.88  &  \nodata  &     &  3.059 \cr 
20 &   12   36   45.68    &   62   14  49.61  &    5.44    
 &   0.53     &   10.22     &   (a) 4.2(0.8)     &  101.0(2.5)  & 
      12       36 46.08  &        62       14 48.5  &  22.43  &  \nodata  & 
3.167  &  1.517 \cr 
         &                      &              &              &            & 
           &   (b) 5.3(1.1)     &  30.80(2.3)  &        12       36 44.03  & 
      62       14 50.5  &  21.52  &  2.095  &  1.754  &  2.819 \cr 
21  &   12   36   22.26    &   62   16  22.37  &    5.16      & 
 0.76     &    6.75     &   9.3(1.4)         &  26.10(6.1)  & 
      12       36 22.10  &        62       16 15.9  &  23.92  &  \nodata  & 
\nodata  &  3.600 \cr 
22 &   12   37  01.35  &   62   11  46.10  &    5.04      & 
 0.65     &    7.79     &   4.8(0.6)         &  95.20(5.6)  & 
      12       37 1.593  &        62       11 46.4  &  20.57  &  1.73  &  1.868
 &  1.791 \cr 
23   &   12   36   27.57    &   62   12  18.03  &    4.89      &   0.70   
 &    6.98     &                    &  17.50(2.5)  &        12       36 27.55
 &        62       12 17.9  &  24.74  &  \nodata  &  \nodata  &  3.536 \cr 
24   &   12   37   19.11    &   62   12  17.80  &    4.87      &   0.78   
 &    6.25     &                    &  16.00(2.7)  &        12       37 18.96
 &        62       12 17.5  &  23.13  &  \nodata  &  \nodata  &    \cr 
         &                      &              &              &            & 
           &                    &  12.60(2.5)  &        12       37 19.59  & 
      62       12 20.7  &  21.89  &  \nodata  &    &    \cr 
25   &   12   37   12.41    &   62   10  37.76  &    4.66      &   0.74   
 &    6.32     &                    &  23.00(2.4)  &        12       37 12.48
 &        62       10 35.6  &  23.51  &  \nodata  &  \nodata  &  3.029 \cr 
26   &   12   37  02.38  &   62   14  27.82  &    4.62      &   0.53     & 
  8.69     &                    &  18.20(2.5)  &        12       37 2.600  & 
      62       14 26.9  &  22.25  &  3.214  &  2.949  &  3.340 \cr 
27   &   12   37   11.95    &   62   12  11.66  &    4.62      &   0.74   
 &    6.25     &   5.3(0.9)         &  33.00(2.4)  &        12       37 12.07
 &        62       12 11.9  &  22.07  &  2.914  &  2.738  &  2.633 \cr 
28 &   12   37   13.92    &   62   18  25.38  &    4.49      & 
 0.82     &    5.45     &                    &  625.8(4.8)  & 
      12       37 13.89  &        62       18 26.2  &  24.41  &  \nodata  & 
\nodata  &  0.745 \cr 
29  &   12   36   16.30    &   62   15  15.32  &    4.38      & 
 0.77     &    5.72     &   3.4(0.6)         &  36.90(2.6)  & 
      12       36 16.10  &        62       15 13.8  &  22.24  &  2.578  &  2.940
 &  2.344 \cr 
30   &   12   36   12.18    &   62   10  30.18  &    4.34      &   0.97   
 &    4.46     &                    &  26.50(5.3)  &        12       36 11.51
 &        62       10 33.8  &  21.49  &  \nodata  &  2.461  &  2.531 \cr 
31 &   12   36   36.50    &   62   11  57.93  &    4.04      & 
 0.70     &    5.74     &                    &  24.30(6.2)  & 
      12       36 36.14  &        62       11 54.3  &  22.87  &  \nodata  & 
0.628  &  2.712 \cr 
32 &   12   36   28.85    &   62   10  45.69  &    4.04      & 
 0.78     &    5.15     &   7.7(0.9)         &  48.10(8.2)  & 
      12       36 28.90  &        62       10 45.2  &  22.96  &  \nodata  &   
 &  2.766 \cr 
33 &   12   36   58.67    &   62  09  30.47  &    4.00      & 
 0.76     &    5.28     &   4.6(0.6)         &  27.50(2.6)  & 
      12       36 58.55  &        62 09 31.4  &  24.13  &  \nodata  &  \nodata
 &  2.915 \cr 
34 &   12   36   21.20    &   62   17  07.02  &    4.00    
 &   0.79     &    5.05     &    (a) 3.4(0.6)    &  164.6(5.0)  & 
      12       36 21.28  &        62       17 8.49  &  21.99  &  \nodata  &     & 
1.402 \cr 
         &                      &              &              &            & 
           &    (b) 3.5(0.7)    &  44.90(2.7)  &        12       36 20.98  & 
      62       17 9.80  &  21.64  &  1.988  & 2.016   &  2.086 \cr 
35   &   12   36   36.77    &   62  08  52.67  &    3.89      &   0.86   
 &    4.54     &                    &  79.70(5.1)  &        12       36 37.03
 &        62 08 52.4  &  22.01  &  \nodata  &  2.139  &  1.868 \cr 
36   &   12   36   58.36    &   62   14  51.60  &    3.83      &   0.53   
 &    7.28     &                    &  30.40(2.6)  &        12       36 57.82
 &        62       14 55.0  &  19.78  &  0.849  &  0.822  &  2.137 \cr 
37 &   12   36   56.35    &   62   12  05.29  &    3.82    
 &   0.59     &    6.51     &    4.5(0.8)        &  25.60(6.4)  & 
      12       36 55.85  &        62       12 1.29  &  21.52  &  2.737  &  1.904
 &  2.583 \cr 
38   &   12   37   11.71    &   62   11  19.34  &    3.75      &   0.75   
 &    5.03     &                    &   $<$6.8 &    &    &     &    &     &    \cr 
39   &   12   36   31.07    &   62  09  57.24  &    3.68      &   0.80   
 &    4.58     &                    &  148.8(4.8)  &        12       36 31.27
 &        62 09 57.6  &  21.74  &  \nodata  &  \nodata  &  1.445 \cr 
40   &   12   37   19.08    &   62   10  21.32  &    3.62      &   0.78   
 &    4.61     &                    &  23.80(2.5)  &        12       37 19.55
 &        62       10 21.2  &  24.10  &  \nodata  &     &  2.743 \cr 
41   &   12   36   24.05    &   62   10  13.63  &    3.56      &   0.82   
 &    4.36     &                    &  38.50(2.6)  &        12       36 24.32
 &        62       10 17.1  &  24.31  &  \nodata  &  \nodata  &  2.154 \cr 
42   &   12   36   35.69    &   62   14  29.29  &    3.49      &   0.61   
 &    5.68     &                    &  82.40(4.8)  &        12       36 35.59
 &        62       14 24.0  &  20.16  &  2.005  &  2.046  &    \cr 
         &                      &              &              &            & 
           &                    &  37.70(4.8)  &        12       36 35.87  & 
      62       14 36.0  &  21.11  &  1.018  &  1.151  &    \cr 
43   &   12   36  08.51  &   62   14  36.20  &    3.48      &   0.84     & 
  4.14     &                    &  41.20(2.5)  &        12       36 8.598  & 
      62       14 35.4  &  22.66  &  \nodata  &  \nodata  &    \cr 
         &                      &              &              &            & 
           &                    &  22.00(2.7)  &        12       36 8.920  & 
      62       14 30.7  &  21.68  &  0.849  &  0.822  &    \cr 
44  &   12   37  00.09  &   62  09  06.98  &    3.45      & 
 0.80     &    4.29     &   3.1(0.6)         &  297.2(5.0)  & 
      12       37 0.270  &        62 09 9.70  &  21.85  &  \nodata  &  \nodata
 &  0.892 \cr 
45   &   12   36   34.42    &   62   12  39.14  &    3.04      &   0.66   
 &    4.65     &                    &  185.8(5.0)  &        12       36 34.51
 &        62       12 40.9  &  19.98  &  1.224  &  1.247  &  1.097 \cr 
46   &   12   36   34.81    &   62   16  26.06  &    2.82      &   0.69   
 &    4.07     &                    &  52.70(6.2)  &        12       36 34.87
 &        62       16 28.4  &  19.80  &  0.847  &  0.893  &  1.911 \cr 
47   &   12   36   48.70    &   62   12  16.73  &    2.63      &   0.58   
 &    4.57     &                    &  22.30(2.8)  &        12       36 48.62
 &        62       12 15.7  &  20.97  &  \nodata  &  2.238  &  2.437 \cr 
48   &   12   37  09.53  &   62   14  35.73  &    2.51      &   0.61     & 
  4.15     &                    & $<6.2$   &    &    &     &    &     &    \cr 
49   &   12   36   57.15    &   62   14  08.71  &    2.00      &   0.50   
 &    4.02     &                    &  28.20(5.2)  &        12       36 57.40
 &        62       14 7.80  &  21.07  &  \nodata  &  1.461  &  1.966 \cr 
\enddata
\label{tab1}
\tablecomments{In order for the table to fit on the page, some changes from
the journal version had to be made.
The catalog names in Column~1 should all begin with the prefix ``CDFN''.  The alternative
names are as follows:  CDFN3 (GN20); CDFN4 (850-6); CDFN11 (850-5/GN10);
CDFN13 (850-36); CDFN14 (850-9/GN19); CDFN16 (850-3/GN6); CDFN17 (850-1/GN14);
CDFN20 (850-11/GN12); CDFN21 (850-2/GN9); CDFN22 (GN17); CDFN28 (GN4);
CDFN29 (850-7); CDFN31 (GN11); CDFN32 (850-17); CDFN33 (850-16a);
CDFN34 (850-15/GN7); CDFN37 (850-12/GN15); CDFN44 (850-16b/GN16).
In Column~13, the photometric redshifts for sources CDFN6, CDFN9, CDFN14,
CDFN15(b), CDFN18, CDFN26, CDF34(b), CDFN35, and CDFN42
are from Rafferty et al.\ (2011). 
}
\end{deluxetable*}
\clearpage

\begin{center}
\begin{deluxetable*}{cccccccc}
\setcounter{table}{2}
\tablecaption{SMA Sources}
\tabletypesize{\scriptsize}
\tablehead{
{Name} & SMA R.A. & SMA Decl. & SMA & SMA Ref. & VLA & $z_{\rm spec}$ & Redshift Ref. \\ 
&  J2000.0 & J2000.0 & 860~$\mu$m & & 1.4~GHz &  \\
& (deg) & (deg) & (mJy) & & (mJy) &  & \\
(1)& (2) & (3) & (4)  & (5) & (6) & (7) & (8)}
\startdata
GN20 & 189.29875  & 62.37000     & $23.9 \pm2.5$    &   Iono et al.\ (2006) & $87.5 \pm15.5$    &  4.055 & Daddi et al.\ (2009b)  \cr
CDFN1 & 188.98271 & 62.37761 & $17.0\pm1.9$ & this paper & $56.6\pm7.6$ & \nodata & \nodata \cr
850-6  & 189.37833  &   62.21639   &  $14.9 \pm0.9$    &  Barger et al.\ (2012) & $126.0   \pm4.9$      & \nodata & \nodata \cr
CDFN2
& 188.96399  &   62.36314  &   $13.7 \pm2.8$   & this paper  &   $51.9  \pm 7.3$    & \nodata & \nodata  \cr
850-5  & 189.13937  &   62.23575  &   $12.0 \pm1.4$    &  Wang et al.\ (2007) & $33.2   \pm5.6$     & 4.042 & Daddi et al.\ (2009a) \cr
CDFN5  & 189.11333  &   62.10153  &   $11.5 \pm0.7$  &   this paper &   $34.2   \pm2.9$     & \nodata & \nodata  \cr
850-2  & 189.09212   &  62.27103   &  $ 9.3 \pm1.4$    &  Barger et al.\ (2012) &  $26.1   \pm6.1$     & \nodata & \nodata  \cr
850-1  & 189.21661  &   62.20716    &  $7.8 \pm1.0$   &  Cowie et al.\ (2009) & $12.5   \pm2.4$     & 5.183 & Walter et al.\ (2012) \cr
850-17 & 189.12018  &   62.17925  &    $7.7 \pm0.9$  &  Barger et al.\ (2012)  & $48.5   \pm9.4$    &  \nodata & \nodata  \cr
850-3  & 189.07637   &  62.26411    &  $7.2 \pm0.7$  &  Barger et al.\ (2012) &  $163.6   \pm4.9$    &  2.000 & Pope et al.\ (2008) \cr
CDFN15a & 189.14693  &   62.32248   &   $7.2 \pm1.7$   & this paper  &  \nodata  & \nodata &   \nodata \cr
CDFN10 & 189.13301  &   62.28742    &  $7.1 \pm0.5$   &   this paper &  $21.6   \pm2.5$     & \nodata & \nodata  \cr
CDFN18 & 189.42142  &   62.20567   &   $7.1\pm1.8$  & this paper & $27.4 \pm2.7$ & \nodata & \nodata  \cr
850-9 & 189.28004  &   62.23564    &  $7.1 \pm1.4$  &  Barger et al.\ (2012) &  $25.9   \pm2.5$     & 2.490  & Swinbank et al.\ (2004) \cr
CDFN15b & 189.14975  &   62.32236   &   $6.9\pm1.7$ &  this paper  & \nodata & \nodata & \nodata \cr
850-36 & 189.29716  &   62.22533   &  $6.7 \pm0.6$  &  this paper &  $123.8   \pm5.4$     & 1.995 & Bothwell et al.\ (2010) \cr 
CDFN13 & 189.18367  &   62.32722   &   $6.4 \pm1.1$  & this paper   &  $25.4   \pm2.9$     & \nodata & \nodata  \cr
850-13c & 189.30000  &   62.20341   &   $5.3 \pm0.9$   & Wang et al.\ (2011) &   $33.0   \pm2.4$    &  2.914 & Chapman et al.\ (2005) \cr
850-11b & 189.18324  &   62.24742   &   $5.3 \pm1.1$   &  Wang et al.\ (2011) & $30.8   \pm2.3$     & 2.095 & Reddy et al.\ (2006)  \cr
GN17 & 189.25667   &  62.19611             &   $4.8 \pm0.6$  &  this paper & $95.2  \pm5.6$     & 1.73  & Pope et al.\ (2008) \cr
850-16a & 189.24390  &   62.15878   &   $4.6 \pm0.6$    &   this paper & $27.5   \pm2.6$     & \nodata & \nodata \cr  
850-12 & 189.23300  &   62.20053  &    $4.5 \pm0.8$       &  Barger et al.\ (2012) & $25.6   \pm6.4$     & 2.737 & Barger et al.\ (2008)  \cr
850-11a & 189.19205  &   62.24683  &   $ 4.2 \pm0.8$    &  Wang et al.\ (2011) & $101.0   \pm2.5$   &  \nodata & \nodata \cr
850-13b & 189.30943  &   62.20225   &  $ 4.1 \pm0.7$   &   Wang et al.\ (2011) & $25.5   \pm2.6$    &  3.157 & Barger et al.\ (2008) \cr
850-15b & 189.08875  &   62.28558  &    $3.5 \pm0.7$    &  Barger et al.\ (2012) & $164.6   \pm5.0$   &   1.992 &  Swinbank et al.\ (2004)  \cr
850-15a & 189.08792  &   62.28600   &   $3.4 \pm0.6$   &   Barger et al.\ (2012) & $44.9   \pm2.7$    &  \nodata & \nodata \cr
850-7 & 189.06712   &  62.25383  &    $3.4 \pm0.6$       &  Barger et al.\ (2012) & $36.9   \pm2.6$     & 2.578  & Chapman et al.\ (2005) \cr
850-13a & 189.30847 &    62.19900   &   $3.2 \pm0.9$     &  Wang et al.\ (2011) & $22.1   \pm2.5$   &   \nodata & \nodata  \cr
850-16b & 189.25125 &    62.15275   &   $3.1 \pm0.6$  &  this paper &  $297.2   \pm5.0$    &  \nodata & \nodata
\enddata
\label{tab2}
\end{deluxetable*}
\end{center}

\end{document}